\documentclass[prd,12pt,nofootinbib,preprint]{revtex4}

\usepackage[T1]{fontenc}
\usepackage{amsmath,amssymb}
\usepackage{epsfig}
\usepackage{url}
\usepackage{subfigure}
\usepackage{graphicx}
\usepackage{grffile}
\usepackage[usenames,dvipsnames]{color}
\usepackage{slashed}
\usepackage{array}
\usepackage{multirow}
\usepackage[colorlinks,citecolor=blue]{hyperref}
\usepackage{color}
\usepackage{cancel}
\usepackage{gensymb}
\usepackage{soul}

\def \met{\slashed{E}_T }

\def\a {\alpha}
\def\b {\beta}

\def\l {\lambda}

\def\bar {\overline}

\def\be {\begin{equation}}
\def\ee {\end{equation}}
\def\beq {\begin{equation}}
\def\eeq {\end{equation}}
\def\bea {\begin{eqnarray}}
\def\eea {\end{eqnarray}}

\newcommand{\besub}{\begin{subequations}}
\newcommand{\eesub}{\end{subequations}}

\def\beq{\begin{equation}}
\def\eeq{\end{equation}}
\def\barr{\begin{array}}
\def\earr{\end{array}}

\begin{document}
\title{Probing the $H^\pm W^\mp Z$ interaction at the high energy upgrade of the LHC}

\author{Amit Adhikary}
\email{amitadhikary@iisc.ac.in}
\affiliation{Centre for High Energy Physics, Indian Institute of Science, C.V.Raman Avenue,
Bangalore 560012, India}

\author{Nabarun Chakrabarty}
\email{chakran@iisc.ac.in, nabarunc@iitk.res.in}
\affiliation{Centre for High Energy Physics, Indian Institute of Science, C.V.Raman Avenue,
Bangalore 560012, India, Department of Physics, Indian Institute of Technology Kanpur, Kanpur-208016, Uttar Pradesh, India}

\author{Indrani Chakraborty}
\email{indranic@iitk.ac.in}
\affiliation{Department of Physics, Indian Institute of Technology Kanpur, Kanpur-208016, Uttar Pradesh, India} 

\author{Jayita Lahiri}
\email{jayitalahiri@rnd.iitg.ac.in
}
\affiliation{Regional Centre for Accelerator-based Particle Physics, Harish-Chandra Research Institute, HBNI,
Chhatnag Road, Jhunsi, Allahabad - 211019, India, Department of Physics, Indian Institute of Technology Guwahati, North Guwahati,  Assam - 781039, India} 

\begin{abstract} 
An $H^\pm W^\mp Z$ interaction at the tree level is a common feature of new physics models that feature scalar triplets.
In this study, we aim to probe the strength of the aforementioned interaction in a model-agnostic fashion  at the futuristic 27 TeV proton-proton collider. We assume that the $H^\pm$ couples dominantly to ($W^\pm,Z$) and ($t,b$) and
specifically study the processes that involve the $H^\pm W^\mp Z$ vertex at the production level, that is, $p p \to H^\pm j j$ and $p p \to Z H^\pm$. Moreover, we look into both $H^\pm \to W^\pm Z,~t b$ decays for either production process. Our investigations reveal that the $H^\pm j j$ production process has a greater reach compared to $Z H^\pm$. Moreover, the discovery potential of a charged Higgs improves markedly with respect to the earlier studies corresponding to lower centre-of-mass energies. Finally, we recast our results in the context of the popular Georgi-Machacek model. 
\end{abstract} 
\maketitle

\section{Introduction} 

With the discovery of a Higgs boson of mass 125 GeV \cite{Aad:2012tfa,Chatrchyan:2012ufa}, the particle spectrum of the Standard Model (SM) is complete. Moreover, the properties
of the discovered boson are found to be increasingly consistent with that of the SM Higgs. Despite this success, the SM remains far from being the complete framework. Certain issues on both theoretical and experimental fronts advocate for additional dynamics beyond the SM. Interestingly, extending just the scalar sector of the SM can suffice to address such issues. This puts forth extended Higgs sectors as prototypes of 
new physics (NP). 

A subset of extended Higgs sectors features additional $SU(2)_L$ scalar multiplets and all of them predict at least one singly charged scalar $H^+$.
While the tree level fermionic couplings of $H^+$ are generally proportional to the fermion mass for all $SU(2)_L$ representations, it is the coupling
to the gauge bosons that can differ. For instance, the $H^+ W^- Z$ interaction vanishes at the tree level for a two-Higgs doublet model (2HDM)~\cite{Deshpande:1977rw,Branco:2011iw,PhysRevD.22.1725,Gunion:1989we}. This
coupling is generated at one-loop~\cite{Kanemura:1997ej,Kanemura:1999tg,Abbas:2018pfp,Arhrib:2007rm} and hence is generally small in magnitude. And this is true regardless the number of such scalar doublets. 
On the contrary, the aforementioned coupling is non-zero at the tree level itself in case of $SU(2)_L$ scalar triplets~\cite{PhysRevD.22.2227,Magg:1980ut,Lazarides:1980nt,PhysRevD.22.2860} and is proportional to the
vacuum expectation value (VEV) acquired by the neutral component of the triplet. The triplet VEV is however tightly constrained by measurements of the $\rho$-parameter for the simplistic Higgs triplet model (HTM) (that employs a complex scalar triplet over and above the doublet) thereby obscuring the observability of the $H^+ W^- Z$ coupling. That said, there exist non-minimal extensions of the HTM where this problem can be circumvented. These involve combining a real scalar triplets with a complex ones in a custodially symmetric manner such that the $\rho$-parameter is unity for all values of the triplet VEV. The most popular of such extensions arguably is the Georgi-Machacek (GM) model (see \cite{Georgi:1985nv,Chanowitz:1985ug,Gunion:1989ci,Aoki:2007ah,Chiang:2012cn,Hartling:2014zca,Hartling:2014aga,Chiang:2014bia,Blasi:2017xmc,Logan:2015xpa,Keeshan:2018ypw,Ghosh:2019qie,Banerjee:2019gmr,Azevedo:2020mjg,Moultaka:2020dmb,Ismail:2020kqz,Zhou:2018zli,Sun:2017mue,Degrande:2017naf,Chang:2017niy} for a partial list of references) that features an $SU(2)_L \times SU(2)_R$ global symmetry in the scalar potential. This framework predicts two charged Higgses both of which couple to 
$W^+,Z$. In fact, they also get to interact with the SM fermions by virtue of mixings induced by the scalar potential. Given that the strengths of such Yukawa interactions are proportional to the corresponding fermion masses, the one with the $(t,b)$ pair is of foremost importance here. The charged Higgses emerging from the more non-minimal extensions also naturally feature sizeable couplings to ($W,Z$) and ($t,b$). Therefore, owing to the fact that the $H^+ W^- Z$ interaction features in a plethora of new physics (NP) scenarios, there is the enticing possibility of probing the same in a model-independent fashion.

A singly charged Higgs lighter than the $t$-quark has been looked for at the LHC via the $t \to H^+ b$
decay. On the other hand, a heavier one is searched in processes where it is produced in association with a $t$ and subsequently decays to a pair of fermions. Such channels include $H^+ \to t \bar{b}$~\cite{Aaboud:2018cwk,Aad:2015typ,Khachatryan:2015qxa}, $H^+ \to \tau^+ \nu_\tau$~\cite{Aaboud:2018gjj,Aad:2014kga}, $H^+ \to c \bar{b}$~\cite{Sirunyan:2018dvm} and $H^+ \to c \bar{s}$~\cite{Khachatryan:2015uua}. These searches mainly rely on the strength of the Yukawa interactions of $H^+$. In addition, bosonic decays of the form $H^+ \to W^+ X$, where $X$ denotes a scalar
have also been of substantial interest~\cite{Patrick:2017ele,Patrick:2016rtw,Li:2016umm,Moretti:2016jkp,Moretti:2016jkp,Coleppa:2014cca,Coleppa:2019cul}.
On the other hand, the $H^+ W^- Z$ interaction has
been probed in $p p \to H^\pm j j, H^\pm \to W^\pm Z$~\cite{Sirunyan:2017sbn,Aad:2015nfa} in context of the GM model and scalar triplet extensions of the minimal supersymmetric standard model (MSSM)~\cite{Bandyopadhyay:2014vma,Bandyopadhyay:2015oga,Bandyopadhyay:2015ifm,Bandyopadhyay:2017klv}. A recent review on charged Higgs phenomenology in 2HDM is \cite{Akeroyd:2016ymd}.

In this work, we set out to study the $H^+ W^- Z$ interaction at the proposed 27 TeV energy upgrade~\cite{Azzi:2019yne} of the Large Hadron Collider (LHC), the so-called high-energy LHC (HE-LHC). We adopt a generic but simplistic framework comprising $H^+ W^- Z$ and $H^+ t b$ interactions only. We look for the $H^+$ in processes
that involve the $H^+ W^- Z$ coupling at the production vertex. Two such important processes are $WZ$ fusion and the $W$-mediated $p p \to Z H^\pm$. Following production, the $H^\pm \to W^\pm Z, t b$ decay modes are looked at and bounds on the corresponding branching fractions are derived corresponding to 15 ab$^{-1}$ integrated luminosity. The generality of the adopted framework makes the obtained results applicable to a wide variety of models that consist of at least one charged Higgs.  

The study is organised as follows. We propose the theoretical framework in section \ref{model}. The existing LHC constraints and our analysis strategy are discussed in section \ref{limits}. Section \ref{collider} contains the analyses for the proposed channels. The results so obtained are recasted as favoured regions in the model parameter space in section \ref{combined}. Finally we summarise and conclude in section \ref{summary}.

\section{Theoretical framework}\label{model}  

The following Lagrangian describes the dimension $\leq$ 4 interactions of an $H^\pm$ in a generic fashion~\cite{Barger:1989fj,Cen:2018okf,DiazCruz:2001tn} \footnote{Momentum dependent interactions borne out of higher dimensional operators \cite{DiazCruz:2001tn} have not been considered in this study.}:
\bea
\mathcal{L} &=& g M_W F g^{\mu\nu} H^+ W^{-}_{\mu} Z_\nu
- \frac{\sqrt{2}}{v} H^+ \bar{t}(M_t A_{tb} P_L + M_b B_{tb} P_R) b  + {\rm h.c.}
\eea

In the above, $F,~A_{tb}$ and $~B_{tb}$ are  dimensionless parameters quantifying the strengths of the $H^{\pm}W^{\mp}Z$ and $H^{\pm}tb$ interactions respectively. To generate some perspective, we comment here on the typical values for $F$, $A_{tb}$ and $B_{tb}$ for two specific classes of models. In scalar sectors containing $SU(2)_L$ doublets alone, $F$ is generated radiatively at the one-loop level. The largest $|F|$ for a 2HDM is 
reported to be $\simeq$ 0.01 \cite{Moretti:2015tva} with further enhancement expected with increasing the number of scalar doublets. For example, addition of an additional \emph{inert} (color-octet) doublet elevates the maximum value of $|F|$ to $\simeq$ 0.03 (0.025) \cite{Moretti:2015tva,Chakrabarty:2020msl}. While this trend is encouraging, one must note that arbitrarily increasing the number of scalar doublets not only is limited by the experimental constraints such as that of the electroweak $T$-parameter and the diphoton signal strength, the model becomes aesthetically unattractive and loses predictive power. Therefore, $|F|$ is in the $\mathcal{O}(10^{-2})$ ballpark for realistic multi-doublet extensions of the SM. The other parameter of interest, $A_{tb}$ equals cot$\beta$ for a 2HDM and its inert doublet and color-octet doublet extensions. Here, tan$\b = \frac{v_2}{v_1}$ with $v_1$ and $v_2$ denoting the VEVs of the two \emph{active} doublets. Since tan$\beta < 1$ is ruled out by flavour constraints, one has $A_{tb} < 1$ for a 2HDM. In fact, tan$\beta$ $\in$ [1,10] are typical values allowed for the pure 2HDM (see \cite{Chowdhury:2017aav} for a global analysis of the 2HDM parameter space) leading to $A_{tb}$ $\in$ [0.1,1]. On the other hand, $B_{tb}$ equals $-$cot$\beta$ (tan$\beta$) for a Type-I (Type-II) 2HDM.

The other class of models are characterised by $SU(2)_L$ scalar triplets where $F$ is non-zero at the tree level itself. One derives $F \simeq \frac{g v_\Delta}{c_W M_W}$ for a single complex triplet. Here, $v_\Delta,~M_W,c_W$ and $g$ denote
the triplet VEV, $W$-mass, cosine of the Weinberg mixing angle and the $SU(2)_L$ gauge coupling respectively. In this case however, the 
$\rho$-parameter constraint dictates $v_\Delta \lesssim 5$ GeV~\cite{Tanabashi:2018oca} leading to $F \lesssim 0.017$. The singly charged scalar $H^+$ couples to quarks by virtue of doublet-triplet mixing due to which $A_{tb} = B_{tb} \simeq \frac{2 v_\Delta}{v_\phi}$ where $v_\phi$ denotes the VEV of the SM-like doublet. And the maximum value permitted by the $\rho$-parameter constraint is $A_{tb} \simeq 0.030$. The small $v_\Delta$ for this model therefore renders both 
$|F|,~|A_{tb}| \simeq \mathcal{O}(0.01)$. The same is not true in case of the GM model where $v_\Delta$ is allowed to be larger by virtue of the global $SU(2)_L \times SU(2)_R$ symmetry of the scalar potential. A global analysis of the minimal GM model in \cite{Chiang:2018cgb} gives the bound $v_\Delta \lesssim$ 35 GeV thereby indicating $F \lesssim 0.32$ and $A_{tb} \lesssim 0.56$.

Other possible bosonic and fermionic interactions of $H^\pm$ are not included in the present study. Motivated by the some NP scenarios such as the Type-I 2HDM and the GM model, we further take $M_b B_{tb} << M_t A_{tb}$ in this work for the subsequent analysis. Therefore, with the framework being describable by the two interactions, the total decay width of $H^+$ reads
\bea
\Gamma_{H^+} &=& \Gamma_{H^+ \to W^+ Z} + \Gamma_{H^+ \to t \bar{b}}.
\eea
The branching ratios read
\besub
\bea
\text{BR}(H^+ \to W^+ Z) &=& \frac{\Gamma_{H^+ \to W^+ Z}}{\Gamma_{H^+ \to W^+ Z} + \Gamma_{H^+ \to t \bar{b} }}, \\
\text{BR}(H^+ \to t \bar{b})  &=& \frac{\Gamma_{H^+ \to t \bar{b} }}{\Gamma_{H^+ \to W^+ Z} + \Gamma_{H^+ \to t \bar{b} }}.
\eea
\eesub
Finally, we give the expressions for the corresponding partial widths below.
\besub
\bea
\Gamma_{H^+ \to W^+ Z} &=& \frac{M_{H^+}}{16 \pi} 
\sqrt{\l(1, x_W, x_Z)}|F|^2 g^2 \Big[\frac{(1 - x_W - x_Z)^2}{4 x_Z} + 2 x_W\Big], \\
\Gamma_{H^+ \rightarrow t \bar{b}} &=& \frac{3 g^2 A^2_{tb} M^2_t M_{H^+}}{32 \pi M^2_W}\sqrt{\lambda(1,x_t,x_b)}(1 - x_t - x_b), \\
\text{with}~~~\lambda(x,y,z) &=& x^2 + y^2 + z^2 - 2xy - 2yz - 2zx.
\eea
\eesub
Here, $x_P = \frac{M_P^2}{M^2_{H^+}}$ for $P = t,b,W,Z$ and $M_{H^+}, M_t, M_b, M_Z$ are the masses of the charged Higgs, top-quark, bottom-quark and $Z$-boson respectively. 
Henceforth, $F, A_{tb}$ and $M_{H^+}$ are counted as the free parameters of the current framework.
   
\section{Analysis strategy and existing LHC limits}
\label{limits}
\begin{figure}[htpb!]{\centering
\subfigure[]{
\includegraphics[scale=0.33]{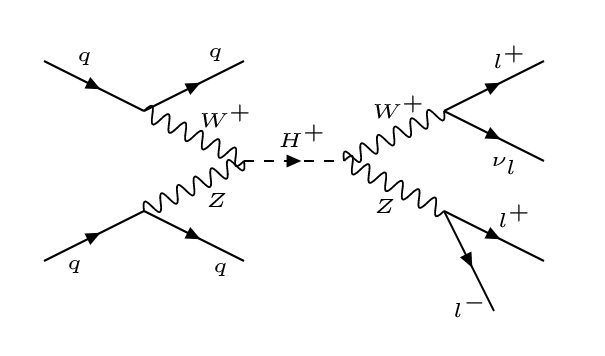}} 
\subfigure[]{
\includegraphics[scale=0.35]{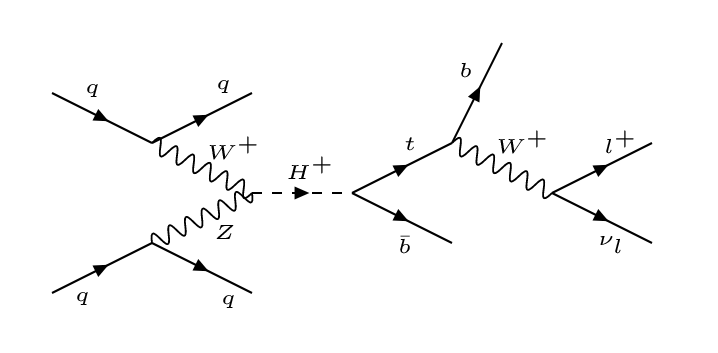}} \\
\subfigure[]{
\includegraphics[scale=0.34]{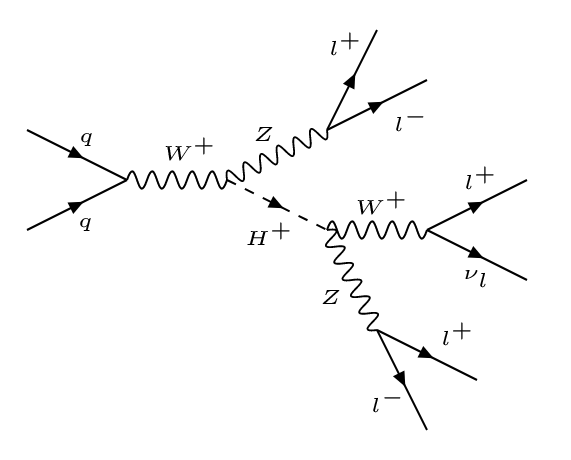}}  
\subfigure[]{
\includegraphics[scale=0.34]{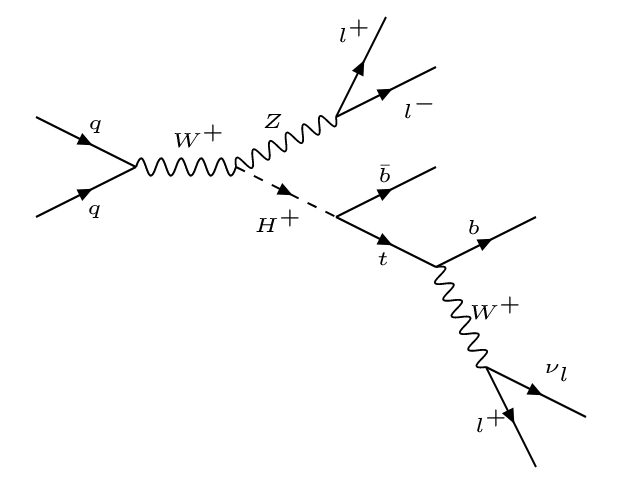}}}
\caption{Feynman diagrams representing the processes
(a) $p p \to H^\pm j j \to W^\pm Z j j \to 3l + 2j + \met$ ,
(b) $p p \to H^\pm j j \to t b j j \to  2b + 1 l + 2j + \met$, 
(c) $p p \to H^\pm Z \to W^\pm Z Z \to 5l + \met$ and
(d) $p p \to H^\pm Z \to t b Z \to 3 l + 2b + \met$. In the diagram, $j$ represents the a light quark jet, while $l^\pm$ denote $e,\mu$. }
\label{feyn}
\end{figure}

Two kinematically distinct topologies for $H^\pm$ production in $pp$- collisions that involve the 
$H^\pm W^\mp Z$ interaction are $p p \to H^\pm j j$ and $p p \to Z H^\pm$. In the former, an $H^\pm$ is produced by the fusion of $W^\pm$ and $Z$ and two forward jets are emitted. This is essentially vector boson-fusion (VBF)
that has close semblance to similar processes for $h$-production, where, $h$ denotes the SM-like Higgs of mass 125 GeV. On the other hand, the $Z H^\pm$ process is 
a $W$-mediated $s$-channel topology. In this work, we aim to probe the $H^+ \to W^+ Z,~t \bar{b}$ decays for both the aforementioned production processes. Leptonic decays of $t,W^\pm$ and $Z$ gives rise to the following four cascades:
(a) $p p \to H^\pm j j \to W^\pm Z j j \to 3l + 2j + \met$,
(b) $p p \to H^\pm j j \to t b j j \to   2b + 1 l + 2j + \met$, 
(c) $p p \to H^\pm Z \to W^\pm Z Z \to 5l + \met$ and
(d) $p p \to H^\pm Z \to t b Z \to 2b + 3 l + \met$, where $l = e,\mu$.
The corresponding Feynman diagrams are shown in Fig.\ref{feyn}.

We reiterate for clarity that both processes (a) and (b) 
describe $H^\pm$-production via VBF with subsequent decays
to $W^\pm Z$ and $t b$  modes respectively. Similarly,
processes (c) and (d) correspond to the $W^\pm Z$ and 
$tb$ decays of $H^\pm$ produced via 
$p p \to Z H^\pm$.

As mentioned before, the framework can be described by $M_{H^+}$, and the couplings 
$F$ and $A_{tb}$. We aim to explore the discovery potential of a 27 TeV $pp$-machine with integrated luminosity $\mathcal{L}$ = 
15 ab$^{-1}$ through a detailed signal and background analysis of the aforementioned signal channels. Further, we supplement the conventional cut-based analyses (CBA) with the more sophisticated multivariate analyses (MVA) using the Boosted Decision Tree Decorrelated (BDTD) algorithm~\cite{Roe:2004na}. An overview of BDTD analysis will be given in subsection \ref{subsec:4A}.


The performances of the proposed channels can be compared by drawing the exclusion and discovery contours corresponding to each in the $F-A_{tb}$ plane for a given $M_{H^+}$.    

It is imperative to discuss possible exclusions on 
\{$M_{H^+}, F, A_{tb}$\} from previous collider searches. The proposed process (a) itself has been searched at the LHC by the ATLAS and CMS collaborations \cite{Aad:2015nfa,Sirunyan:2017sbn}.
These searches have placed upper limits on 
$\sigma_{\text{VBF}} \times \text{BR}(H^+ \to W^+ Z)$ as a function of $M_{H^+}$. The CMS analysis for 
$\sqrt{s}$ = 13 TeV and integrated luminosity = 15.2 fb$^{-1}$~\cite{Aad:2015nfa} predicts a stronger bound compared to the one by ATLAS analysis for $\sqrt{s}$ = 8 TeV with 20.3 fb$^{-1}$ of integrated luminosity~\cite{Sirunyan:2017sbn} except for 300 GeV $\leq M_{H^+} \leq$ 400 GeV. We thus choose to work with the CMS bound in this work. A limit on 
$\sigma_{\text{VBF}} \times \text{BR}(H^+ \to W^+ Z)$ can be obtained as,
\bea
|F|^2~\text{BR}(H^+ \to W^+ Z)|_{M_{H^+}}  \leq  \frac{\Big[ \sigma_{\text{VBF}} \times \text{BR}(H^+ \to W^+ Z)\Big]^{\text{CMS}}_{M_{H^+}}}{\sigma_{|F| = 1}|
_{M_{H^+}}}.
\eea 
Here $\sigma_{|F| = 1}|
_{M_{H^+}}$ denotes the $p p \to H^\pm j j$ cross section in our framework for $|F| = 1$ for a given $M_{H^+}$ in absence of kinematical cuts. 

 Another pertinent search by ATLAS at 13 TeV and 36.2 fb$^{-1}$ of integrated luminosity is $p p \to t b H^\pm, H^\pm \to t b$~\cite{Aaboud:2018cwk}. 
A more stringent bound comes from the recent search at 139 fb$^{-1}$ data~\cite{ATLAS:2020jqj}.
Adopting the more recent bound, a limit on $\sigma_{t b H^{\pm}} \times \text{BR}(H^+ \to t \bar{b})$ is derived as :
\bea
|A_{tb}|^2~\text{BR}(H^+ \to t \bar{b} )|_{M_{H^+}}  \leq  \frac{\Big[ \sigma_{t b H^{\pm}} \times \text{BR}(H^+ \to t \bar{b} )\Big]^{\text{ATLAS}}_{M_{H^+}}}{\sigma^{\prime}_{|A_{tb}| = 1}|,
_{M_{H^+}}}
\eea 
where, $\sigma^{\prime}_{|A_{tb}| = 1}|
_{M_{H^+}}$ denotes the $p p \to t b H^\pm$ cross section for $|A_{tb}| = 1$ for a given $M_{H^+}$ in absence of kinematical cuts. 

\section{Collider analysis}
\label{collider}
We choose to perform analyses for $M_{H^+}$ = 200 GeV, 300 GeV, 500 GeV and 1 TeV which we tag as BP1, BP2, BP3 and BP4 respectively.
The signal and background samples are generated using \texttt{MG5aMC@NLO}~\cite{Alwall:2014hca} at the leading order (LO). The \texttt{NN23LO1}
Parton Distribution Function (PDF) set and default hadronization and factorization scales are used. The parton level events are passed on to \texttt{pythia8}~\cite{Sjostrand:2014zea} for showering and hadronisation and subsequently to \texttt{Delphes-3.4.1} for detector simulation. Specifically, we have throughout used the default ATLAS detector simulation card that comes with
\texttt{Delphes-3.4.1}~\cite{deFavereau:2013fsa}. The multivariate analysis is done using the \texttt{TMVA} package~\cite{2007physics3039H}. The signal significance $S$ is derived using 
$S = \sqrt{2\Big[(N_S + N_B) \log\Big(\frac{N_S + N_B}{N_B}\Big)- N_S\Big]}$ \cite{Cowan:2010js}, with $N_S (N_B)$ denoting the number of signal (background) events surviving the kinematical cuts\footnote{The number of signal ($N_S$) and the background ($N_B$) events can be calculated as:
\bea
N_{S(B)} = \sigma_{S(B)} \times \mathcal{L} \times \epsilon_{S(B)} \,,
\eea
where $\sigma_{S(B)},~\mathcal{L},~\epsilon_S (\epsilon_B)$ denote the signal (background) cross section, integrated luminosity and signal (background) cut-efficiency respectively.}. 

In order to estimate the effects on the final signal significance by assuming a systematic uncertainty of $\sigma_{sys\_un}$, the signal significance formula changes as:
\begin{equation}
S_\text{sys} = \sqrt{2 \left((N_S+N_B) \log \left(\frac{(N_S+N_B)(N_B+\sigma_{B}^{2})}{N_B^{2}+(N_S+N_B)\sigma_{B}^{2}}\right)-\frac{N_B^{2}}{\sigma_{B}^{2}}\log \left(1+\frac{\sigma_{B}^{2}N_S}{N_B(N_B+\sigma_{B}^{2})} \right) \right)} \,,
\label{S_sys}
\end{equation}
where $\sigma_{B}=\sigma_{sys\_un}\times B$.

We set out to analyse the various channels following the aforementioned strategy. 
For processes (a),(c) and (b),(d) in Fig.\ref{feyn}, ($F$,BR($H^+ \to W^+ Z$)) and ($F$,BR($H^+ \to t \bar{b}$)) are respectively treated as the free parameters. That said, for a given signal, it is important to compare the results from CBA to those from MVA for specific reference values of $F$ and the branching ratios. Therefore, we take ($F$,BR($H^+ \to W^+ Z$)) = (0.4,0.4) and ($F$,BR($H^+ \to t b$)) = (0.4,0.4) as the reference for processes (a),(c) and (b),(d) respectively for the ensuing analysis.

\subsection{The $3l + 2 j + \met$ channel}
\label{subsec:4A}
This subsection discusses the VBF production of $H^\pm$ followed by the $H^\pm \to W^\pm Z$ decay. The VBF topology
always leads to a couple of light jets in the forward and backward directions with negligible hadronic activity in the intervening rapidity gap. Since the two jets ($j_1, j_2$) reside in different $\eta$-hemisphere for VBF production, the forward (backward) jet is identified by 
$\eta > 0$ ($\eta < 0$). We therefore demand $\eta_{j_1} \eta_{j_2} < 0$ to tag the two leading light jets as forward and backward ones. The  
$H^\pm \to W^\pm Z$ decay ultimately leads to three leptons, two of which are of same flavor and opposite sign (SFOS) that come from $Z$ and the third one comes from $W^\pm$. In addition, a neutrino also originates from $W^\pm$. We choose to treat the 
$H^{\pm} \to W^{\pm} Z$ branching ratio as a free parameter at this level. The signal cross sections for different $M_{H^+}$ are given in Table \ref{vbftab:1}. 

The most dominant SM backgrounds corresponding to signal come from $W^\pm Z j j$ and $Z Z j j$ production followed by leptonic cascades. These two background processes involve jets at the production level as does the signal.
Therefore, we demanded
$p_T^j$ > 20 GeV, $M_{j_1 j_2}$ > 500 GeV, 
$|\eta_{j}|$ < 5 while generating these in order to pick out the phase space kinematically relevant to the VBF topology. The resulting LO cross sections of these two background processes along with their leptonic decay chains are given in Table \ref{vbftab:1}. Sub-leading contributions to the backgrounds come from $W^+ W^- Z, W^\pm Z Z, ZZZ,
t \bar{t} W^\pm, t \bar{t} Z$. We do not impose cuts at the generation level for the sub-leading ones since they do not involve jets at the production level, and, the corresponding cross sections are again listed in Table \ref{vbftab:1}. A couple of observations then emerge. First, $W^\pm Z Z \to 4l + 2j, Z Z Z$ have negligible cross sections compared to the leading backgrounds. Secondly, though it might naively appear that the other sub-leading channels offer rates comparable to, or, even higher than $ZZjj$, one must remember that their cross sections are computed in the absence of kinematical cuts. Therefore, stringent VBF selection cuts almost nearly obliterate $t \bar{t} W^\pm, t \bar{t} Z, W^\pm Z Z \to 
3 l + 2j + \met$. In all, it is only the $W^\pm Z j j$ and $Z Z j j$ channels that are kinematically relevant to 
the analysis.

Events are selected by demanding at least two light jets ($N_j \geq 2$) and three leptons after vetoing $b$-jets and $\tau$-jets in the final state. Of the two hardest light jets, one is demanded to be forward and the other as  the backward one as mentioned earlier.
We first initiate the cut-based approach. The following trigger-level cuts are applied:
\bea
p_T^l > 10~\text{GeV}, |\eta_{j,l}| < 2.5, \Delta R_{l_1 l_2} > 0.2, \Delta R_{jl} > 0.2, \Delta R_{j_1 j_2} > 0.4.
\eea

\begin{table}[htpb!]
\begin{center}\scalebox{0.95}{
\begin{tabular}{|c|c|c|}
\hline
Signal / Backgrounds  & Process & Cross section $\sigma$ (LO) (fb) \\ \hline \hline
 Signal &  & \\ 
 BP1 ($M_{H^+} = 200$ GeV) & & 5.49\\
 BP2 ($M_{H^+} = 300$ GeV)& $p p \rightarrow H^\pm j j \rightarrow W^\pm Z j j \rightarrow 3l + 2j + \met$ & 3.43\\
 BP3 ($M_{H^+} = 500$ GeV)& & 1.64\\
 BP4 ($M_{H^+} = 1$ TeV)&  & 0.46 \\ \hline \hline
Backgrounds & $p p \to W^\pm Z j j \to 3l + 2j + \met$ & 120.0\\ 
& $p p \to  Z Z j j \to 4l + 2j $ & 4.57 \\
& $p p \to W^\pm Z Z \to 3l + 2j + \met$ & 1.19 \\
& $p p \to W^\pm W^\mp Z \to 3l + 2j + \met$ & 4.80 \\
& $p p \to t \bar{t} Z \to 4l + 2b + \met$ &  10.15 \\
& $p p \to t \bar{t} W^\pm \to 3l + 2b + \met$ & 11.50 \\ 
& $p p \to ZZZ \to 4l + 2j $ &  8.25 $\times 10^{-3}$\\
& $p p \to W^\pm Z Z \to 4l + 2j$ &  8.75 $\times 10^{-2}$\\
\hline
\end{tabular}}
\end{center}
\caption{The LO cross sections for signal and backgrounds for the process $p p \rightarrow H^\pm j j \rightarrow W^\pm Z j j \rightarrow 3l + 2j + \met$. The signal cross sections are computed for the ($F$,BR($H^+ \to W^+ Z$)) = (0.4,0.4) reference point.}
\label{vbftab:1}
\end{table}

To start with, we shall compute the signal significance using the cut-based method. In addition to the trigger cuts, certain specific kinematic variables are identified to extract the signal with higher efficiency. They are $M_{j_1 j_2},|\Delta \eta_{j_1 j_2}|, 
{M_{\text{inv}}^{WZ}}, M_T^{W  Z}$. We denote the corresponding cuts to be $A_1, A_2, A_3, A_4$ respectively and they are optimised by looping over a few configurations and selecting the one that yields the maximum statistical significance. The cut-flow for the signal and background processes corresponding to the chosen BPs are displayed in Table~\ref{vbftab:3}-\ref{vbftab:5b}. The following discussion motivates these variables and compares the signal and backgrounds in their terms.  

\begin{figure}[htpb!]{\centering
\subfigure[]{
\includegraphics[height = 6.5 cm, width = 7.5 cm]{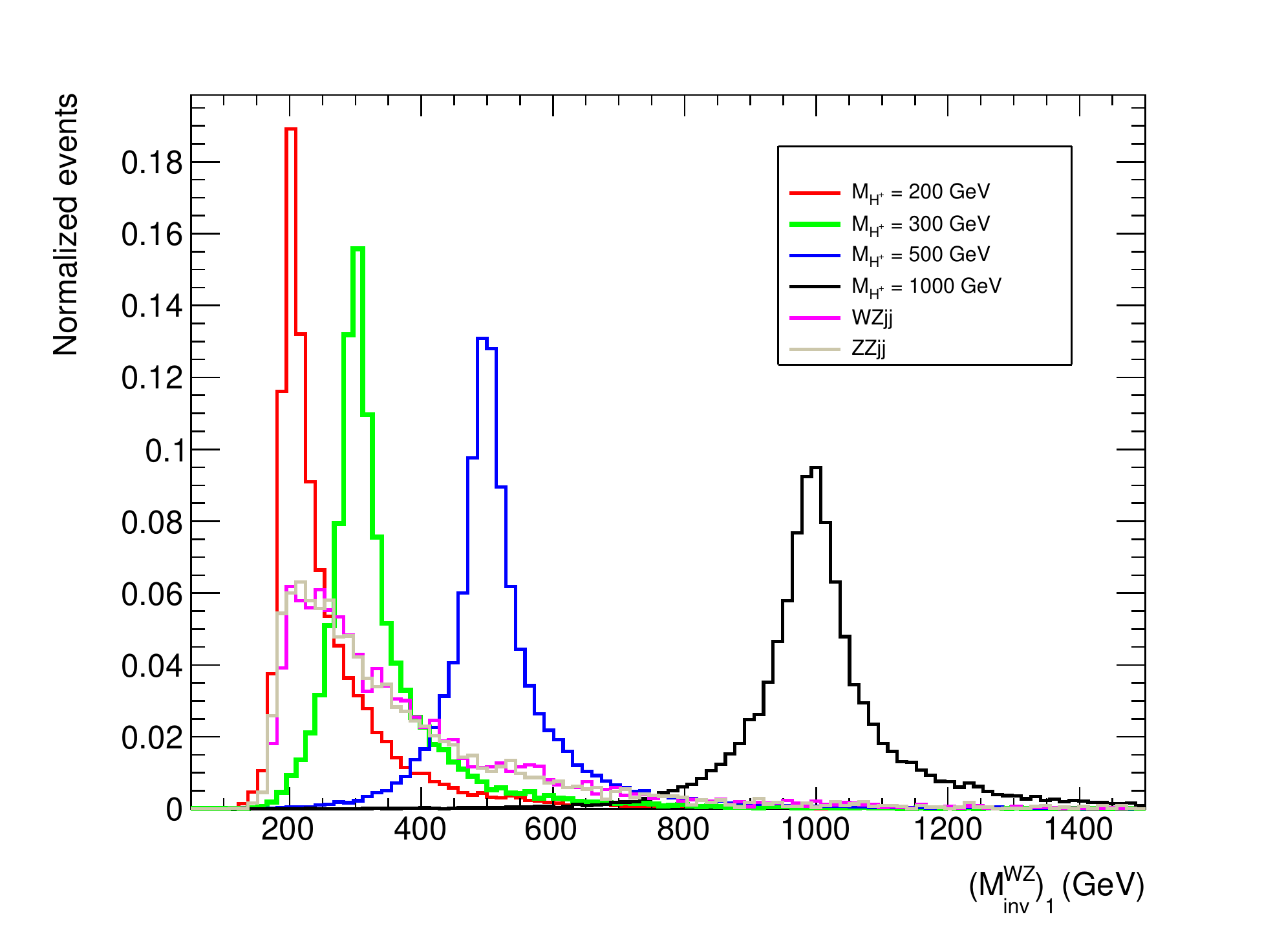}}
\subfigure[]{
\includegraphics[height = 6.5 cm, width = 7.5 cm]{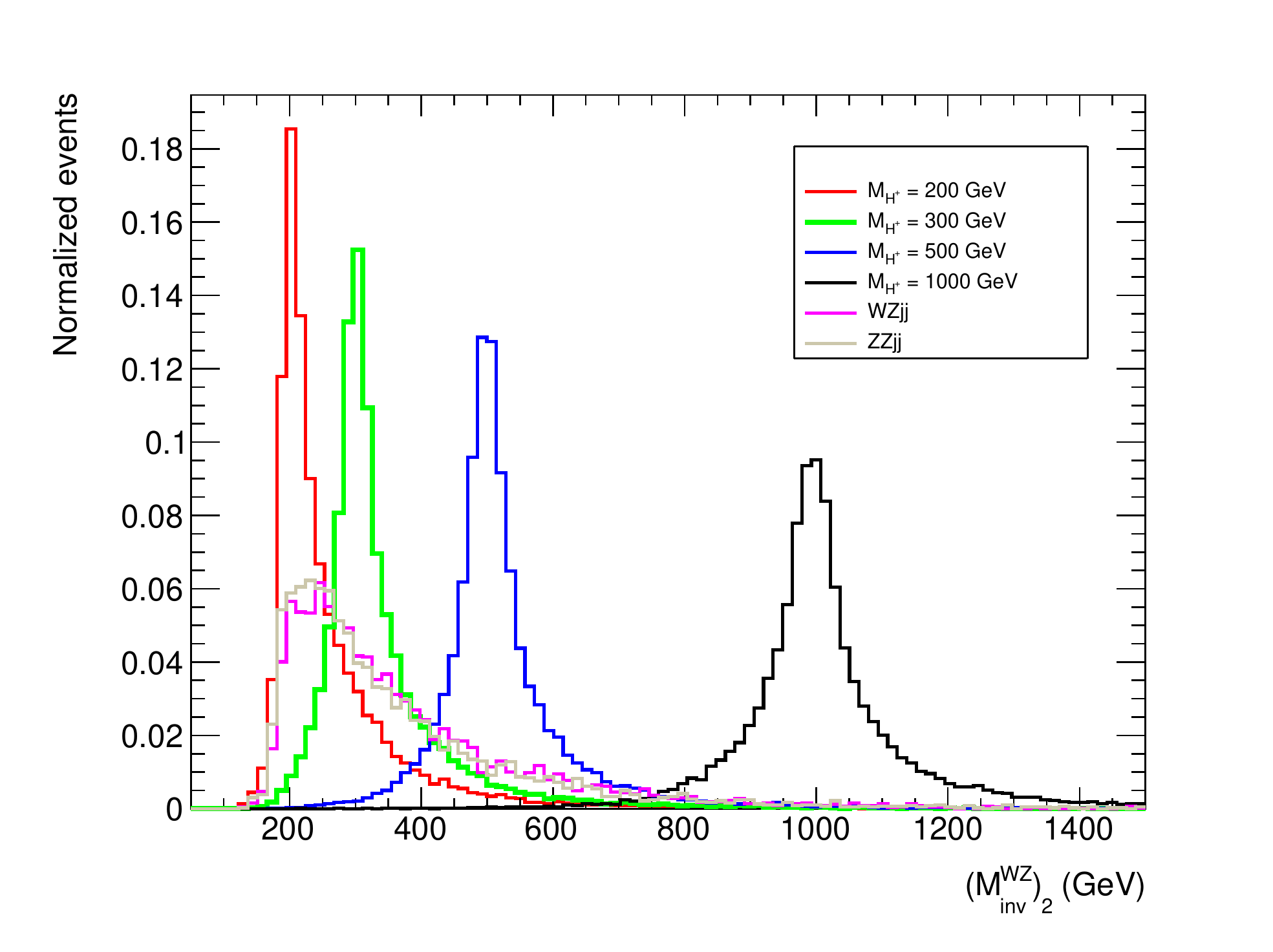}} \\
\subfigure[]{
\includegraphics[height = 6.5 cm, width = 7.5 cm]{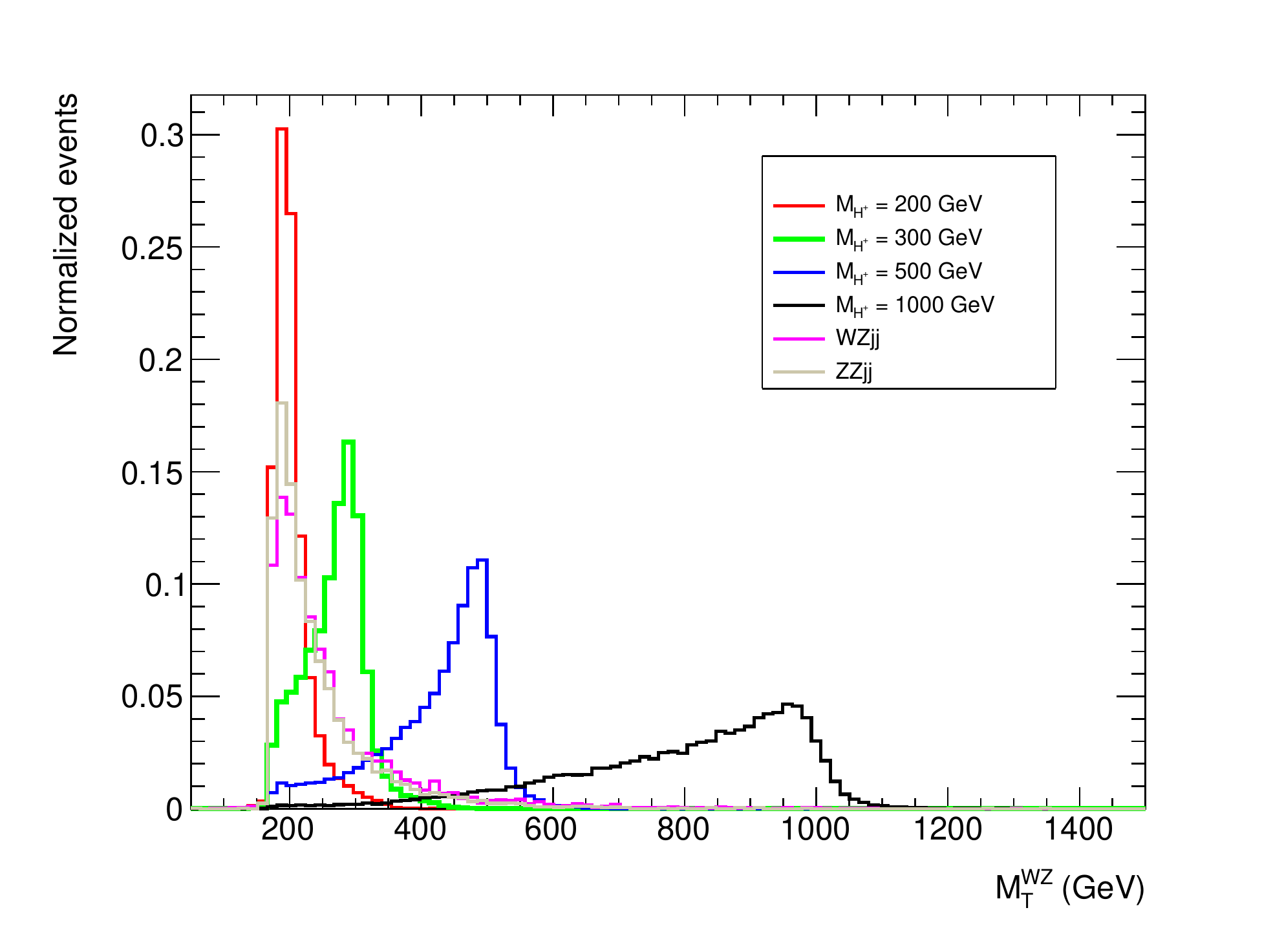}} 
\subfigure[]{
\includegraphics[height = 6.5 cm, width = 7.5 cm]{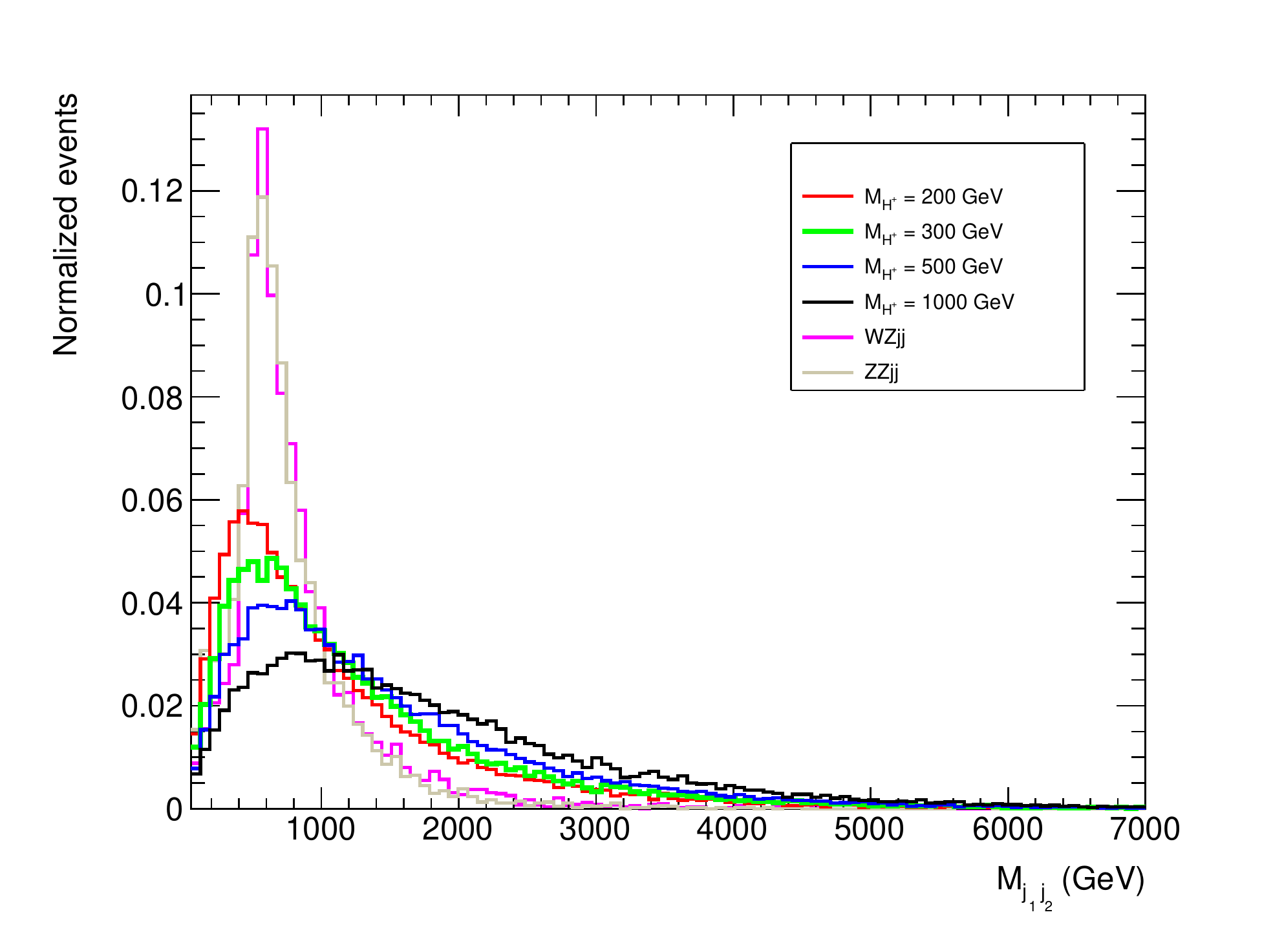}} \\
\subfigure[]{
\includegraphics[height = 6.5 cm, width = 7.5 cm]{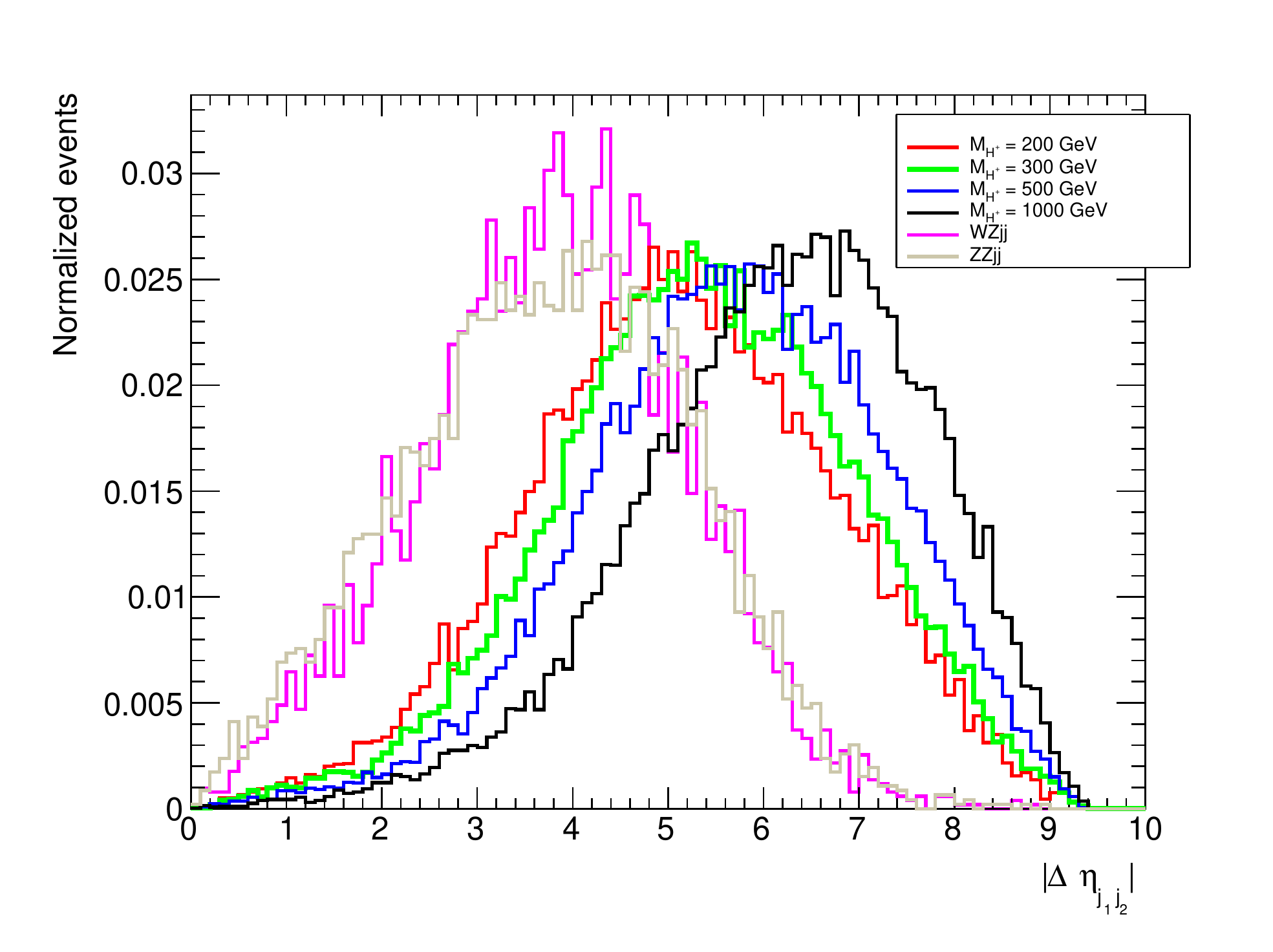}}}
\caption{Normalised distributions of $(M_{\rm inv}^{WZ})_1$, $(M_{\rm inv}^{WZ})_2$, $M_T^{WZ}$, $M_{j_1 j_2}$ and $|\Delta \eta_{j_1 j_2}|$ for the $ 3l + 2j + \slashed{E_T}$ channel.}
\label{fig:1}
\end{figure}

\begin{itemize}

\item \textbf {$A_1$} and \textbf {$A_2$}: A VBF topology demands a high invariant mass of the two forward jet system ($M_{j_1 j_2}$) and a large separation between their pseudo rapidities ($|\Delta \eta_{j_1 j_2}|$). Thus high invariant mass cuts (given in Table \ref{vbftab:2}) are used for all the benchmarks. Strong cuts on $|\Delta \eta_{j_1 j_2}|$ for enhancing the significance for all benchmark points can be found in Table \ref{vbftab:2}. Corresponding normalised distributions for both signal and backgrounds are drawn in Fig.\ref{fig:1}(d) and Fig.\ref{fig:1}(e). It can be observed that the peak of $M_{j_1 j_2}$ distribution shifts towards higher value with increase in $M_{H^+}$. From the distributions, it is evident that these cuts reduce the backgrounds to a large extent. The same is also reflected in the first two columns of Table \ref{vbftab:3}, Table \ref{vbftab:4} and Table \ref{vbftab:5}. 

\item \textbf{$A_3$}: Next comes a cut on the invariant mass of the $3l + \met$ system (${M_{\text{inv}}^{WZ}}$), which must peak at the mass of the decaying $H^+$ in case of the signal. For constructing this variable, we have taken into account two possible combinations comprising of one same flavour opposite sign (SFOS) lepton pair (peaking at $M_Z$) , one isolated lepton coming from $W$-boson and $z$-component of missing transverse energy \footnote{$\slashed{E}_{T,z} = \frac{1}{2 {p_T^{l}}^2} (A_W p_Z^l \pm E^l \sqrt{A_W^2 - 4 {p_T^l}^2 {\slashed{E_T}}^2})$, where $A_W = M_W^2 + 2(p_x^l \slashed{E}_{T,x} + p_y^l \slashed{E}_{T,y}$). $\slashed{E}_{T,x}$ and $\slashed{E}_{T,y}$ are the $x$ and $y$ component of the missing transverse momentum.}. Two solutions of the $z$-component of missing transverse energy lead to two different kinematic variables : $({M_{\text{inv}}^{WZ}})_1$ and $({M_{\text{inv}}^{WZ}})_2$.  Comparing these two variables with $M_{H^+}$, the closest one is chosen to be ${M_{\text{inv}}^{WZ}}$. For clarity we have plotted distributions of $({M_{\text{inv}}^{WZ}})_1$ and $({M_{\text{inv}}^{WZ}})_2$ for both signal and dominant backgrounds before applying any cut. Whereas, for backgrounds since the source of $3l+\met$ is not a single particle, the corresponding distributions of $({M_{\text{inv}}^{WZ}})_1$ and $({M_{\text{inv}}^{WZ}})_2$ in Fig.\ref{fig:1}(a), Fig.\ref{fig:1}(b) do not peak around  $M_{H^+}$\footnote{While drawing the distributions, we only have shown the distribution for two dominant backgrounds, {\em i.e.} $W^\pm Z j j $ and $Z Z j j $.}.

\item \textbf{$A_4$}: Another potentially important variable for this signal is the transverse mass of $W Z$-system. It can be constructed as: 
\bea
M_T^{W  Z} = \sqrt{[E_T (W) + E_T (Z)]^2 - [\vec{p}_T (W) + \vec{p}_T (Z)]^2}
\eea
where $\vec{p}_T (W)$ is computed by adding $\vec{p}^{\rm miss}_T$ and $\vec{p}_T$ of the lepton (coming from $W^\pm$) vectorially and $E_T(W)$ is obtained from the scalar sum of $p^{\rm miss}_T$ and the lepton transverse energy. $\vec{p_T}(Z)$ is constructed by adding the $\vec{p}_T$ of the two leptons of SFOS that peak near the $Z$-mass and $E_T(Z)$ is calculated using the following relation.
\bea
E^2_T(Z) = |\vec{p_T}(Z)|^2 + M_Z^2
\eea

 As the $W^\pm, Z$ coming from heavy $H^\pm$  are highly boosted, peak of the normalised distribution of $M_T^{W  Z}$ will be shifted to the higher end starting from BP1 to BP4 in Fig.\ref{fig:1}(c). The optimised set of cuts on $M_T^{W  Z}$ can be found in Table \ref{vbftab:2}. 

\end{itemize}

\begin{table}[htpb!]
\begin{center}\scalebox{0.95}{
\begin{tabular}{|c|c|c|c|c|}
\hline
$M_{H^+}$ & $A_1$ & $A_2$ & $A_3$ & $A_4$ \\ \hline \hline
200 GeV  & $M_{j_1 j_2}$ > 500 GeV  & $|\Delta \eta_{j_1 j_2}|$ > 5 & $|{M_{\text{inv}}^{WZ}} - M_{H^+}| < 50~\text{GeV}$ & $|M_T^{W  Z} - M_{H^+}| < 100~\text{GeV}$ 
\\
300 GeV  & $M_{j_1 j_2}$ > 500 GeV  & $|\Delta \eta_{j_1 j_2}|$ > 5.5 & $|{M_{\text{inv}}^{WZ}} - M_{H^+}| < 50~\text{GeV}$ & $|M_T^{W  Z} - M_{H^+}| < 50~\text{GeV}$ 
\\
500 GeV  & $M_{j_1 j_2}$ > 500 GeV  & $|\Delta \eta_{j_1 j_2}|$ > 6 & $|{M_{\text{inv}}^{WZ}} - M_{H^+}| < 60~\text{GeV}$ & $|M_T^{W  Z} - M_{H^+}| < 125~\text{GeV}$ 
\\
1 TeV  & $M_{j_1 j_2}$ > 500 GeV  & $|\Delta \eta_{j_1 j_2}|$ > 4.5 & $|{M_{\text{inv}}^{WZ}} - M_{H^+}| < 60~\text{GeV}$ & $|M_T^{W  Z} - M_{H^+}| < 120~\text{GeV}$ 
\\
\hline
\end{tabular}}
\end{center}
\caption{The optimised selection cuts for the $3l + 2j + \met$ channel for the four benchmarks.}
\label{vbftab:2}
\end{table}

\begin{table}[htpb!]
\begin{center}
\begin{tabular}{||c|c|c|c|c||}
\hline
\multicolumn{1}{||c|}{ $M_{H^+} = 200$ GeV} &
\multicolumn{4}{|c||}{NEV ($\mathcal{L}=$ 15 ab$^{-1}$)} \\ 
\cline{2-5}
  & $A_1$ & $A_2$ & $A_3$ & $A_4$ \\
\hline
Signal  &   7170 & 3916 & 3190 & 3183 \\
\hline\hline
$W^{\pm} Z j j$ & 88484 & 11462 & 4662 & 4647 \\
\hline
$Z Z j j$  &  4637 & 622 & 301 & 301 \\
\hline
\multicolumn{1}{||c|}{Signal yield = 3138} & \multicolumn{1}{c|}{Total background yield = 4948} & \multicolumn{3}{c||}{Signal significance ($0\%$ $\sigma_{sys\_un}$) = $40.8$} \\
\hline
\multicolumn{1}{||c|}{} & \multicolumn{1}{c|}{} & \multicolumn{3}{c||}{Signal significance ($2\%~(5\%)$ $\sigma_{sys\_un}$) = $22.3~(10.3)$} \\
\hline
\hline
\end{tabular}
\caption{Signal and background yields after applying the selection cuts and projected signal significance for BP1 with 15 ab$^{-1}$ integrated luminosity. The signal yields are computed for the ($F$,BR($H^+ \to W^+ Z$)) = (0.4,0.4) reference point.}
\label{vbftab:3}
\end{center}
\end{table}

\begin{table}[htpb!]
\begin{center}
\begin{tabular}{||c|c|c|c|c||}
\hline
\multicolumn{1}{||c|}{ $M_{H^+} = 300$ GeV} &
\multicolumn{4}{|c||}{NEV ($\mathcal{L}=$ 15 ab$^{-1}$)} \\ 
\cline{2-5}
  & $A_1$ & $A_2$ & $A_3$ & $A_4$ \\
\hline
Signal  &   5529 & 2433 & 2057 & 1591 \\
\hline\hline
$W^{\pm} Z j j$ & 89118 & 6300 & 2934 & 1494 \\
\hline
$Z Z j j$  &  4599 & 316 & 135 & 68 \\
\hline
\multicolumn{1}{||c|}{Signal yield = 1591} & \multicolumn{1}{c|}{Total background yield = 1562} & \multicolumn{3}{c||}{Signal significance ($0\%$ $\sigma_{sys\_un}$) = $35.3$} \\
\hline
\multicolumn{1}{||c|}{} & \multicolumn{1}{c|}{} & \multicolumn{3}{c||}{Signal significance ($2\%~(5\%)$ $\sigma_{sys\_un}$) = $26.3~(14.5)$} \\
\hline
\hline
\end{tabular}
\caption{Signal and background yields after applying the selection cuts and projected signal significance for BP2 with 15 ab$^{-1}$ integrated luminosity. The signal yields are computed for the ($F$,BR($H^+ \to W^+ Z$)) = (0.4,0.4) reference point.}
\label{vbftab:4}
\end{center}
\end{table}

\begin{table}[htpb!]
\begin{center}
\begin{tabular}{||c|c|c|c|c||}
\hline
\multicolumn{1}{||c|}{ $M_{H^+} = 500$ GeV} &
\multicolumn{4}{|c||}{NEV ($\mathcal{L}=$ 15 ab$^{-1}$)} \\ 
\cline{2-5}
  & $A_1$ & $A_2$ & $A_3$ & $A_4$ \\
\hline
Signal  &   3283 & 1254 & 1050 & 942 \\
\hline\hline
$W^{\pm} Z j j$ & 88484 & 2912 & 489 & 172 \\
\hline
$Z Z j j$  &  4637 & 157 & 26 & 6 \\
\hline
\multicolumn{1}{||c|}{Signal yield = 942} & \multicolumn{1}{c|}{Total background yield = 178} & \multicolumn{3}{c||}{Signal significance ($0\%$ $\sigma_{sys\_un}$) = $47.3$} \\
\hline
\multicolumn{1}{||c|}{} & \multicolumn{1}{c|}{} & \multicolumn{3}{c||}{Signal significance ($2\%~(5\%)$ $\sigma_{sys\_un}$) = $44.0~(34.4)$} \\
\hline
\hline
\end{tabular}
\caption{Signal and background yields after applying the selection cuts and projected signal significance for BP3 with 15 ab$^{-1}$ integrated luminosity. The signal yields are computed for the ($F$,BR($H^+ \to W^+ Z$)) = (0.4,0.4) reference point.}
\label{vbftab:5}
\end{center}
\end{table}

\begin{table}[htpb!]
\begin{center}
\begin{tabular}{||c|c|c|c|c||}
\hline
\multicolumn{1}{||c|}{ $M_{H^+} = 1$ TeV} &
\multicolumn{4}{|c||}{NEV ($\mathcal{L}=$ 15 ab$^{-1}$)} \\ 
\cline{2-5}
  & $A_1$ & $A_2$ & $A_3$ & $A_4$ \\
\hline
Signal  &   862 & 654 & 491 & 262 \\
\hline\hline
$W^{\pm} Z j j$  &  88484 & 19602 & 324 & 15 \\
\hline
$Z Z j j$ & 4637 & 1065 & 178 & 0 \\
\hline
\multicolumn{1}{||c|}{Signal yield = 262} & \multicolumn{1}{c|}{Total background yield = 15} & \multicolumn{3}{c||}{Signal significance ($0\%$ $\sigma_{sys\_un}$) = $33.0$} \\
\hline
\multicolumn{1}{||c|}{} & \multicolumn{1}{c|}{} & \multicolumn{3}{c||}{Signal significance ($2\%~(5\%)$ $\sigma_{sys\_un}$) = $32.6~(31.0)$} \\
\hline
\hline
\end{tabular}
\caption{Signal and background yields after applying the selection cuts and projected signal significance for BP4 with 15 ab$^{-1}$ integrated luminosity. The signal yields are computed for the ($F$,BR($H^+ \to W^+ Z$)) = (0.4,0.4) reference point.}
\label{vbftab:5b}
\end{center}
\end{table}

\begin{table}[htpb!]
\begin{center}
\resizebox{16cm}{!}{
\begin{tabular}{|c|c|c|c|c|c|c|}
\hline
 &  \hspace{5mm} {\texttt{NTrees}} \hspace{5mm} & \hspace{5mm} {\texttt{MinNodeSize}} \hspace{5mm} & \hspace{5mm} {\texttt{MaxDepth}}~~ \hspace{5mm} & \hspace{5mm} {\texttt{nCuts}} ~~\hspace{5mm} & \hspace{5mm} {\texttt{KS-score for}}~~\hspace{5mm} & \hspace{5mm} {\texttt BDT Score} \hspace{5mm} \\
 & & & & & {\texttt{Signal(Background)}} & \\
\hline
\hline
\hspace{5mm} BP1 \hspace{5mm} & 180 & 3 \% & 2.0 & 35 & 0.251(0.621)& 0.02 \\ \hline
\hspace{5mm} BP2 \hspace{5mm} & 100 & 4 \% & 2.0 & 60 & 0.409(0.524)& 0.148 \\ \hline
\hspace{5mm} BP3 \hspace{5mm} & 150 & 3 \% & 2.0 & 39 & 0.11(0.541) & 0.232 \\ \hline
\hspace{5mm} BP4 \hspace{5mm} & 200 & 3 \% & 2.0 & 35 & 0.346(0.054)& 0.19 \\ \hline
\end{tabular}}
\end{center}
\caption{Tuned BDT parameters for BP1, BP2, BP3 and BP4 for the $3 l + 2 j + \slashed{E_T}$ channel.}
\label{BDT-param1}
\end{table}
\begin{figure}[tp!]{\centering
\subfigure[]{
\includegraphics[width=3in,height=2.55in]{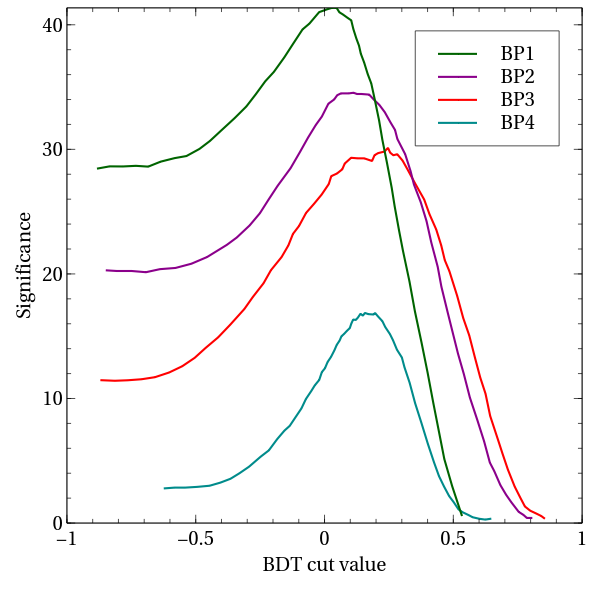}}
\subfigure[]{
\includegraphics[width=2.8in,height=2.55in]{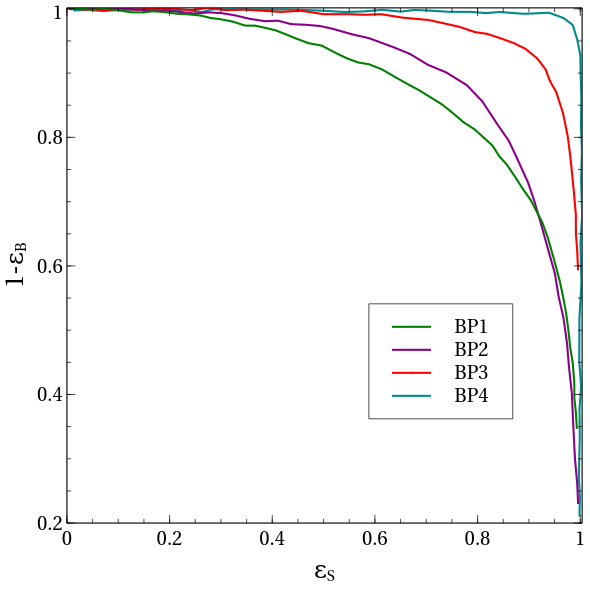}}}
\caption{ 
(a) Variation of significance with BDT-score for $3 l + 2 j + \slashed{E_T}$ channel. (b) ROC curves for chosen benchmark points for $3 l + 2 j + \slashed{E_T}$ channel.}
\label{ROC-BDTScore-3lvjj}
\end{figure}

Overall, the CBA shows that $M_{H^+}$ = 500 GeV has the highest observability. Next we perform an MVA to improve over the results obtained from cut-based analysis. Before going to perform the MVA analysis, let us elaborate on the details of the same.  Here the MVA is performed 
using {\em Decorrelated Boosted Decision Tree} (BDTD) algorithm within the Toolkit for Multivariate Data Analysis
(TMVA) framework. 

First the signal and the background-like events are distinguished using {\em decision trees} as classifier. To segregate the signal and background-like events depending on its {\em purity} {\footnote{The purity $p$ can be defined as : $ p = \frac{S}{S+B}$. An event can be tagged as signal (background) when $p > 0.5$ ($p < 0.5$).}}, each node of the decision tree is associated with one discerning variable on which an optimised cut value is imposed. By modifying the BDTD variable {\texttt NCuts}, one can perform the aforementioned task within TMVA. Starting from the zeroth node or root node, the training of the decision trees is carried on unless a particular depth ({\texttt MaxDepth}) specified by the user is achieved. Finally from the final {\em leaf nodes}, an event can be tagged as signal or background according to the purity.

The decision trees being prone to statistical fluctuations of the training sample, are said to be weak classifiers. This problem can be alleviated by combining a set of weak classifiers into a stronger one  through the modification of the weight of the events. One can thus create new decision trees using this particular procedure termed as {\em Boosting}. Throughout the analysis, we have used {\em Adaptive boost}, compatible for weak classifiers, with input variable transforming in a decorrelated manner. Within TMVA framework, it can be realised as {\texttt Decorrelated AdaBoost}. The BDTD parameters like the number of decision trees {\texttt{NTrees}}, the maximum depth of the decision tree allowed
 {\texttt{MaxDepth}}, the minimum percentage of training events in each leaf node {\texttt{MinNodeSize}} and {\texttt{NCuts}} 
 for four benchmarks of our analysis have been tabulated in Table \ref{BDT-param1}. To get rid of the overtraining of the signal and background 
 samples, the results of the Kolmogorov-Smirnov test, {\em i.e.} Kolmogorov-Smirnov score (KS-score) should be > 0.1 as well as stable. In fact, KS-score > 0.01 can also avoid the overtraining if it remains stable even after changing the 
internal parameters of the algorithm. Now the BDTD algorithm makes a ranking list out of the kinematic variables that are fed into the algorithm, according to the performance of variables in separating signal from backgrounds. Several kinematic variables might be highly correlated.  We select a reasonable number of top ranking variables among them for providing best possible separation between the signal and backgrounds. Thus for this channel following kinematic variables with maximum discerning abilities are proposed for MVA:
\bea
&& \eta_{j_1}, \eta_{j_2}, \Delta R_{j_1 j_2}, p_T^{l_1}, p_T^{l_3}, \Delta R_{l_1 j_1}, \Delta R_{l_2 j_1}, \Delta R_{l_1 j_2}, \Delta R_{l_2 j_2}, \nonumber \\ 
&& p_T^{j_1 j_2}, p_T^{l_1 l_2 l_3}, M_{j_1 j_2}, 
 M_{j_1 j_2 l_1 l_2 l_3}, \met, M_{\rm inv}^{W Z}, p_{T, vec}^{\rm tot}, M_T^{WZ}
\eea
Here, $\Delta R_{ij}$ denotes the distance in the $\eta-\phi$ plane between the $i$th and $j$th particle. While $p_T^{ijk} (p_{T, vec}^{ijk})$ is the scalar (vector) sum of the transverse momentum of the system of $i$th, $j$th and $k$th particle, $ M_{j_1 j_2 l_1 l_2 l_3}$ represents invariant mass of "$j_1, j_2, l_1, l_2, l_3$"-system. The rest of the variables have been defined earlier in the text. According to the BDT ranking, among these variables, the five most important variables to differentiate the signal from backgrounds are : $p_T^{j_1 j_2}, M_{\rm inv}^{W Z},  M_{j_1 j_2 l_1 l_2 l_3}, M_{j_1 j_2}, M_T^{WZ}$.

Fig.\ref{KSscore-3lvjj} (a),(b),(c),(d) in Appendix \ref{Plots} depict the KS-scores for signal and backgrounds for all four benchmarks. KS-scores have also been tabulated in Table \ref{BDT-param1} for convenience. To maximise the significance, one can adjust {\em BDT cut value} or {\em BDT score}. The BDT cut values adjusted for four benchmarks can be found in Table \ref{BDT-param1}. In Fig.\ref{ROC-BDTScore-3lvjj}(a), we have presented the variation of the signal significances with BDT score. Here it can be easily inferred that the significance attains a maximum value at a particular BDT score \footnote{If the curves show any fluctuations, then we put the BDT cut value just before the curve starts fluctuating}. Next we draw Receiver's Operative Characteristic (ROC) curves \footnote{ROC curve is a plot of signal efficiency ($\epsilon_S$) vs. efficiency of 
rejecting the backgrounds $(1-\epsilon_B)$, $\epsilon_B$ being the background efficiency.} for all benchmarks in Fig.\ref{ROC-BDTScore-3lvjj}(b), to estimate the degree of rejecting the backgrounds with respect to the signal. From Fig.\ref{ROC-BDTScore-3lvjj}, it can be inferred that the degree of rejecting backgrounds reduces with decrease in $M_{H^+}$. But owing to large signal cross section of BP1 compared to the rest of the three BPs, BP1 fares the best in probing this channel.

The yields for signal (for $M_{H^+} = 200 ~{\rm GeV}, 300 ~{\rm GeV}, 500 ~{\rm GeV}, 1 ~{\rm TeV} $) and corresponding backgrounds after optimisation through BDTD-analysis are given in Table \ref{3lvjj_BDTD}. The numbers events are computed taking $F = 0.4$ and ${\rm BR}(H^+ \rightarrow W^+ Z) = 40 \%$ at $\mathcal{L}= 15$ ab$^{-1}$. BP1 is seen to offer the highest observability through MVA. And this observability decreases upon increasing $M_{H^+}$. Importantly, the MVA is seen to yield higher statistical significance than the CBA for all the values of $M_{H^+}$. The maximum improvement of significance (with respect to the cut-based analysis) is $ \simeq 35 \%$ which occurs for $M_{H^+} = 300$ GeV. Besides, we have computed the significances taking into account $2\%$ and $5\%$ systematic uncertainties in Table \ref{3lvjj_BDTD}.  

\begin{figure}[htpb!]
\centering 
\includegraphics[height = 7 cm, width = 7.5 cm]{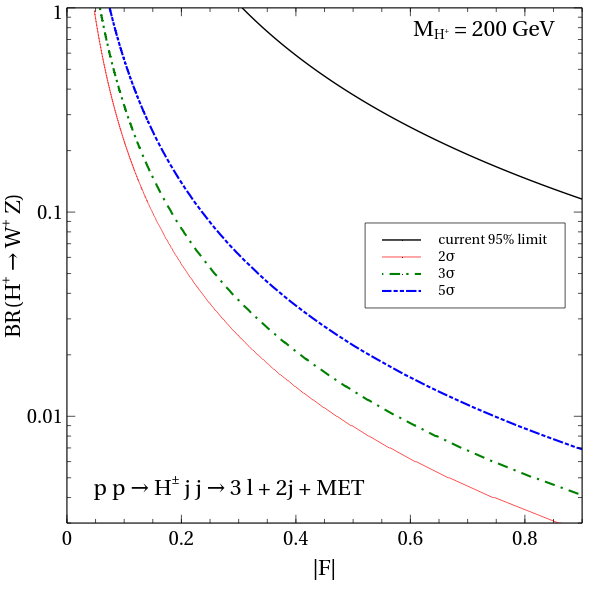}~~~~~~ 
\includegraphics[height = 7 cm, width = 7.5 cm]{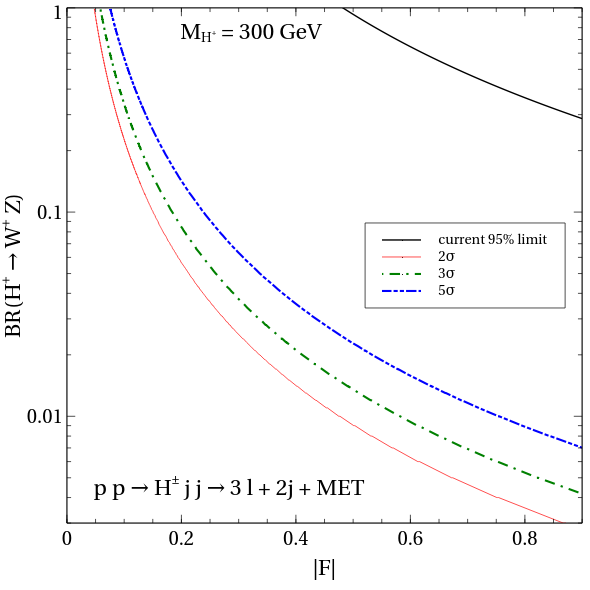} \\
\includegraphics[height = 7 cm, width = 7.5 cm]{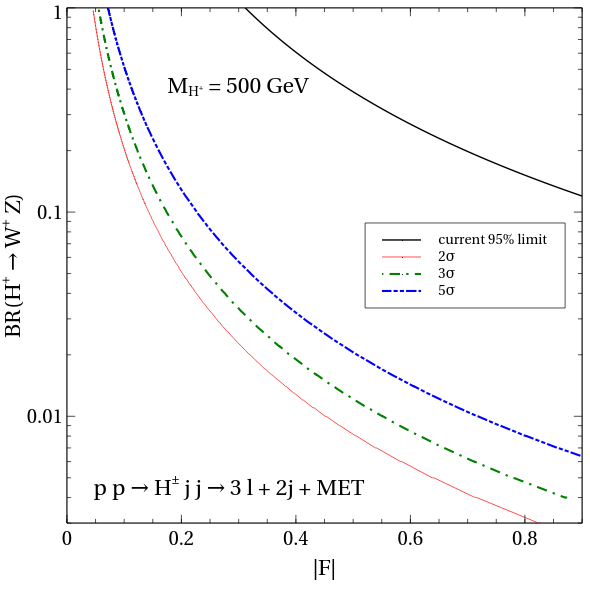}~~~~~~ 
\includegraphics[height = 7 cm, width = 7.5 cm]{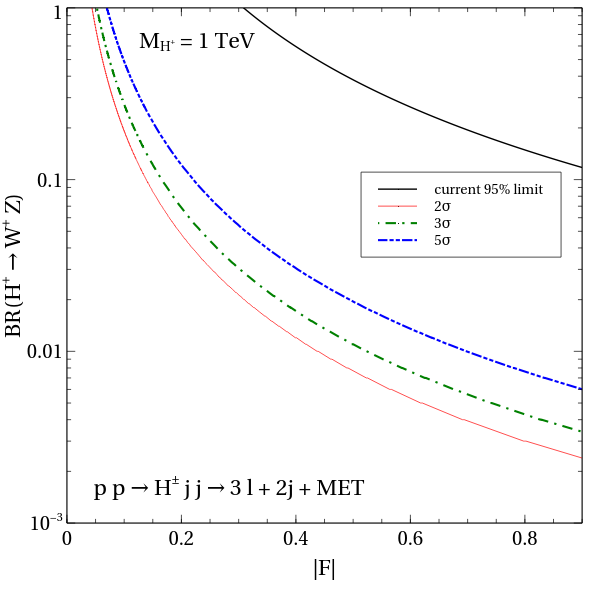}
\caption{The 2$\sigma$, 3$\sigma$ and 5$\sigma$ contours
for the $p p \to H^\pm j j,~H^+ \to W^+ Z$ signal in the $|F|$-BR($H^+ \to W^+ Z$) plane for different $M_{H^+}$. The region above the black curve is ruled out at 95$\%$ confidence level (CL) by the CMS search for $pp \to H^\pm j j,~H^+ \to W^+ Z$ channel.}
\label{fig:contour_3lvjj}
\end{figure}

We remark here that $N_S$ for each BP can be extrapolated for arbitrary $F$ and BR($H^+ \to W^+ Z$ by appropriately scaling the number corresponding to the ($F$,BR($H^+ \to W^+ Z$)) = (0.4,0.4) reference point.
Given the MVA outperforms the CBA for all the BPs, the numbers from the former scaled to draw contours in the 
BR($H^+ \to W^+ Z$) vs $|F|$ plane as shown in Fig.~\ref{fig:contour_3lvjj}. 


\begin{center}
\begin{table}[htb!]
\centering
\scalebox{0.8}{%
\begin{tabular}{|c|c|c|}\hline
\multicolumn{3}{|c|}{$M_{H^+}=200$ GeV} \\ \hline
 & Process  & Yield at 15 ab $^{-1}$ \\ \hline \hline
\multirow{3}{*}{Background}   
 & $W^\pm Z j j$             & $3502$ \\ 
 & $Z Z j j$            & $130$ \\ \cline{2-3} 
 & Total                      & $3632$ \\ \hline
\multicolumn{2}{|c|}{Signal ($pp\to H^\pm j j \to  3l + 2j +\met$)} & $3506$ \\\hline 
\multicolumn{2}{|c|}{Significance ($0\%$ $\sigma_{sys\_un}$)} & $51.3$ \\ \hline  
\multicolumn{2}{|c|}{Signal significance ($2\%~(5\%)$ $\sigma_{sys\_un}$)}           & $30.5~(14.6)$  \\\hline
\end{tabular}}
\quad
\scalebox{0.8}{%
\begin{tabular}{|c|c|c|}\hline
\multicolumn{3}{|c|}{$M_{H^+}=300$ GeV} \\ \hline
 & Process  & Yield at 15 ab $^{-1}$ \\ \hline \hline
\multirow{3}{*}{Background}   
 & $W^\pm Z j j$             & $1119$ \\ 
 & $Z Z j j$            & $33$ \\ \cline{2-3}
 & Total                      & $1152$ \\ \hline
\multicolumn{2}{|c|}{Signal ($pp\to H^\pm j j \to  3l + 2j +\met$)} & $1955$ \\\hline 
\multicolumn{2}{|c|}{Significance ($0\%$ $\sigma_{sys\_un}$)} & $47.5$ \\ \hline  
\multicolumn{2}{|c|}{Signal significance ($2\%~(5\%)$ $\sigma_{sys\_un}$)}           & $36.8~(21.2)$  \\\hline
\end{tabular}}
\bigskip

\scalebox{0.8}{%
\begin{tabular}{|c|c|c|}\hline
\multicolumn{3}{|c|}{$M_{H^+}=500$ GeV} \\ \hline
 & Process  & Yield at 15 ab $^{-1}$ \\ \hline \hline
\multirow{3}{*}{Background}   
 & $W^\pm Z j j$             & $320$ \\ 
 & $Z Z j j$            & $9$ \\ \cline{2-3}
 & Total                      & $329$ \\ \hline
\multicolumn{2}{|c|}{Signal ($pp\to H^\pm j j \to  3l + 2j +\met$)} & $1179$ \\\hline 
\multicolumn{2}{|c|}{Significance ($0\%$ $\sigma_{sys\_un}$)} & $47.3$ \\ \hline  
\multicolumn{2}{|c|}{Signal significance ($2\%~(5\%)$ $\sigma_{sys\_un}$)}           & $42.4~(30.3)$  \\\hline
\end{tabular}}
\quad
\scalebox{0.8}{%
\begin{tabular}{|c|c|c|}\hline
\multicolumn{3}{|c|}{$M_{H^+} = 1$ TeV} \\ \hline
 & Process  & Yield at 15 ab $^{-1}$ \\ \hline \hline
\multirow{3}{*}{Background}   
 & $W^\pm Z j j$             & $14$ \\ 
 & $Z Z j j$            & $1$ \\ \cline{2-3} 
 & Total                      & $15$ \\ \hline
\multicolumn{2}{|c|}{Signal ($pp\to H^\pm j j \to  3l + 2j +\met$)} & $311$ \\\hline 
\multicolumn{2}{|c|}{Significance ($0\%$ $\sigma_{sys\_un}$)} & $37.2$ \\ \hline  
\multicolumn{2}{|c|}{Signal significance ($2\%~(5\%)$ $\sigma_{sys\_un}$)}           & $36.7~(34.6)$  \\\hline
\end{tabular}}
\caption{The signal and background yields at $15~{\rm ab}^{-1}$ for $M_{H^+}$ = 200 GeV, 300 GeV, 500 GeV and 1 TeV along with signal significances for the $3l + 2j + \met$ channel after the BDTD analysis. The signal yields are computed for the ($F$,BR($H^+ \to W^+ Z$)) = (0.4,0.4) reference point.}
\label{3lvjj_BDTD}
\end{table}
\end{center}

\subsection{The $2b + 1l + 2j+\met$ channel}
\label{sec:2b2jlnu}

In the previous subsection we explored VBF production of the charged Higgs $H^{\pm}$ where one forward and one backward light jet is produced along with $H^\pm$ and $H^\pm$ further decays into $W^\pm$ and $Z$. In this subsection, the $H^\pm \to t b$ decay is probed with the $t$ decaying leptonically leading to a final state containing two $b$-tagged jets, two light jets, one isolated lepton (leptons include electron, muon) and 
$\met$. Forward and backward jet-tagging has been done using the criteria mentioned earlier. Here, the dominant background contribution comes from top pair production. There are two possibilities, \textit{viz.} semi-leptonic decay of $t \bar t$, where one of the top quarks decays hadronically or full leptonic decay of $t \bar t$, where both the top quarks decay leptonically. The next dominant contribution comes from QCD-QED $2bl\nu jj$ ($l=e,\mu$) production. We also generate the subdominant backgrounds, \textit{viz.} $t\bar{t}h,~t\bar{t}Z,~t\bar{t}W$ and $W$-boson associated single top production or $tW$. Among these the $tW$ background is generated with one extra parton in the final state and here the $W$-bosons are decayed in all possible channels which can give rise to the final state of our interest. The cross sections for the chosen signal benchmarks and background processes are given in Table~\ref{crosssecvbf}. We impose the following generation-level cuts while generating these backgrounds to better handle the statistics for large backgrounds : 
\bea
p_{T}^{j,b}> 20~\text{GeV},~ p_{T}^l > 10~\text{GeV}, ~|\eta_{j,b,l}|< 5.0, ~\Delta R_{j,b,l}> 0.2
\eea

\begin{table}[htpb!]
\begin{center}\scalebox{0.95}{
\begin{tabular}{|c|c|c|}
\hline
Signal / Backgrounds  & Process & Cross section $\sigma$ (fb) \\ \hline \hline
 Signal &  & \\ 
 BP1 ($M_{H^+} = 200$ GeV) & & 123.37\\
 BP2 ($M_{H^+} = 300$ GeV)& $p p \rightarrow H^\pm j j \rightarrow t b j j \rightarrow 2b + 1 l + 2j + \met$ & 76.93\\
 BP3 ($M_{H^+} = 500$ GeV)& & 37.04\\
 BP4 ($M_{H^+} = 1$ TeV)& & 10.62 \\ \hline \hline
Backgrounds & $p p \to t \bar t$ (semileptonic) (NNLO) & 906843.5\\ 
& $p p \to t \bar t$ (leptonic) (NNLO) & 245010.26 \\
& $p p \to 2 b l \nu j j$ (LO) &  69691\\
& $p p \to tW+j$ (LO) &  19733.25 \\
& $p p \to t \bar t h$ (NLO) & 2860 \\
& $p p \to t \bar t Z$ (NLO) & 3477.02 \\
& $p p \to t \bar{t} W$ (LO) &  986.58 \\
\hline
\end{tabular}}
\end{center}
\caption{The cross sections of signal and backgrounds for the process $p p \rightarrow H^\pm j j \rightarrow t b j j \rightarrow 2b + 1 l + 2j + \met$. The signal cross sections are computed for the ($F$,BR($H^+ \to t \bar{b}$)) = (0.4,0.4) reference point.}
\label{crosssecvbf}
\end{table}

In addition to the generation-level cuts we add stronger trigger-level cuts on $|\eta_{j,b,l}|$ as : $|\eta_{j,b,l}| < 2.5 $. To first perform a rectangular CBA , a set of relevant kinematic variables are constructed and their distributions are observed. Here we mostly present those distributions that illustrate the features of VBF topology and shall apply suitable cuts on the variables for maximising the significance.

\begin{figure}
\centering
\includegraphics[scale=0.37]{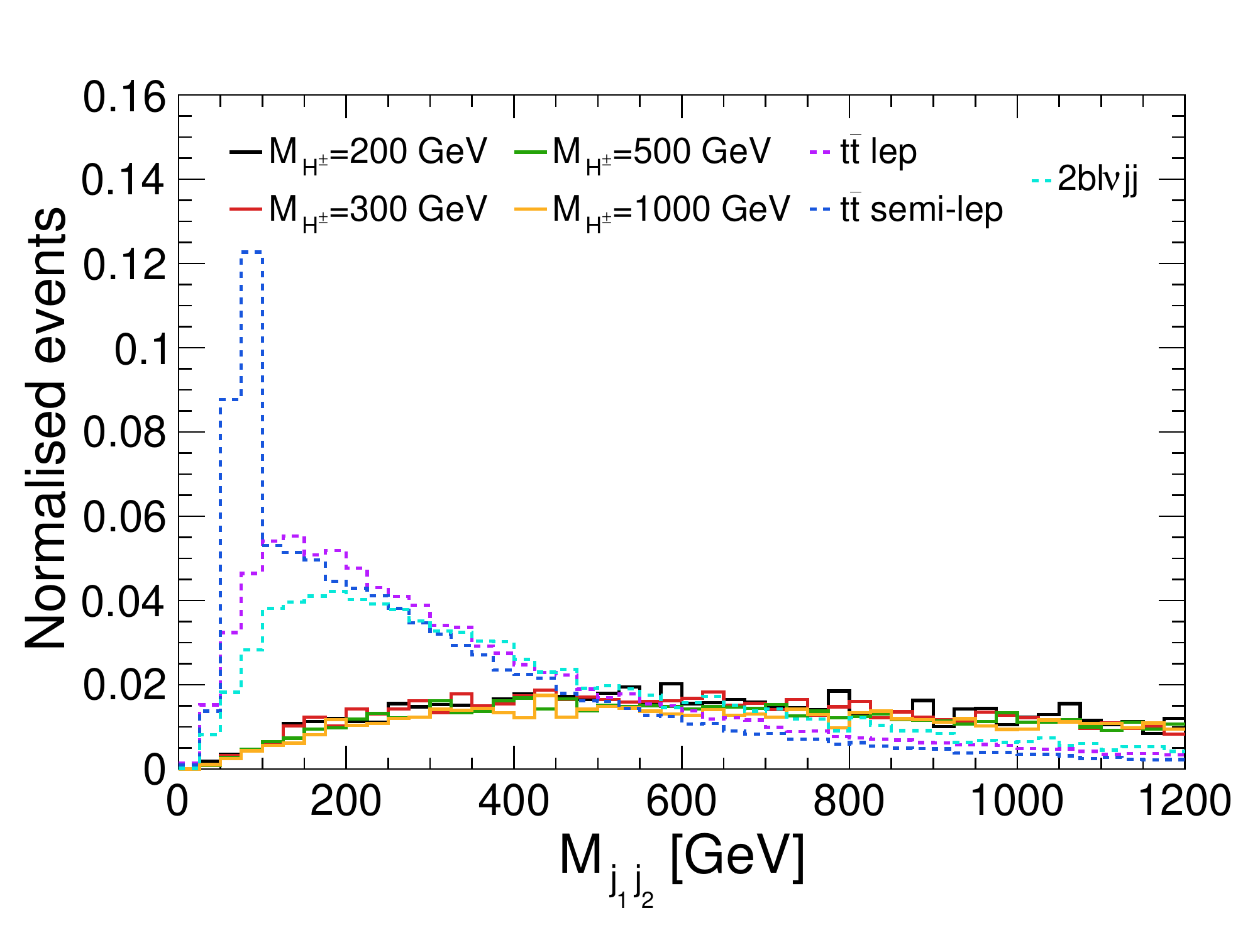}
\includegraphics[scale=0.37]{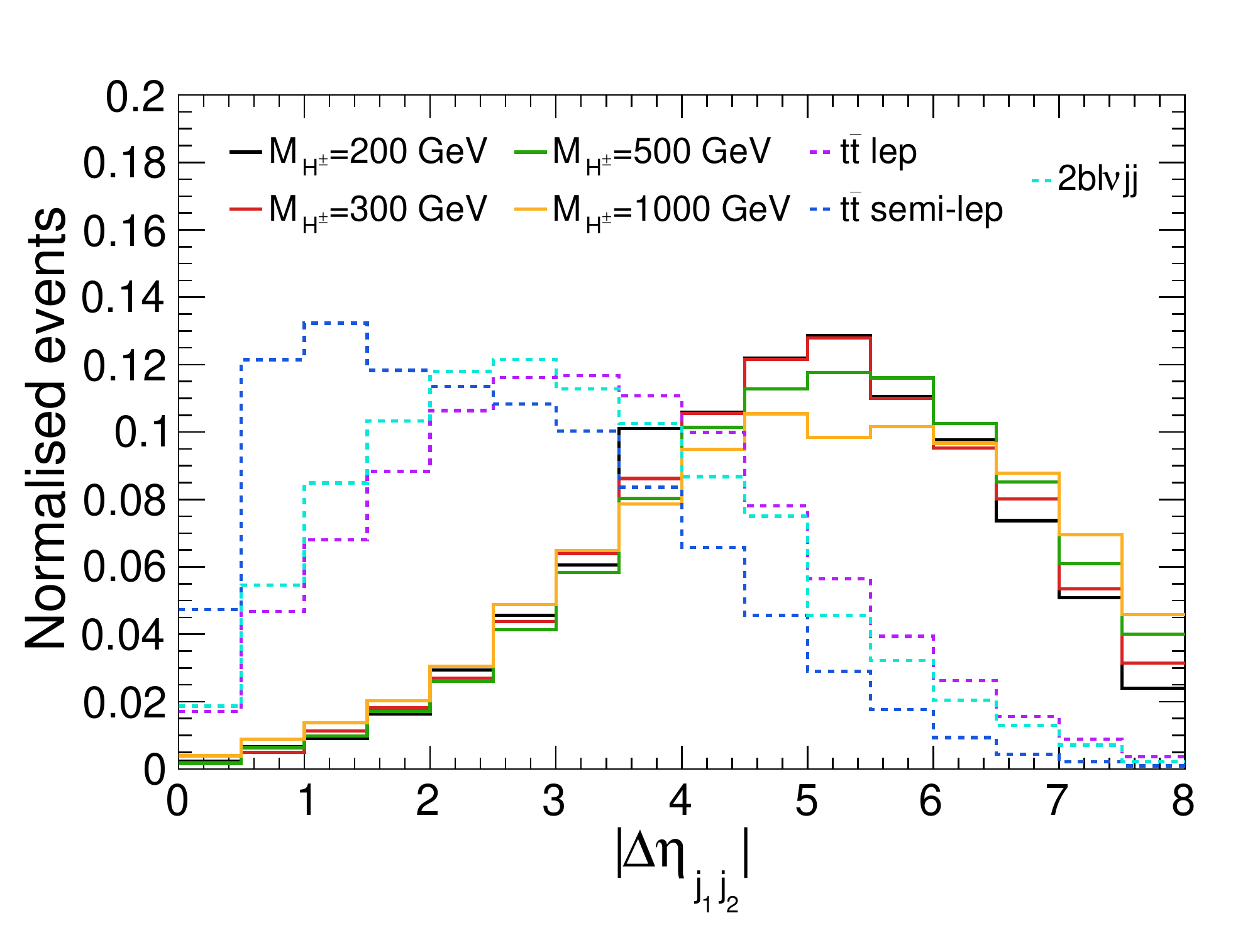}\\
\includegraphics[scale=0.37]{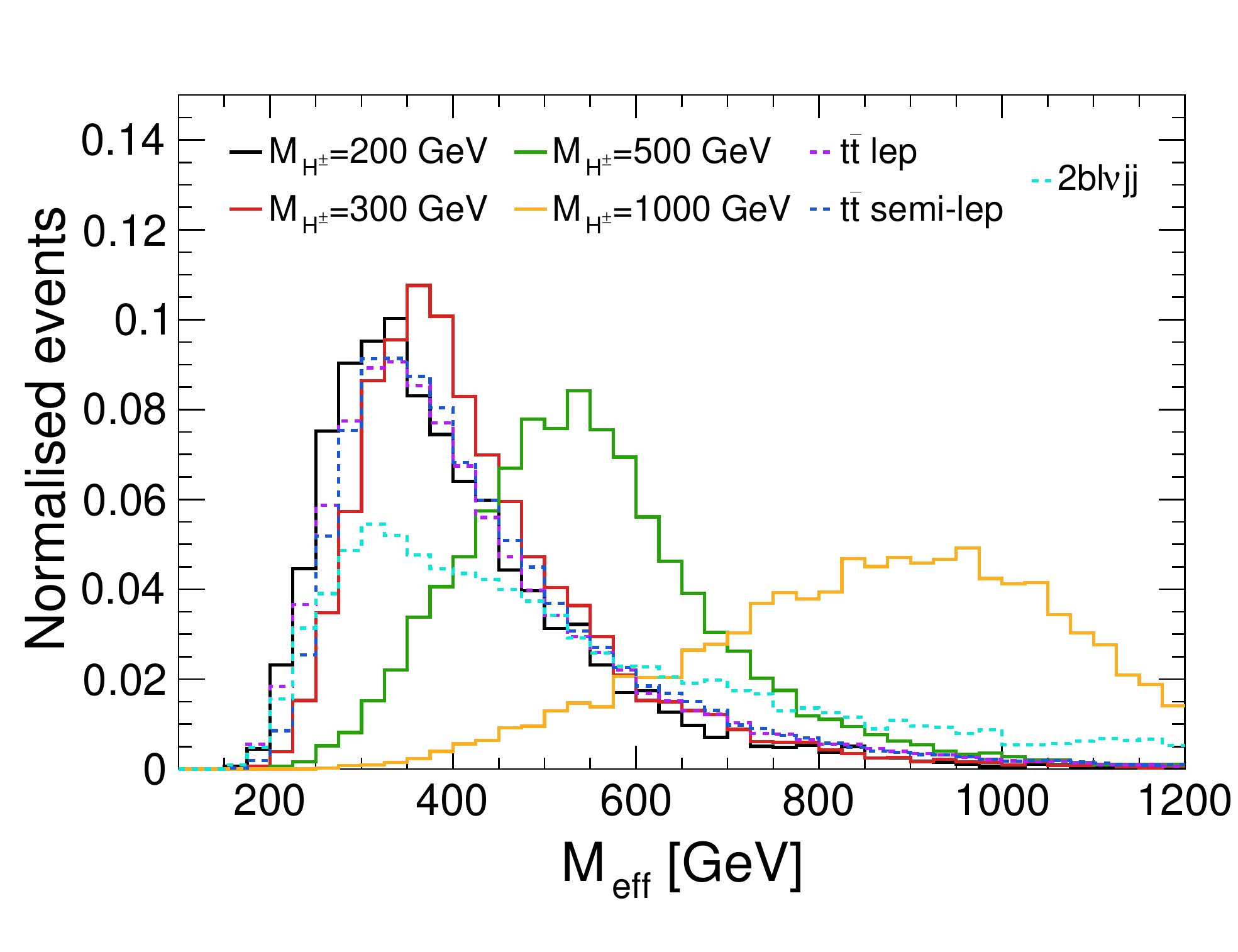}
\includegraphics[scale=0.37]{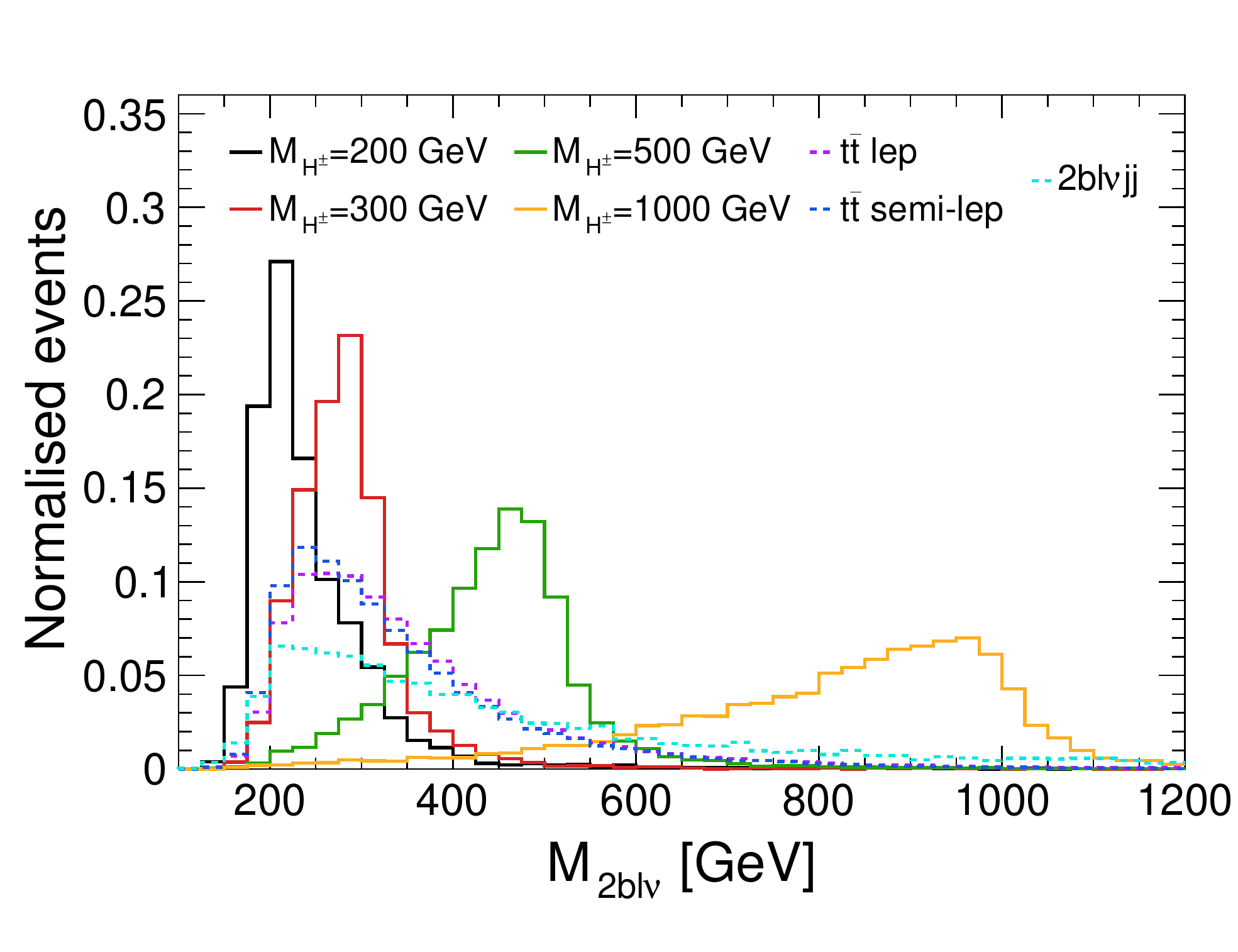} \\
\includegraphics[scale=0.37]{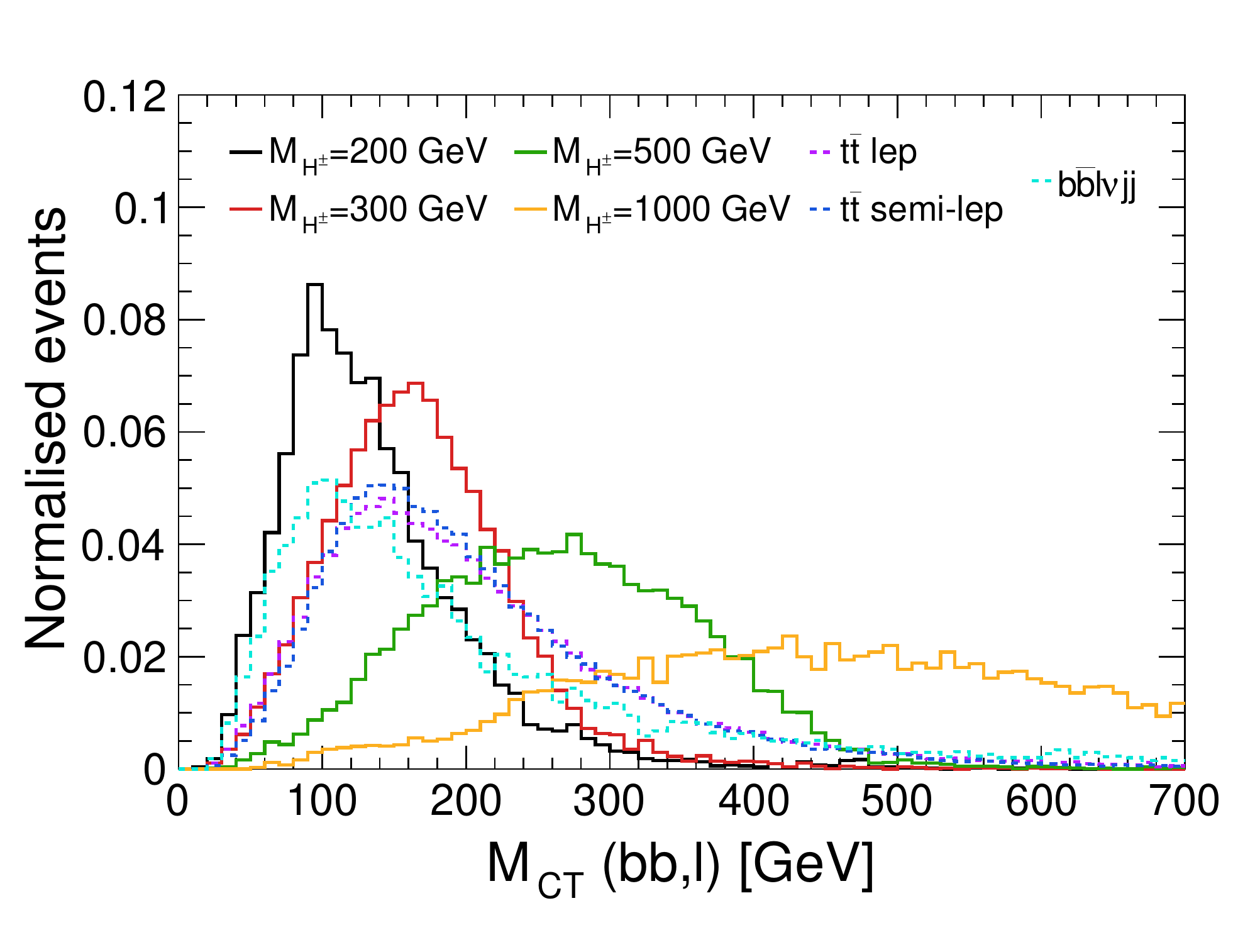}
\caption{ Normalised distributions of $M_{j_1j_2}$, $|\Delta \eta_{j_1j_2}|$, $M_{eff}$, $M_{2bl\nu}$ and $M_{CT}(bb,l)$ for the $2b + 1l + 2j+\slashed{E_T}$ channel.}
\label{fig:2b2jlnu}
\end{figure}

\begin{itemize}
\item $B_1$ and $B_2$ : The absence of hadronic activity in the central region is evident from the distributions of $M_{j_1j_2}, |\Delta \eta_{j_1j_2}|$. The distribution of $|\Delta \eta_{j_1j_2}|$ peaks at a larger value for the signal as compared to the background processes. Consequently $M_{j_1j_2}$ is distributed towards higher mass values. The more separated the two jets are, the larger will be their invariant mass. We also explicitly put a central jet-veto condition, implying the veto on the additional production
of jets in the central region (between two forward and backward jets).

\item $B_3$ : Similar to the previous channel, the charged Higgs mass can be reconstructed for this case too since there is only one neutrino in the final state that comes from the decay of the $W$. First, we reconstruct the four top masses from two possible choices of pairing with bottom quark and the two possible solution corresponding to the $z$-component of $\met$ ($\slashed{E}_{T,z}$) as described in section \ref{subsec:4A}. 
 Here we choose those combinations for which the reconstructed top mass is closest to $M_t = 173$ GeV. After that we construct the invariant mass of whole $2bl\nu$-system $M_{2bl\nu}$, which should peak around the charged Higgs mass ($M_{H^+}$) for the signal events. This particular variable plays a crucial role in signal background separation, since for backgrounds the reconstructed mass will not peak around $M_{H^+}$. 

\item $B_4$ : We construct an observable called $M_{eff}$ which is defined as the scalar sum of transverse momenta of all visible particles and $\met$. This variable helps us separate signal from background especially in the high $M_{H^{+}}$ region, because the decay products of a heavy charged Higgs will naturally be boosted rendering higher $M_{eff}$ for the whole system.

\item $B_5$ : We also utilise the contransverse mass variable, $M_{CT}(a,b)$~\cite{Tovey:2008ui} which is an invariant quantity for two
 objects $a$ and $b$ having contra-linear (opposite in direction) and equal magnitude boost. This variable is defined as, $$M^2_{CT}(a,b)=M^2(a
)+M^2(b)+2(E_T(a)E_T(b)+\vec{p_T}(a).\vec{p_T}(b)).$$ Here, $M(a)$ and $\vec{p_T}(a)$ corresponds to the invariant mass and transverse
 momentum vector respectively for the object $a$, whereas the transverse energy, $E_T$ is defined as $\sqrt{|\vec{p}_T|^2 + M^2}$. Since in
 case of the signal $b \bar b$ and $l$ come from the decay of charged Higgs, the end point of $M_{CT}(b \bar b, l)$ distribution in this
 case will indicate the charged Higgs mass in the large $M_{H^{+}}$ limit. The reason is the following :  $M_{CT}(b \bar b, l)$ is
 constructed from the transverse momenta of the $b \bar{b}$ and $l$ system where both of them are bounded from above by $M_{H^{+}}/2$ by
 construction. Therefore, $M_{CT}(b \bar b, l)$ can take a maximum value of $M_{H^{+}}$, the remaining terms being negligible compared to
 $M_{H^{+}}$. We can see from Fig.~\ref{fig:2b2jlnu}, this variable possesses good discriminating power. Although we have not used this
 $M_{CT}(b \bar b, l)$ in the cut-based analysis since it becomes redundant after applying cuts on the other discriminating variables,
 because of its correlation with variables such as $M_{2bl\nu}$, we have utilised this variable as an input for our BDTD analysis which
 will be discussed shortly.
 \end{itemize}

The optimised cuts on the aforementioned variables for each of the benchmark points are given in Table.~\ref{bblnujjcuts}. These cuts are applied to  select  the  events  over  and  above  the  basic  generation-level and trigger-level cuts discussed above.

\begin{table}[htpb!]
\begin{center}\scalebox{1.0}{
\begin{tabular}{|c|c|c|c|c|}
\hline
$M_{H^+}$ & $B_1$ & $B_2$ & $B_3$ & $B_4$ \\ \hline \hline
200 GeV  & $|\Delta \eta_{j_1 j_2}| > 3$  & $M_{j_1 j_2} > 500$ GeV & 100 GeV $< M_{H^+} <$ 300 GeV & -
\\
300 GeV  & $|\Delta \eta_{j_1 j_2}| > 3$  & $M_{j_1 j_2} > 500$ GeV & 100 GeV $< M_{H^+} <$ 300 GeV &  -
\\
500 GeV  & $|\Delta \eta_{j_1 j_2}| > 3$  & $M_{j_1 j_2} > 500$ GeV & $M_{H^+} > 350 $GeV & $M_{eff} > 400$ GeV
\\
1 TeV  & $|\Delta \eta_{j_1 j_2}| > 3$   & $M_{j_1 j_2} > 500$ GeV  & $M_{H^+} > 650 $GeV & $M_{eff} > 750$ GeV
\\
\hline
\end{tabular}}
\end{center}
\caption{The optimised selection cuts for the $2b + 1l + 2j+\met$ channel.}
\label{bblnujjcuts}
\end{table}
\begin{table}[htpb!]
\resizebox{16cm}{!}{
\begin{tabular}{|c|c|c|c|c|c|c|}
\hline
 &  \hspace{5mm} {\texttt{NTrees}} \hspace{5mm} & \hspace{5mm} {\texttt{MinNodeSize}} \hspace{5mm} & \hspace{5mm} {\texttt{MaxDepth}}~~ \hspace{5mm} & \hspace{5mm} {\texttt{nCuts}} ~~\hspace{5mm} & \hspace{5mm} {\texttt{KS-score for}}~~\hspace{5mm}  & \hspace{5mm} {\texttt{BDT Score}} \hspace{5mm} \\
 & & & & & {\texttt{Signal(Background)}} &\\
\hline
\hline
\hspace{5mm} BP1 \hspace{5mm} & 40 & 3 \% & 2 & 40 & 0.31(0.145) & 0.4 \\ \hline
\hspace{5mm} BP2 \hspace{5mm} & 100 & 3 \% & 2 & 20 & 0.396(0.236) & 0.23 \\ \hline
\hspace{5mm} BP3 \hspace{5mm} & 100 & 3 \% & 2 & 20 & 0.335(0.43)  & 0.25 \\ \hline
\hspace{5mm} BP4 \hspace{5mm} & 100 & 3 \% & 2 & 20 & 0.029(0.044) & 0.32 \\ \hline
\end{tabular}}
\caption{Tuned BDT parameters for BP1, BP2, BP3 and BP4 for the $2b + 1l + 2j+\met$ channel.}
\label{BDT-param2}
\end{table}
\begin{figure}[tp!]{\centering
\subfigure[]{
\includegraphics[width=3in,height=2.55in]{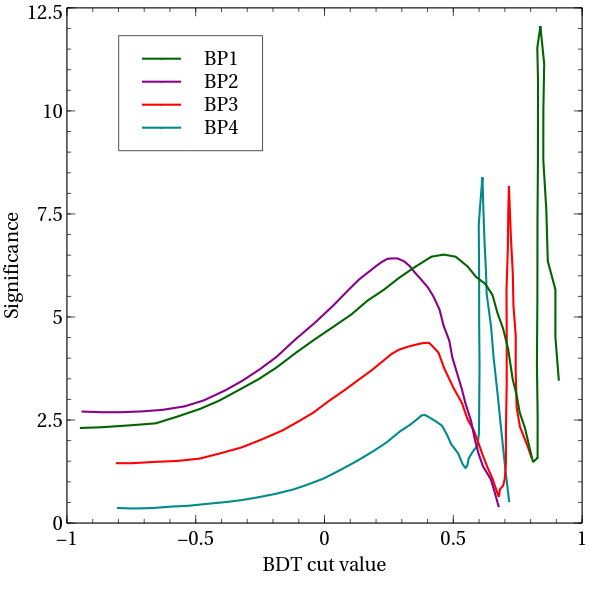}}
\subfigure[]{
\includegraphics[width=2.8in,height=2.55in]{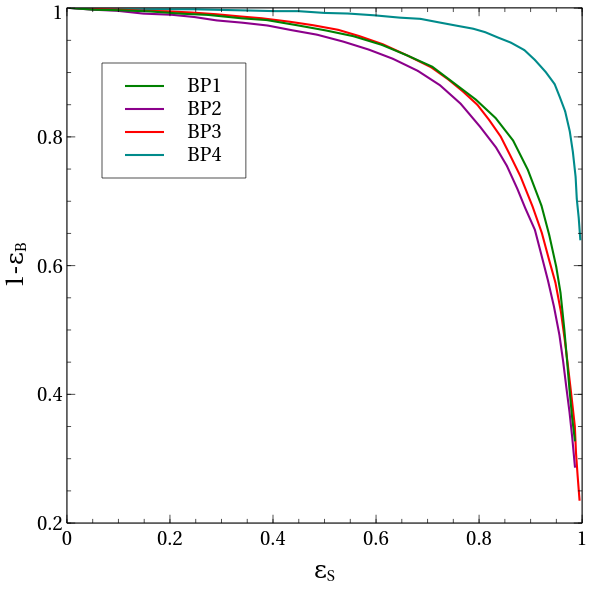}}}
\caption{ 
(a) Variation of significance with BDT-score for $2b + 1l + 2 j + \slashed{E_T}$ channel. (b) ROC curves for chosen benchmark points for $2 b + 1l + 2 j + \slashed{E_T}$ channel.}
\label{ROC-BDTScore-2b1l2jv}
\end{figure}

\begin{table}[htpb!]
\begin{center}
\scalebox{0.9}{
\begin{tabular}{||c|c|c|c||}
\hline
\multicolumn{1}{||c|}{ $M_{H^+} = 200$ GeV} &
\multicolumn{3}{|c||}{NEV ($\mathcal{L}=$ 15 ab$^{-1}$)} \\ 
\cline{2-4}
  & $B_1$ & $B_2$ & $B_3$  \\
\hline
Signal  &   33079 & 27362 & 24870  \\
\hline\hline
$t \bar t$ semileptonic  &  4.9$\times 10^7$  & 1.7$\times 10^7$ & 9.6$\times 10^6$  \\
\hline
$t \bar t$ leptonic  &  3.6$\times 10^7$ & 1.5$\times 10^7$ & 7.7$\times 10^6$  \\
\hline
$2b l \nu jj$  &  5.2$\times 10^6$ & 2.9$\times 10^6$ & 811360  \\
\hline
$tW$  &  188553 & 65018 & 19506  \\
\hline
$t \bar t h$  &  30598 & 12777 & 4371 \\
\hline
$t \bar t Z$  &  73202 & 37124 & 13595  \\
\hline
$t \bar t W$  &  16703 & 4898 & 2261  \\
\hline
\multicolumn{1}{||c|}{Signal yield = 24870} & \multicolumn{1}{c|}{Total background yield = 1.8$\times 10^7$} & \multicolumn{2}{c||}{Signal significance ($0\%$ $\sigma_{sys\_un}$) = $5.9$} \\
\hline
\multicolumn{1}{||c|}{} & \multicolumn{1}{c|}{} & \multicolumn{2}{c||}{Signal significance ($2\%~(5\%)$ $\sigma_{sys\_un}$) = $0.07~(0.03)$ } \\
\hline
\hline
\end{tabular}}
\caption{Signal and background yields after applying the selection cuts and projected signal significance for BP1 with 15 ab$^{-1}$ integrated luminosity. The signal yields are computed for the ($F$,BR($H^+ \to t \bar{b}$)) = (0.4,0.4) reference point.} 
\label{bp1}
\end{center}
\end{table}

\begin{table}[htpb!]
\begin{center}
\scalebox{0.9}{
\begin{tabular}{||c|c|c|c||}
\hline
\multicolumn{1}{||c|}{ $M_{H^+} = 300$ GeV} &
\multicolumn{3}{|c||}{NEV ($\mathcal{L}=$ 15 ab$^{-1}$)} \\ 
\cline{2-4}
  & $B_1$ & $B_2$ & $B_3$ \\
\hline
Signal  &   48351 & 39091 & 33816  \\
\hline\hline
$t \bar t$ semileptonic  &  4.9$\times 10^7$  & 1.7$\times 10^7$ & 9.6$\times 10^6$  \\
\hline
$t \bar t$ leptonic  &  3.6$\times 10^7$ & 1.5$\times 10^7$ & 7.7$\times 10^6$  \\
\hline
$2b l \nu jj$  &  5.2$\times 10^6$ & 2.9$\times 10^6$ & 811360  \\
\hline
$tW$  &  188553 & 65018 & 19506  \\
\hline
$t \bar t h$  &  30598 & 12777 & 4371 \\
\hline
$t \bar t Z$  &  73202 & 37124 & 13595  \\
\hline
$t \bar t W$  &  16703 & 4898 & 2261  \\
\hline
\multicolumn{1}{||c|}{Signal yield = 33816} & \multicolumn{1}{c|}{Total background yield = 1.8$\times 10^7$} & \multicolumn{2}{c||}{Signal significance ($0\%$ $\sigma_{sys\_un}$) = $8.0$} \\
\hline
\multicolumn{1}{||c|}{} & \multicolumn{1}{c|}{} & \multicolumn{2}{c||}{Signal significance ($2\%~(5\%)$ $\sigma_{sys\_un}$) = $0.09~(0.04)$} \\
\hline
\hline
\end{tabular}}
\caption{Signal and background yields after applying the selection cuts and projected signal significance for BP2 with 15 ab$^{-1}$ integrated luminosity. The signal yields are computed for the ($F$,BR($H^+ \to t \bar{b}$)) = (0.4,0.4) reference point.} 
\label{bp2}
\end{center}
\end{table}

\begin{table}[htpb!]
\begin{center}
\scalebox{0.9}{
\begin{tabular}{||c|c|c|c|c||}
\hline
\multicolumn{1}{||c|}{ $M_{H^+} = 500$ GeV} &
\multicolumn{4}{|c||}{NEV ($\mathcal{L}=$ 15 ab$^{-1}$)} \\ 
\cline{2-5}
  & $B_1$ & $B_2$ & $B_3$ & $B_4$ \\
\hline
Signal  &   25195 & 21194 & 19484 & 17586 \\
\hline\hline
$t \bar t$ semileptonic  &  3.1$\times 10^7$  & 1.1$\times 10^7$ & 5.6$\times 10^6$ & 5.1$\times 10^6$ \\
\hline
$t \bar t$ leptonic  &  1.6$\times 10^7$ & 6.5$\times 10^6$ & 3.9$\times 10^6$ & 3.1$\times 10^6$ \\
\hline
$2b l \nu jj$  &  5.2$\times 10^6$ & 2.9$\times 10^6$ & 2.3$\times 10^6$ & 1.9$\times 10^6$ \\
\hline
$tW$  &  188553 & 65018 & 52015 & 45513 \\
\hline
$t \bar t h$  &  30598 & 12777 & 9079 & 7734 \\
\hline
$t \bar t Z$  &  73202 & 37124 & 27189 & 23529  \\
\hline
$t \bar t W$  &  16703 & 4898 & 3140 & 2763 \\
\hline
\multicolumn{1}{||c|}{Signal yield = 17586} & \multicolumn{1}{c|}{Total background yield = 1.0$\times 10^7$} & \multicolumn{3}{c||}{Signal significance ($0\%$ $\sigma_{sys\_un}$) = $5.6$} \\
\hline
\multicolumn{1}{||c|}{} & \multicolumn{1}{c|}{} & \multicolumn{3}{c||}{Signal significance ($2\%~(5\%)$ $\sigma_{sys\_un}$) = $0.09~(0.04)$} \\
\hline
\hline
\end{tabular}}
\caption{Signal and background yields after applying the selection cuts and projected signal significance for BP3 with 15 ab$^{-1}$ integrated luminosity. The signal yields are computed for the ($F$,BR($H^+ \to t \bar{b}$)) = (0.4,0.4) reference point.}
\label{bp3}
\end{center}
\end{table}

\begin{table}[htpb!]
\begin{center}
\scalebox{0.9}{
\begin{tabular}{||c|c|c|c|c||}
\hline
\multicolumn{1}{||c|}{ $M_{H^+} = 1$ TeV} &
\multicolumn{4}{|c||}{NEV ($\mathcal{L}=$ 15 ab$^{-1}$)} \\ 
\cline{2-5}
  & $B_1$ & $B_2$ & $B_3$ & $B_4$ \\
\hline
Signal  &   6466 & 5553 & 5136 & 4149 \\
\hline\hline
$t \bar t$ semileptonic  &  3.1$\times 10^7$  & 1.1$\times 10^7$ & 1.8$\times 10^6$ & 679093 \\
\hline
$t \bar t$ leptonic  &  1.6$\times 10^7$ & 6.5$\times 10^6$ & 872734 & 269754 \\
\hline
$2b l \nu jj$  &  5.2$\times 10^6$ & 2.9$\times 10^6$ & 1.5$\times 10^6$ & 1.0$\times 10^6$ \\
\hline
$tW$  &  188553 & 65018 & 39011 & 26007 \\
\hline
$t \bar t h$  &  30598 & 12777 & 4371 & 3362 \\
\hline
$t \bar t Z$  &  73202 & 37124 & 10458 & 5752 \\
\hline
$t \bar t W$  &  16703 & 4898 & 1381 & 879 \\
\hline
\multicolumn{1}{||c|}{Signal yield = 4149} & \multicolumn{1}{c|}{Total background yield = 1.9$\times 10^6$} & \multicolumn{3}{c||}{Signal significance ($0\%$ $\sigma_{sys\_un}$) = $3.0$} \\
\hline
\multicolumn{1}{||c|}{} & \multicolumn{1}{c|}{} & \multicolumn{3}{c||}{Signal significance ($2\%~(5\%)$ $\sigma_{sys\_un}$) = $0.11~(0.04)$} \\
\hline
\hline
\end{tabular}}
\caption{Signal and background yields after applying the selection cuts and projected signal significance for BP4 with 15 ab$^{-1}$ integrated luminosity. The signal yields are computed for the ($F$,BR($H^+ \to t \bar{b}$)) = (0.4,0.4) reference point.}
\label{bp4}
\end{center}
\end{table}

Tables~\ref{bp1}-\ref{bp4} show the cut-flow for the signal and the background processes, yielding
a fair indication of the efficiency of each cut. We have also calculated the projected significance for each benchmark point at 27 TeV LHC with 15 ab$^{-1}$ luminosity. 
 
We can see in Tables~\ref{bp1}-\ref{bp4} that BP2 fares best in terms of signal significance, followed by BP1, BP3 and BP4. Therefore, BP1 which has the lowest mass of charged Higgs, has lowest discriminating capacity between the signal and background. However, its large cross-section enables us to achieve substantial signal significance. On the other hand, BP4, which corresponds to the heaviest charged Higgs has the best discriminating power among all the BPs. But its meagre production cross-section makes it difficult to obtain a good significance. However, as we can rely on various good separation variables in case of BP4, we expect significant improvement over the cut-based analysis when we perform multivariate analysis. Evidently, for BP2 and BP3 there are good optimisations between production cross section and signal-background separation and consequently they yield considerable significance even with the rectangular CBA.


Next we proceed to perform the MVA using BDTD algorithm with the hope of improving the significance with respect to that obtained from CBA. Details of BDTD algorithm along with the definitions of the BDTD parameters are already discussed in subsection \ref{subsec:4A}. Thus we shall avoid repeating that here. Let us move on to Table \ref{BDT-param2}, where all the tuned BDT parameters like {\texttt NTrees}, {\texttt MinNodeSize}, {\texttt MaxDepth}, {\texttt nCuts}, {\texttt KS-scores} (both for signal and backgrounds) and {\texttt BDT score} for four benchmarks are tabulated. The signal and background distributions along with their KS-scores are depicted in Fig.\ref{KSscore-2b1l2jv} in Appendix \ref{Plots}. From this figure it can be inferred that the best possible signal background separation occurs for BP4. The variation of significance with BDT cut value or BDT score, depicted in Fig.\ref{ROC-BDTScore-2b1l2jv}(a) shows that the significance becomes highest for a particular value of BDT score. While computing the final signal significance, we use the BDT cut values for which the significances are maximised for four benchmarks. Fig.\ref{ROC-BDTScore-2b1l2jv}(b) represents the ROC curves for four benchmarks. it can be clearly seen that the background rejection efficiency for BP4 is best compare to the rest. Inspite of having the best background rejection efficiency, BP4 has the lowest signal significance owing to small signal cross section (Table \ref{crosssecvbf}). According to the ability to differentiate between the signal and background, let us choose following thirteen kinematic variables:

\bea
&& M_{bb},~p_T^{bb},~\Delta R_{bb},~M_{j_1j_2},~\eta_{j_1j_2},~|\Delta \eta_{j_1j_2}| ,~M_{eff},~\Delta\phi_{l\met}, \nonumber \\
&& \Delta\phi_{bb,l j_1j_2},~
M_{CT}(bb,l), ~M_{bbj_1},~M_{bbj_2},~M_{2bl\nu}.
\label{2b2jlnu:eq1}
\eea


where $\eta_{j_1j_2}$ and $|\Delta \eta_{j_1j_2}|$ are the pseudorapidity of the system of two leading light jets and $\Delta\eta_{j_1j_2}$ between the two leading jets respectively. 
$\Delta\phi_{bb,l jj}$ is the azimuthal angle separation between the two $b$-jet system and the system of lepton with two leading light jets, and the rest of the variables have their usual meaning. The most important variables to distinguish between the signal and backgrounds are $\Delta R_{bb},~M_{j_1j_2},~\eta_{j_1j_2}, ~M_{eff},~|\Delta \eta_{j_1j_2}|, ~M_{bbj_1},~M_{bbj_2},~M_{2bl\nu}$. While doing the analysis, we have set $F = 0.4$ and we fix the branching ratio of the decay $H^+\to t \bar{b}$ to be $40\%$. The normalised distributions of the best kinematic variables for  different $M_{H^+}$ are shown in Fig.~\ref{fig:2b2jlnu}. After doing the BDTD analysis, we summarise the results in Table ~\ref{2b2jlnu:tab1}.


\begin{center}
\begin{table}[htb!]
\centering
\scalebox{0.75}{%
\begin{tabular}{|c|c|c|}\hline
\multicolumn{3}{|c|}{$M_{H^+}=200$ GeV} \\ \hline
 & Process & Yield at 15 ab$^{-1}$ \\ \hline \hline
\multirow{8}{*}{Background}   
 & $t\bar{t}$ lep              & $2914397$ \\ 
 & $t\bar{t}$ semi-lep         & $8977751$ \\ 
 & $2bl\nu jj$        & $797091$ \\
 & $tW$                        & $55272$\\  
 & $t\bar{t}h$                 & $16688$ \\  
 & $t\bar{t}Z$                 & $13821$ \\ 
 & $t\bar{t}W$                 & $1339$ \\ \cline{2-3} 
 & Total               & $12776359$ \\ \hline
\multicolumn{2}{|c|}{Signal }           & $23261$ \\\hline 
\multicolumn{2}{|c|}{Signal significance ($0\%$ $\sigma_{sys\_un}$)}         &  $6.5$ \\\hline
\multicolumn{2}{|c|}{Signal significance ($2\%~(5\%)$ $\sigma_{sys\_un}$)}           & $0.09~(0.04)$  \\\hline
\end{tabular}}
\quad
\scalebox{0.75}{%
\begin{tabular}{|c|c|c|}\hline
\multicolumn{3}{|c|}{$M_{H^+}=300$ GeV} \\ \hline
 & Process  & Yield at 15 ab$^{-1}$ \\ \hline \hline
\multirow{8}{*}{Background}   
 & $t\bar{t}$ lep             & $6617115$ \\ 
 & $t\bar{t}$ semi-lep        & $18207150$ \\ 
 & $2bl\nu jj$       & $1301479$ \\
 & $tW$                       & $108774$\\  
 & $t\bar{t}h$                & $37645$ \\  
 & $t\bar{t}Z$                & $30693$ \\ 
 & $t\bar{t}W$                & $2952$ \\ \cline{2-3} 
 & Total                      & $26305808$ \\ \hline
\multicolumn{2}{|c|}{Signal }           & $52583$ \\\hline 
\multicolumn{2}{|c|}{Signal significance ($0\%$ $\sigma_{sys\_un}$)}           & $10.2$ \\\hline
\multicolumn{2}{|c|}{Signal significance ($2\%~(5\%)$ $\sigma_{sys\_un}$)}          & $0.1~(0.04)$  \\\hline
\end{tabular}}

\bigskip

\scalebox{0.75}{%
\begin{tabular}{|c|c|c|}\hline
\multicolumn{3}{|c|}{$M_{H^+}=500$ GeV} \\ \hline
 & Process  & Yield at 15 ab$^{-1}$ \\ \hline \hline
\multirow{8}{*}{Background}   
 & $t\bar{t}$ lep             & $3914039$ \\ 
 & $t\bar{t}$ semi-lep        & $10555658$ \\ 
 & $2bl\nu jj$       & $1547140$ \\
 & $tW$                       & $95951$\\  
 & $t\bar{t}h$                & $36787$ \\  
 & $t\bar{t}Z$                & $35596$ \\ 
 & $t\bar{t}W$                & $2568$ \\ \cline{2-3} 
 & Total                      & $16187739$ \\ \hline
\multicolumn{2}{|c|}{Signal }           & $54387$ \\\hline 
\multicolumn{2}{|c|}{Signal significance ($0\%$ $\sigma_{sys\_un}$)}           & $13.5$ \\\hline
\multicolumn{2}{|c|}{Signal significance ($2\%~(5\%)$ $\sigma_{sys\_un}$)}           & $0.17~(0.07)$  \\\hline
\end{tabular}}
\quad
\scalebox{0.75}{%
\begin{tabular}{|c|c|c|}\hline
\multicolumn{3}{|c|}{$M_{H^+}=1$ TeV} \\ \hline
 & Process  & Yield at 15 ab$^{-1}$ \\ \hline \hline
\multirow{8}{*}{Background}   
 & $t\bar{t}$ lep             & $356490$ \\ 
 & $t\bar{t}$ semi-lep        & $1197033$ \\ 
 & $2bl\nu jj$       & $899014$ \\
 & $tW$                       & $87108$\\  
 & $t\bar{t}h$                & $11004$ \\  
 & $t\bar{t}Z$                & $11292$ \\ 
 & $t\bar{t}W$                & $451$ \\ \cline{2-3} 
 & Total                      & $2562392$ \\ \hline
\multicolumn{2}{|c|}{Signal }           & $42525$ \\\hline 
\multicolumn{2}{|c|}{Signal significance ($0\%$ $\sigma_{sys\_un}$)}           & $26.5$ \\\hline
\multicolumn{2}{|c|}{Signal significance ($2\%~(5\%)$ $\sigma_{sys\_un}$)}           & $0.82~(0.33)$  \\\hline
\end{tabular}}
\caption{The signal and background yields at $15~{\rm ab}^{-1}$ for $M_{H^+}$ = 200 GeV, 300 GeV, 500 GeV and 1 TeV along with signal significances for the $2b+1l+ 2j+\met$ channel after the BDTD analysis. The signal yields are computed for the ($F$,BR($H^+ \to t \bar{b}$)) = (0.4,0.4) reference point.}
\label{2b2jlnu:tab1}
\end{table}
\end{center}

Comparing the results obtained from BDTD analysis (Table.~\ref{2b2jlnu:tab1}) and those from cut-based analysis (Table~\ref{bp1}-\ref{bp4}) we can see that BP1 is hardly improved by the BDTD analysis. Since the kinematic variables do not offer a good separation in this case, the BDTD will perform marginally better than an optimized cut-based analysis. The results improve slightly in case of BP2. The improvement becomes striking in case of BP3 and BP4, because these benchmarks correspond to observables with significant discriminating power. Considering $2 \%$ and $5 \%$ systematic uncertainties, the significances has also been calculated and tabulated in Table \ref{2b2jlnu:tab1}.

\begin{figure}
\centering
\includegraphics[scale=0.37]{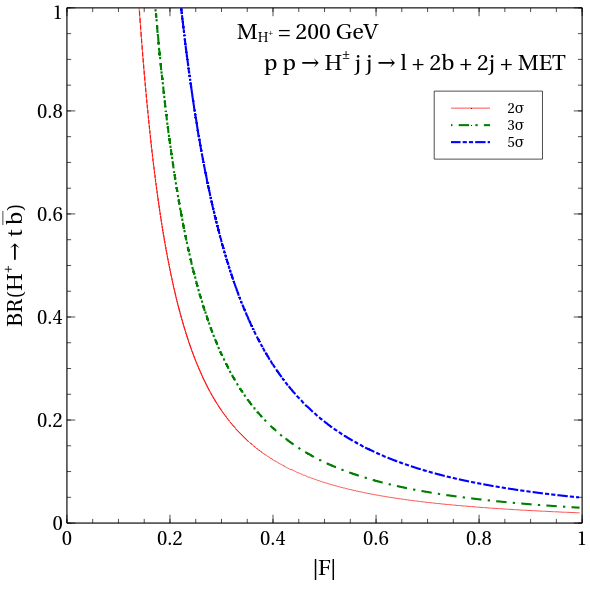}
\includegraphics[scale=0.37]{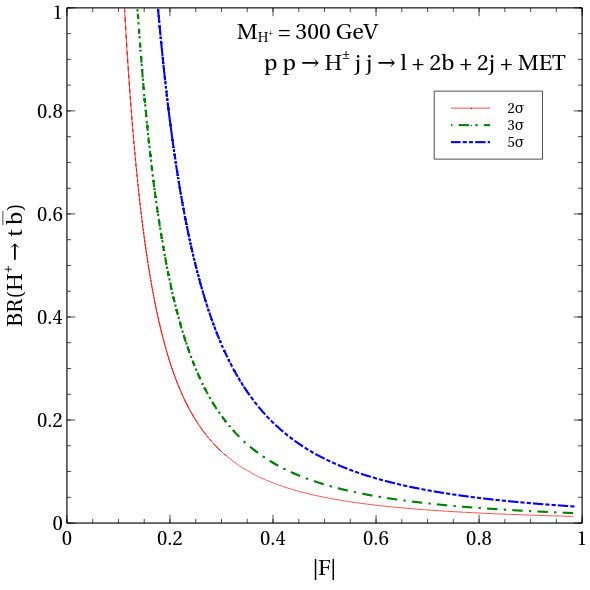}\\
\includegraphics[scale=0.37]{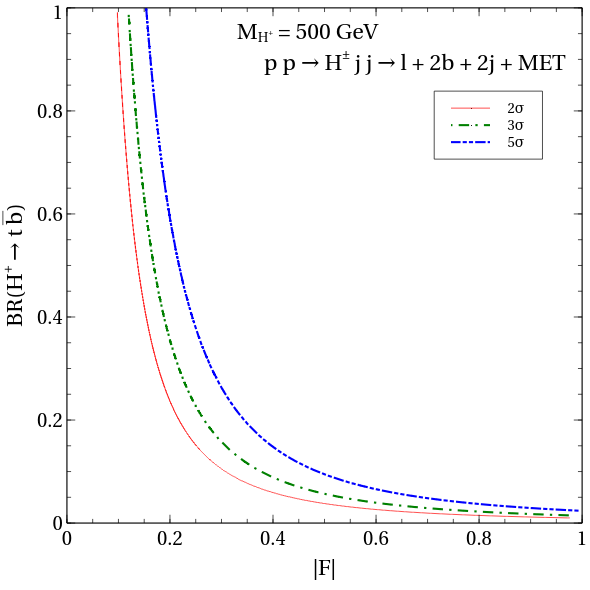}
\includegraphics[scale=0.37]{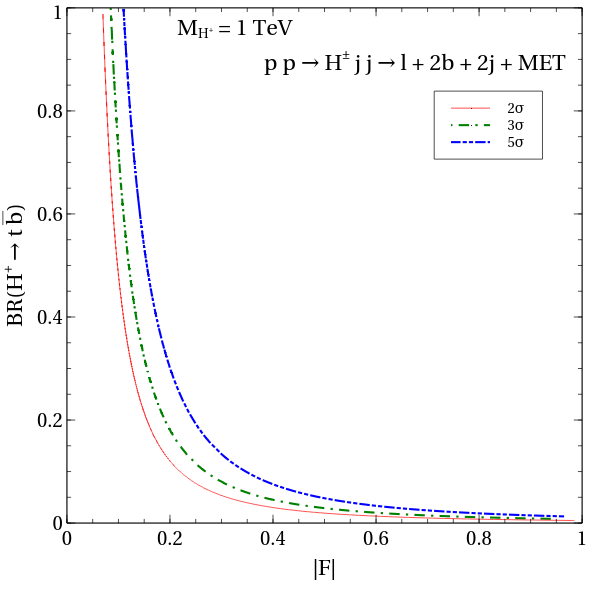} \\
\caption{The 2$\sigma$, 3$\sigma$ and 5$\sigma$ contours
for the $p p \to H^\pm j j,~H^+ \to t \bar{b}$ signal in the $|F|$-BR($H^+ \to t \bar{b}$) plane for different $M_{H^+}$.}
\label{contour:2b2jlnu}
\end{figure}

Similar to what was done for the previous channel, we next display the contours of constant signal significance in the $|F|$-BR$(H^{\pm} \to t \bar{b}$) plane in Fig.~\ref{contour:2b2jlnu}.

\subsection{The $5l + \met$ channel}

With the results of the VBF analyses in hand, we turn our attention to the $p p \to Z H^\pm \to 5l + \met$ channel. To elaborate a bit, the $H^\pm$ here is produced in association with a $Z$-boson via an $s$-channel exchange of the $W$-boson and it subsequently leads to the hadronically quiet $H^\pm \to W^\pm Z \to 3 l + \met$ decay cascade. The other $Z$-boson also decays leptonically. The Feynman diagram corresponding to this signal can be found in Fig.\ref{feyn}(c).
Throughout this analysis, we remain agnostic to the fact that the prospects of observing an $H^+$ in this channel could be low and this would be attributed to a small signal cross section(s).  

Background processes leading to an exactly $5l + \met$ final state are $p p \to W^\pm Z Z$ and 
$p p \to h W^\pm$. Subleading backgrounds come from $p p \to h Z \to 6l$ and $p p \to t \bar{t} h \to 6l + 2b + \met$. Mis-identification of a lepton as missing energy can lead to a $5l + \met$ state originating from the former. As for the latter, obtaining the same final state as the signal would entail mis-identification of one lepton and 2 $b$-jets. 
We show the cross sections for the signal and the backgrounds in Table \ref{cs_5lv}. 

\begin{table}[htpb!]
\begin{center}\scalebox{0.95}{
\begin{tabular}{|c|c|c|}
\hline
Signal / Backgrounds  & Process & Cross section $\sigma$ (fb) \\ \hline \hline
 Signal &  & \\ 
 BP1 ($M_{H^+} = 200$ GeV) & & $2.64 \times 10^{-2}$\\
 BP2 ($M_{H^+} = 300$ GeV)& $p p \rightarrow Z H^\pm \rightarrow W^\pm Z Z \rightarrow 5l + \met$ & $6.54 \times 10^{-3}$ \\
 BP3 ($M_{H^+} = 500$ GeV)& & $9.84 \times 10^{-4}$\\
 BP4 ($M_{H^+} = 1$ TeV)& & $6.26 \times 10^{-5}$ \\ \hline \hline
Backgrounds & $p p \to W^\pm Z Z \to 5l + \met$ & 7.93 
$\times 10^{-2}$ (LO)\\ 
& $p p \to  h W^\pm \to 5l + \met $ & 6.85 $\times 10^{-2}$ (LO)\\
& $p p \to h Z \to 6l$ &  1.13  $\times 10^{-2}$ (LO)\\
& $p p \to t \bar{t} h \to 6l +2b + \met$ &  1.42  $\times 10^{-2}$ (NLO) \\ 
\hline
\end{tabular}}
\end{center}
\caption{Cross sections of signal and backgrounds for the process $p p \rightarrow Z H^\pm \rightarrow W^\pm Z Z \rightarrow 5l + \met$. The signal cross sections are computed for the ($F$,BR($H^+ \to W^+ Z$)) = (0.4,0.4) reference point.}
\label{cs_5lv}
\end{table}

\begin{figure}[htpb!]
\centering 
\includegraphics[height = 8 cm, width = 8 cm]{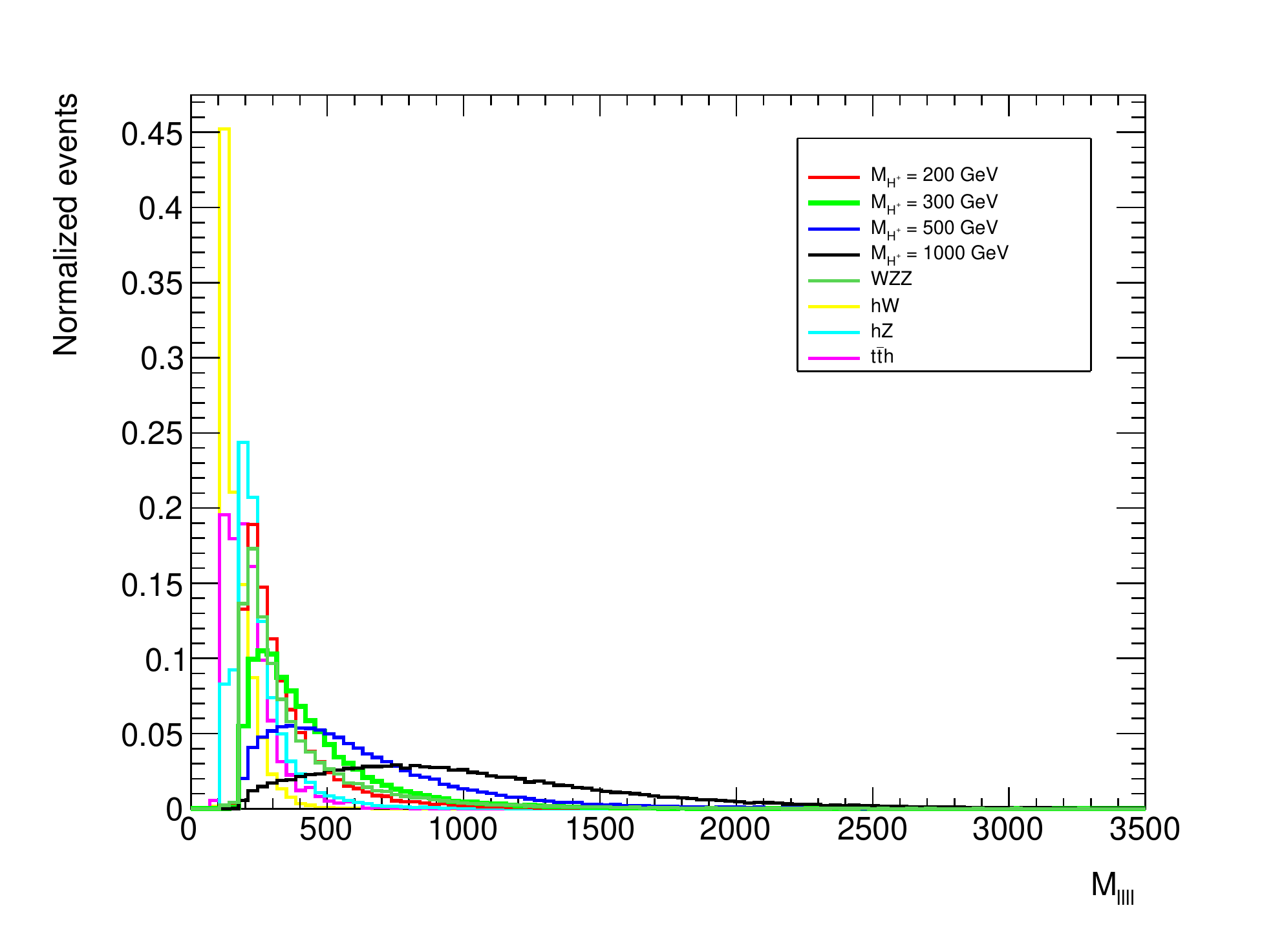}
\caption{Normalised distribution of $M_{llll}$ for the $5l +\met$ channel.}
\label{dist-5lv}
\end{figure}
\begin{table}[htpb!]
\begin{center}
\resizebox{16cm}{!}{
\begin{tabular}{|c|c|c|c|c|c|c|}
\hline
 &  \hspace{5mm} {\texttt{NTrees}} \hspace{5mm} & \hspace{5mm} {\texttt{MinNodeSize}} \hspace{5mm} & \hspace{5mm} {\texttt{MaxDepth}}~~ \hspace{5mm} & \hspace{5mm} {\texttt{nCuts}} ~~\hspace{5mm} & \hspace{5mm} {\texttt{KS-score for}}~~\hspace{5mm} & \hspace{5mm} BDT Score \hspace{5mm}\\
 & & & & & {\texttt{Signal(Background)}}&  \\
\hline
\hline
\hspace{5mm} BP1 \hspace{5mm} & 160 & 4 \% & 2.0 & 30 & 0.023(0.051) & -0.02\\ \hline
\hspace{5mm} BP2 \hspace{5mm} & 180 & 3 \% & 2.0 & 39 & 0.323(0.943) & 0.06 \\ \hline
\hspace{5mm} BP3 \hspace{5mm} & 200 & 3 \% & 2.0 & 40 & 0.981(0.121) &  0.17\\ \hline
\end{tabular}}
\end{center}
\caption{Tuned BDT parameters for BP1, BP2 and BP3 for the $5l + \met$ channel.}
\label{BDT-param3}
\end{table}
\begin{figure}[tp!]{\centering
\subfigure[]{
\includegraphics[width=3in,height=2.55in]{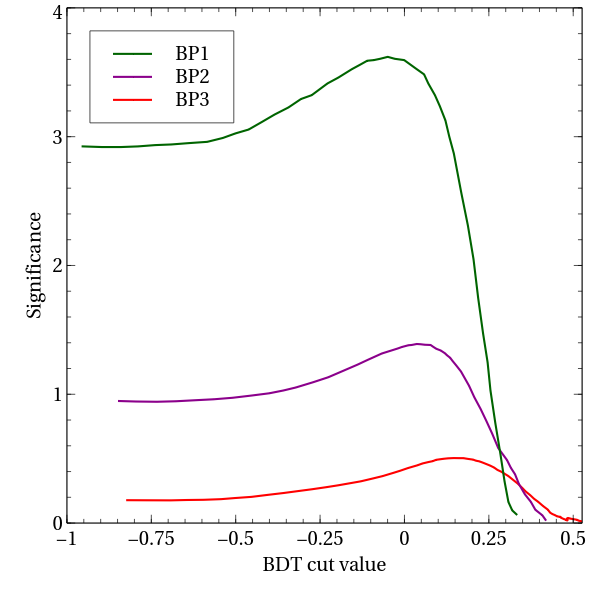}}
\subfigure[]{
\includegraphics[width=2.8in,height=2.55in]{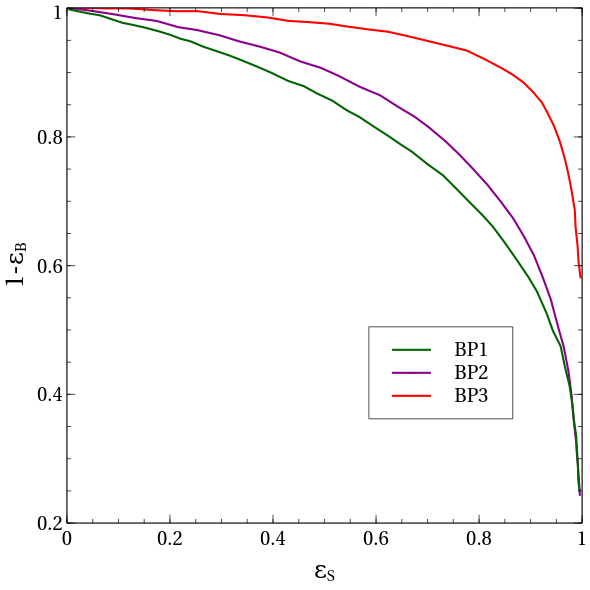}}}
\caption{(a) Variation of significance with BDT-score for $5 l  + \slashed{E_T}$ channel, (b) ROC curves for chosen benchmark points for $5 l + \slashed{E_T}$ channel. }
\label{ROC-BDTScore-5lv}
\end{figure}
BP4 is excluded from the subsequent analysis owing to the small signal cross section. 
We start by vetoing the $b$-jets and $\tau$-jets in the final state, so that the final state (comprising of five leptons) is completely leptonic. Here by lepton we mean $e,\mu$ and have taken all possible combinations of those five leptons leading to two same flavour opposite sign (SFOS) pairs and one isolated lepton. 

Since the MVA was found to outperform the CBA for the two VBF signals, we choose to omit the CBA for this process. Different BDT parameters used to carry out the MVA, are given in Table \ref{BDT-param3}. Fig.\ref{KSscore-5lv} in Appendix \ref{Plots} depicts the distribution for signal and backgrounds along with the KS-scores for this channel. In Fig.\ref{ROC-BDTScore-5lv}(a) and Fig.\ref{ROC-BDTScore-5lv}(b)
 we show the variation of significance with BDT score and ROC curves for three benchmarks respectively. It can be seen that the background rejection efficiency increases with increasing $M_{H^+}$. The following twenty kinematic variables are chosen according to the degree of distinguishing capability to initiate the MVA. 
\bea
&& (M_{ll})_{1}, (M_{ll})_{2}, M_{llll}, M_{\rm inv}^{W Z}, \met, 
\Delta R_{l_1 l_2}, \Delta R_{l_1 l_3}, \Delta R_{l_1 l_4}, \Delta R_{l_1 l_5}, \Delta R_{l_2 l_3}, \Delta R_{l_2 l_4}, \nonumber \\
&& \Delta R_{l_2 l_5}, \Delta R_{l_3 l_4}, \Delta R_{l_3 l_5}, \Delta R_{l_3 l_5}, \Delta \phi_{l_1 ~ \met}, \Delta \phi_{l_2 ~\met}, \Delta \phi_{l_3 ~\met}, \Delta \phi_{l_4 ~ \met}, \Delta \phi_{l_5 ~ \met}
\eea
Here we reconstruct the two $Z$-boson invariant masses $(M_{ll})_1$ and $(M_{ll})_2$ by combinatorially identifying the corresponding two SFOS lepton pairs. Any cut on these is expected to suppress the $h W^\pm$ and $t \bar{t} h$ backgrounds appropriately that do not involve a pair of on-shell $Z$-bosons. Next, the invariant mass of the 
two pairs of SFOS lepton system ($M_{llll}$), originating from the decay of two $Z$-bosons, is constructed. Since for the $h W^\pm$ background, $h$ decays into $Z Z^*$, the distribution of $M_{llll}$ peaks around 125 GeV in Fig.\ref{dist-5lv}.
Thus a veto cut on $M_{llll}$ around 125 GeV for all values of $M_{H^+}$ would help suppressing the aforementioned background to a large extent. Similarly to what was done for the $H^\pm jj \to W^\pm Z j j$ channel, we construct the invariant mass of the $lll\nu$ system ($M_{\rm inv}^{W Z}$) that comes from $H^\pm$. For $W^\pm Z Z$ background, $W^\pm Z$ are not originated from the decay of a single mother particle as in the signal. Therefore one can rely on this variable which has substantial discriminatory power between the signal and the backgrounds. Here $\Delta \phi_{l_i ~\met}$ is the azimuthal angle between $l_i$ and the missing transverse energy vector. Rest of the aforementioned kinematic variables have been defined earlier. Among these twenty variables, $(M_{ll})_{1}, (M_{ll})_{2}, M_{llll}, M_{\rm inv}^{W Z}, \Delta R_{l_1 l_2}, \Delta R_{l_1 l_3}, \met$ turn out to be the best performing having high discerning ability to separate signal and backgrounds. Corresponding yields for signal and backgrounds at $\mathcal{L} = 15~ {\rm ab}^{-1}$ for $M_{H^+} = 200~{\rm GeV}, 300~{\rm GeV}$ and $500 ~{\rm GeV}$ along with the signal significances are given in Table \ref{5lv:BDTD}. The maximum significance of 4.24 is predicted for $M_{H^+} = 200$ GeV following the BDTD analysis. The significance declines for the higher charged Higgs masses. Significances with $2 \%$ and $5 \%$ systematic uncertainties are presented in Table \ref{5lv:BDTD}.

\begin{figure}[htpb!]
\centering 
\includegraphics[height = 7 cm, width = 7.5 cm]{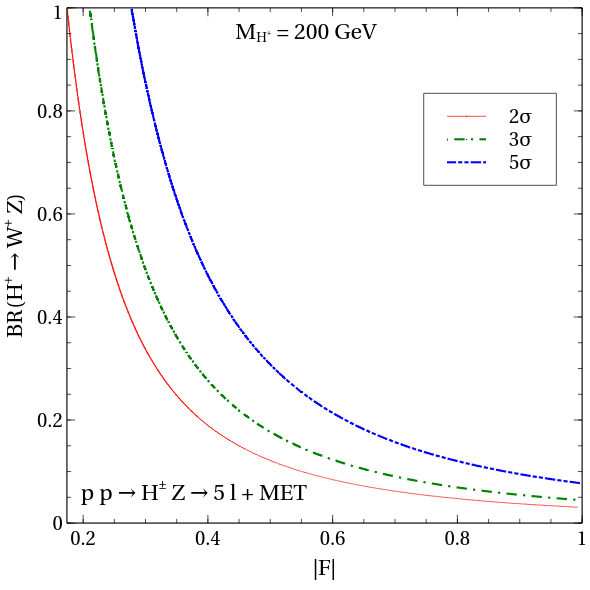}~~~~
\includegraphics[height = 7 cm, width = 7.5 cm]{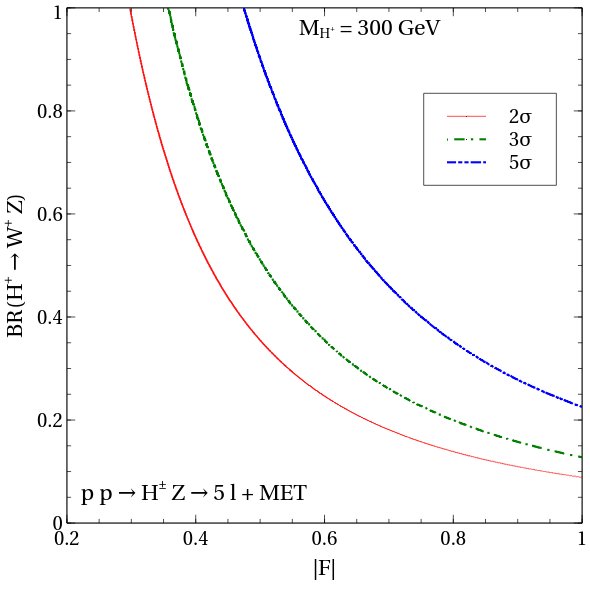} \\
\includegraphics[height = 7 cm, width = 7.5 cm]{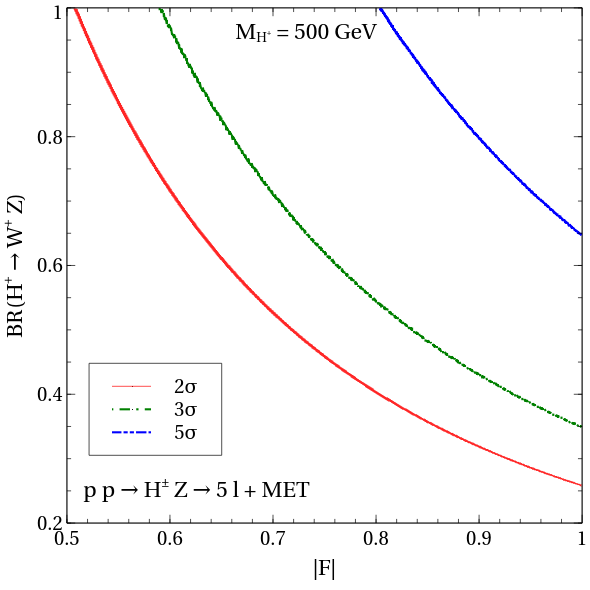}  
\caption{The 2$\sigma$, 3$\sigma$ and 5$\sigma$ contours
for the $p p \to Z H^\pm,~H^+ \to W^+ Z$ signal in the $|F|$-BR($H^+ \to W^+ Z$) plane for different $M_{H^+}$.}
\label{fig:contour_5lv}
\end{figure}

\begin{center}
\begin{table}[htb!]
\centering
\scalebox{0.75}{%
\begin{tabular}{|c|c|c|}\hline
\multicolumn{3}{|c|}{$M_{H^+}=200$ GeV} \\ \hline
 & Process  & Yield at 15 ab$^{-1}$ \\ \hline \hline
\multirow{5}{*}{Background}   
 & $W^\pm Z Z$             & $34$ \\ 
 & $h W^\pm$            & $3$ \\ 
 & $h Z$                & $7$ \\
 & $t\bar{t}h$                & $\sim 0$ \\  \cline{2-3} 
 & Total                      & $44$ \\ \hline
\multicolumn{2}{|c|}{Signal ($pp\to H^\pm Z \to 5l + \met$)} & $31$ \\\hline 
\multicolumn{2}{|c|}{Significance ($0\%$ $\sigma_{sys\_un}$)} & $4.24$\\ \hline  
\multicolumn{2}{|c|}{Signal significance ($2\%~(5\%)$ $\sigma_{sys\_un}$)}           & $4.2~(4.0)$  \\\hline
\end{tabular}}
\quad
\scalebox{0.75}{%
\begin{tabular}{|c|c|c|}\hline
\multicolumn{3}{|c|}{$M_{H^+}=300$ GeV} \\ \hline
 & Process  & Yield at 15 ab$^{-1}$ \\ \hline \hline
\multirow{5}{*}{Background}   
 & $W^\pm Z Z$             & $18$ \\ 
 & $h W^\pm$            & $\sim 0$ \\ 
 & $h Z$                & $1$ \\
 & $t\bar{t}h$                & $\sim 0$ \\  \cline{2-3}
 & Total                      & $19$ \\ \hline
\multicolumn{2}{|c|}{Signal ($pp\to H^\pm Z \to 5l + \met$)} & $7$ \\\hline 
\multicolumn{2}{|c|}{Significance ($0\%$ $\sigma_{sys\_un}$)} & $1.5$ \\ \hline  
\multicolumn{2}{|c|}{Signal significance ($2\%~(5\%)$ $\sigma_{sys\_un}$)}           & $1.5~(1.48)$  \\\hline
\end{tabular}}
\bigskip

\scalebox{0.75}{%
\begin{tabular}{|c|c|c|}\hline
\multicolumn{3}{|c|}{$M_{H^+}=500$ GeV} \\ \hline
 & Process  & Yield at 15 ab$^{-1}$ \\ \hline \hline
\multirow{5}{*}{Background}   
 & $W^\pm Z Z$             & $3$ \\ 
 & $h W^\pm$            & $\sim 0$ \\ 
 & $h Z$                & $\sim 0$ \\
 & $t\bar{t}h$                & $\sim 0$ \\  \cline{2-3}
 & Total                      & $3$ \\ \hline
\multicolumn{2}{|c|}{Signal ($pp\to H^\pm Z \to 5l + \met$)} & $1.3$ \\\hline 
\multicolumn{2}{|c|}{Significance ($0\%$ $\sigma_{sys\_un}$)} & $0.7$ \\ \hline  
\multicolumn{2}{|c|}{Signal significance ($2\%~(5\%)$ $\sigma_{sys\_un}$)}           & $0.7~(0.7)$  \\\hline
\end{tabular}}
\caption{The signal and background yields at $15~{\rm ab}^{-1}$ for $M_{H^+}$ = 200 GeV, 300 GeV and 500 GeV along with signal significances for the $5l + \met$ channel after the BDTD analysis. The signal yields are computed for the ($F$,BR($H^+ \to W^+ Z$)) = (0.4,0.4) reference point.}
\label{5lv:BDTD}
\end{table}
\end{center}
Fig.~\ref{fig:contour_5lv} displays the 2$\sigma$, 3$\sigma$ and 5$\sigma$ contours in the 
$|F|$-BR($H^+ \to W^+ Z$) plane. Upon comparing it with Fig.~\ref{fig:contour_3lvjj}, one concludes that for an $H^+$ decaying to the massive gauge bosons, $Z H^+$ production has a much lower reach compared to VBF production. To quote a few numbers, an $H^+$ of mass 500 GeV that corresponds to 
$|F|$ = 0.4 can be discovered at 5$\sigma$ for BR($H^+ \to W^+ Z$) $\gtrsim$ $3\%$ via the $3l + 2j + \met$ final state. In contrast, the $5l + \met$ final state demands $F\simeq 0.8$ and BR($H^+ \to W^+ Z$) $\simeq$ $100\%$ to attain the same discovery potential. 
\subsection{The $3l + 2b +\met$ channel}
\label{sec:3l2bnu}

This subsection contains the analysis for the signal where following production in association with a $Z$, the charged boson decays to $tb$. In addition to $Z \to ll$ decay, we have the $t \to b l \nu$ leptonic decay in this case. This gives rise to three lepton final state with two $b$-tagged jets and $\met$. The dominant contribution to the background in this final state comes from the $t\bar{t}X$ type production where $X=h,Z,W^\pm$. The $ZZZ$ and $W^\pm ZZ$ production, $Zh,~W^\pm h$ and $b\bar{b} h$ are the sub-dominant backgrounds in this channel owing to their smaller cross section compared to the previous ones. We present the cross sections of chosen signal benchmarks and background processes in Table~\ref{crosssecvh}.

\begin{table}[htpb!]
\begin{center}\scalebox{0.95}{
\begin{tabular}{|c|c|c|}
\hline
Signal / Backgrounds  & Process & Cross section $\sigma$ (fb) \\ \hline \hline
 Signal &  & \\ 
 BP1 ($M_{H^+} = 200$ GeV) & & 0.9\\
 BP2 ($M_{H^+} = 300$ GeV)& $p p \rightarrow H^\pm Z \rightarrow t b Z \rightarrow 2b + 3l + \met$ & 0.224\\
 BP3 ($M_{H^+} = 500$ GeV)& & 0.034\\
 BP4 ($M_{H^+} = 1$ TeV)& & 0.0022 \\ \hline \hline
Backgrounds & $p p \to t \bar t h$ (NLO)~\cite{bkg_twiki_cs}                    & 2860.0\\ 
& $p p \to t \bar t Z$ (NLO)~\cite{Lazopoulos:2008de} & 3477.02 \\
& $p p \to t \bar t W$ (LO) &  986.58 \\
& $p p \to WZZ$ (LO) &  0.5 \\
& $p p \to ZZZ$ (LO) & 0.07 \\
& $p p \to Zh$ (NNLO QCD + NLO EW) \cite{bkg_twiki_cs} & 0.057 \\
& $p p \to Wh$ (NNLO QCD + NLO EW) \cite{bkg_twiki_cs} &  0.185 \\
& $p p \to b \bar b h$ (LO) &  0.005 \\
\hline
\end{tabular}}
\end{center}
\caption{The cross sections of signal and backgrounds for the process $p p \rightarrow H^\pm Z \rightarrow t b Z \rightarrow 3l + 2b +\met$. The signal cross sections are computed for the ($F$,BR($H^+ \to t \bar{b}$)) = (0.4,0.4) reference point.}
\label{crosssecvh}
\end{table}

We generate all of these backgrounds with the application of following generation level cuts :
\bea
p_T^b>20~\text{GeV}, ~p_T^l >10~\text{GeV},~|\eta_{b,l}|<5.0, ~\Delta R_{b,l}>0.2 \,.
\eea

Further, exactly two $b$-tagged jets and three isolated leptons with the aforementioned transverse momenta satisfying $|\eta_{b,l}|<2.5$ are demanded. Out of these three leptons we first construct the invariant mass of the SFOS lepton pair, which must peak at $M_Z$. Then the remaining lepton along with the $\met$ and $b$-jets is used to reconstruct the charged Higgs mass. We apply the same technique to reconstruct $M_{H^+}$ as described in section~\ref{sec:2b2jlnu}. Next, we perform the MVA by adjusting the BDT parameters as mentioned in Table \ref{BDT-param4}. The signal and background distributions and the KS-scores are depicted in Fig.\ref{KSscore-3l2bv} in Appendix \ref{Plots}. The significance vs. BDT score plot has been presented in Fig.\ref{ROC-BDTScore-3l2bv}(a). Degree of rejecting the backgrounds can be identified from the ROC curves for three benchmarks in Fig.\ref{ROC-BDTScore-3l2bv}(b). 
Following the ranking of the kinematic variables provided by BDTD algorithm we choose the following twelve kinematic variables:
\begin{equation}
M_{bb},~\Delta R_{bb},~p_{T}^{l l},~M_{ll},~\Delta,  \phi_{ll},~\Delta\phi_{Z l_3},~p_{T,vis}^{H^\pm},~
p_{T, vis}^{H^\pm Z},~\eta_{H^\pm Z}^{vis}\\~\Delta \phi_{H^\pm Z},~M_{2bl\nu} ,~M_{CT}(H^+ Z).
\label{2b2jlnu:eq1}
\end{equation}

\begin{figure}[htb!]
\centering
\includegraphics[scale=0.37]{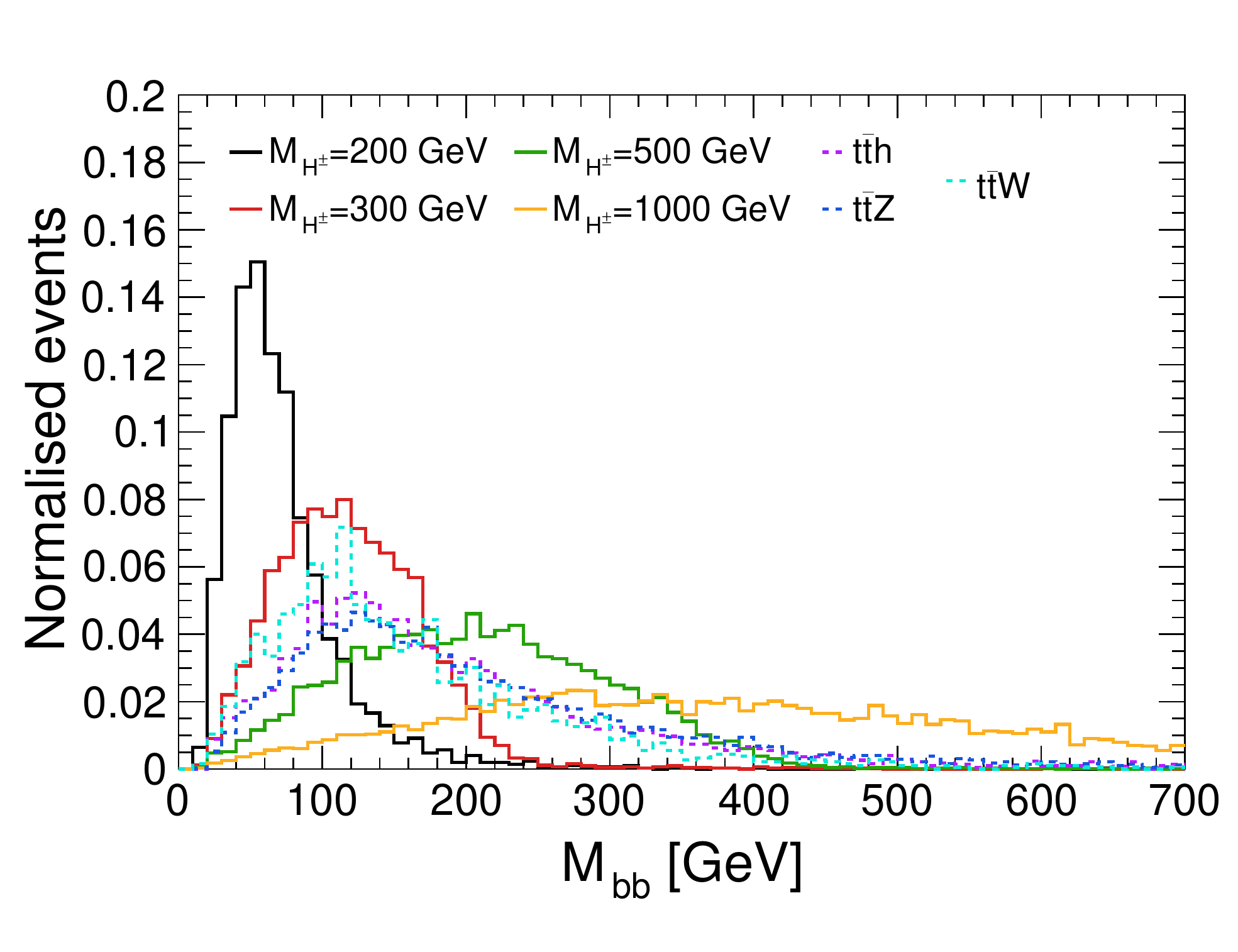}
\includegraphics[scale=0.37]{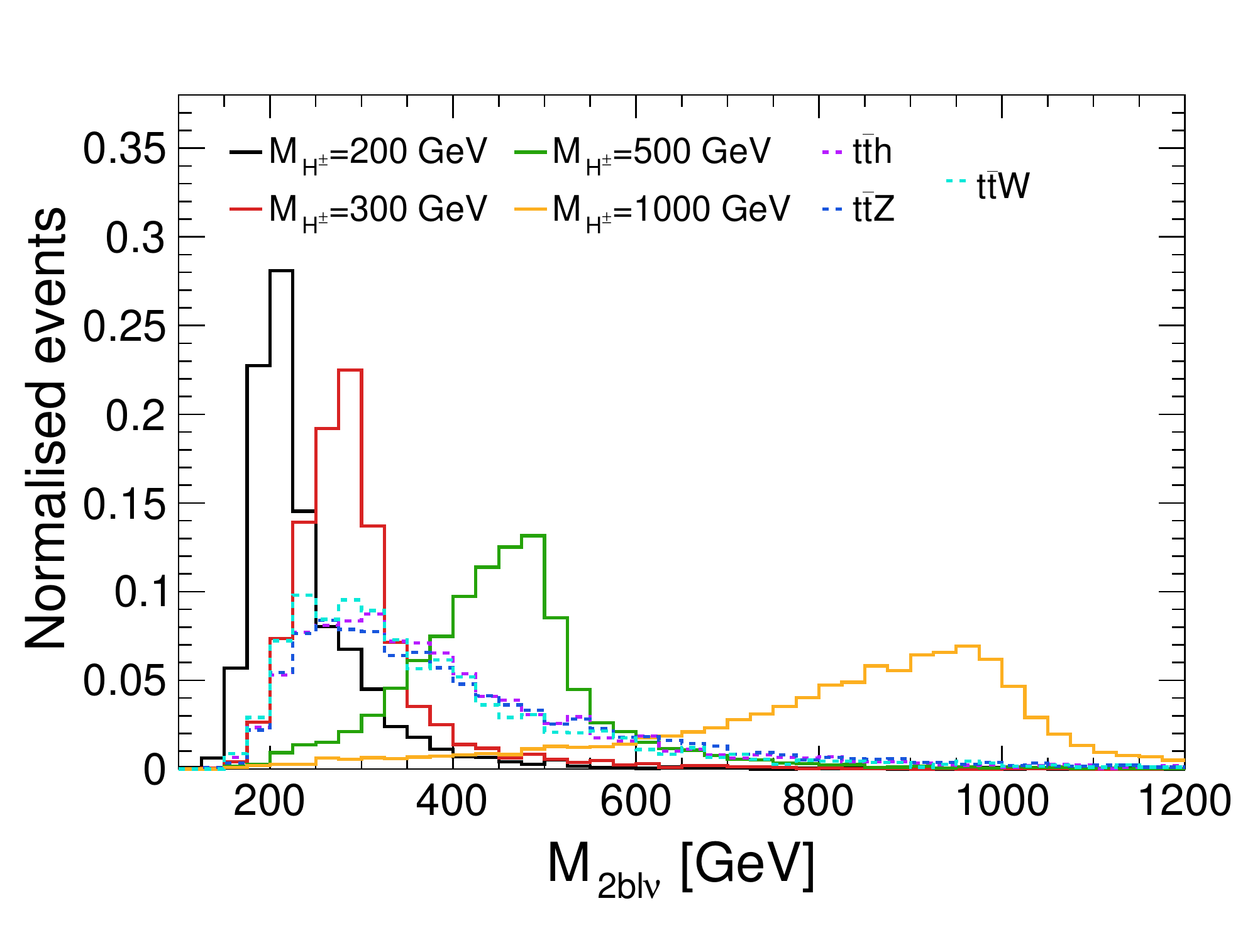}\\
\includegraphics[scale=0.37]{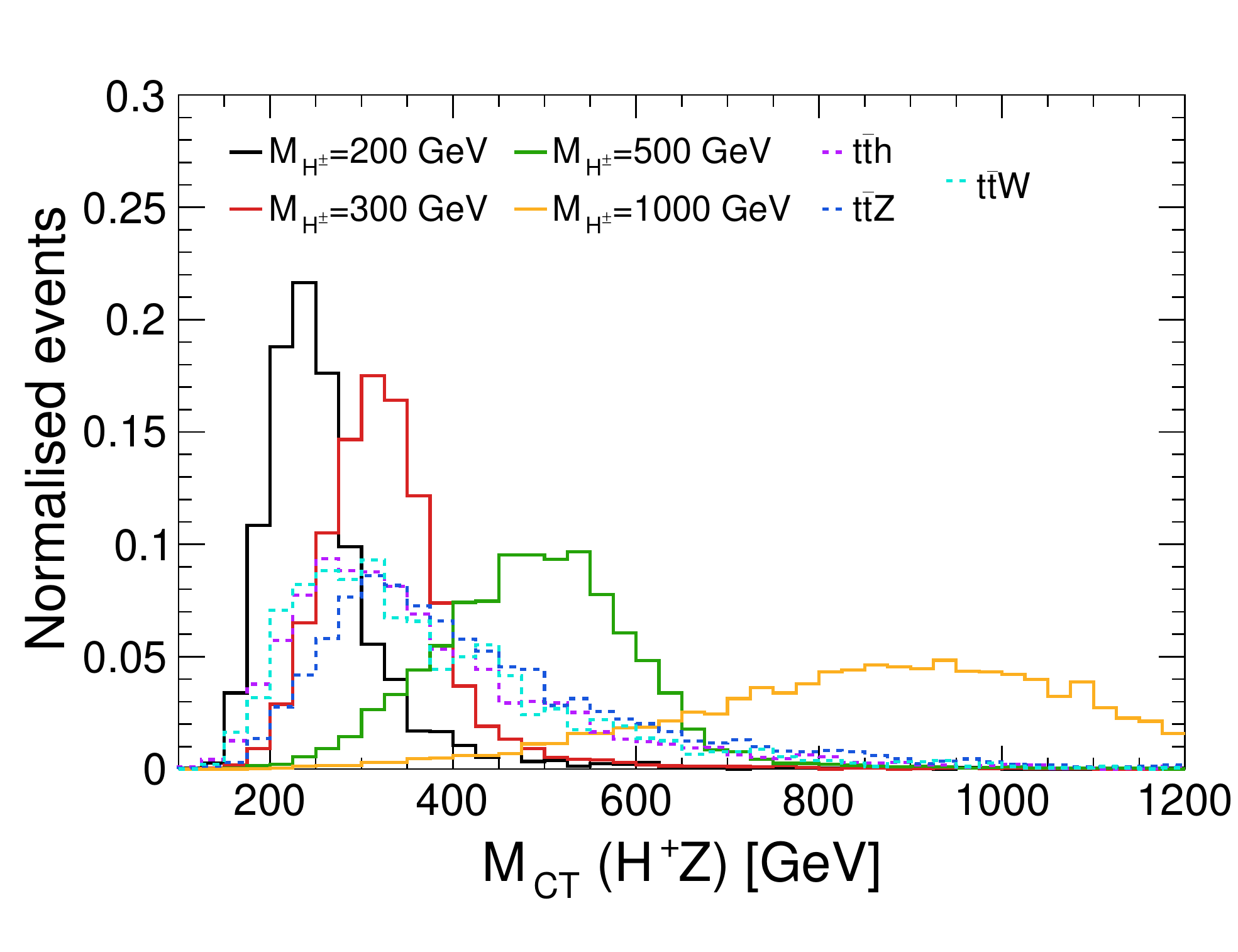}
\includegraphics[scale=0.37]{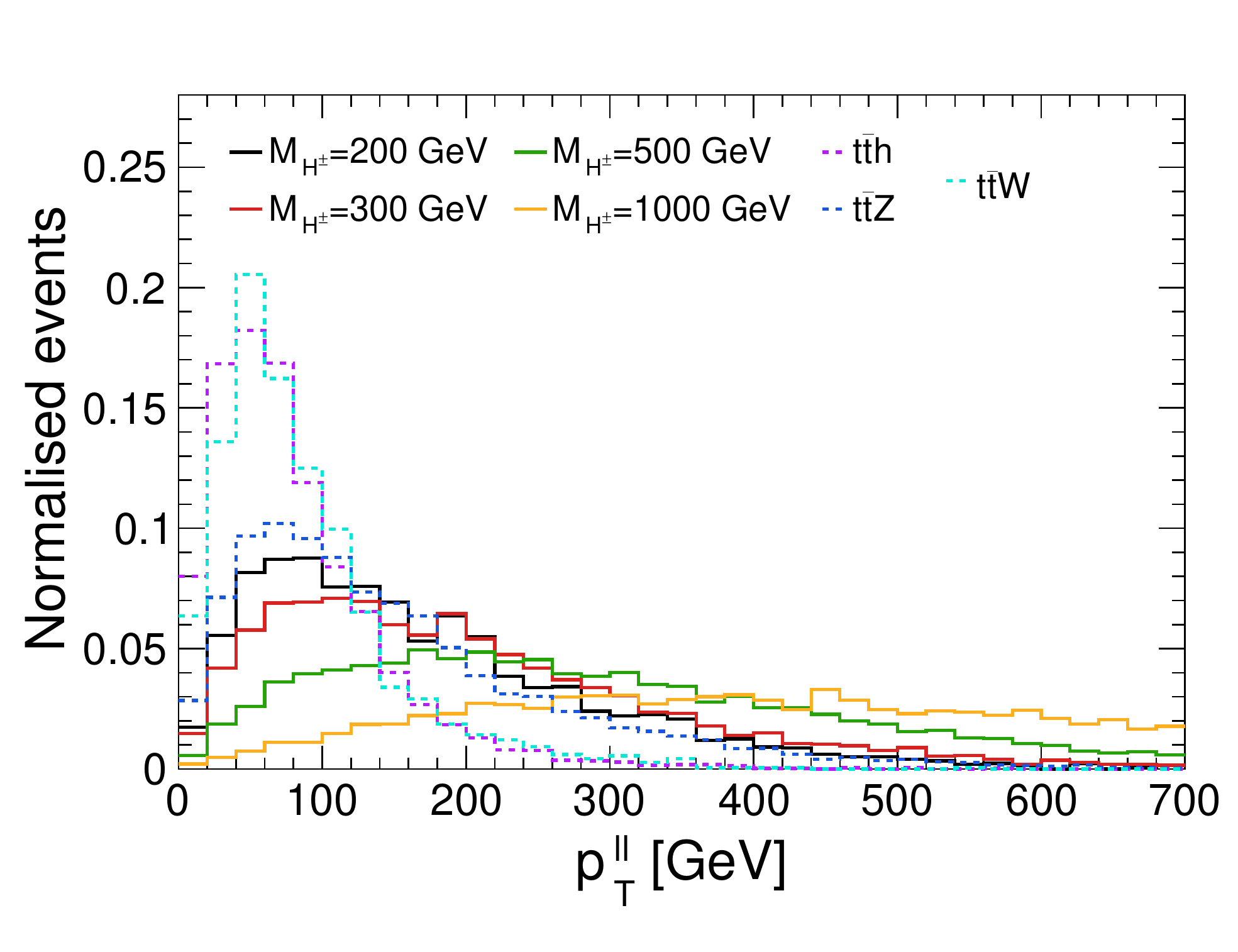}
\caption{ Normalised distributions of $M_{bb}$, $M_{2bl\nu}$, $M_{CT}(H^+ Z)$ and $p_{T}^{ll}$ for the $3l + 2b +\met$ channel.}
\label{3l2bnu:fig1}
\end{figure}

\begin{figure}[tp!]{\centering
\subfigure[]{
\includegraphics[width=3in,height=2.55in]{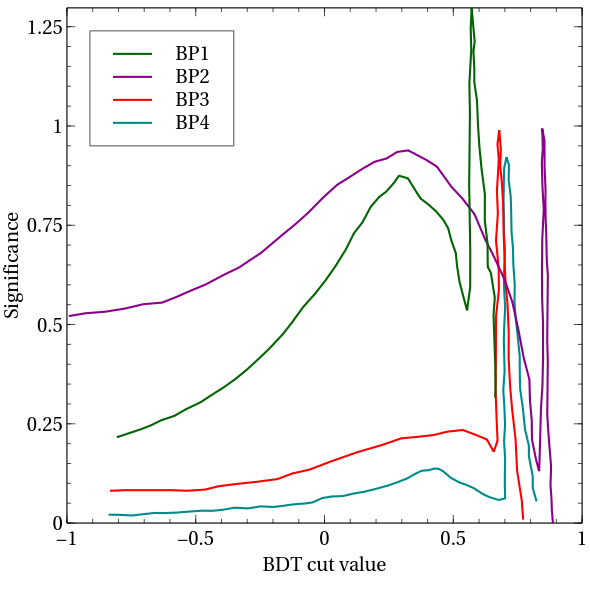}}
\subfigure[]{
\includegraphics[width=2.8in,height=2.55in]{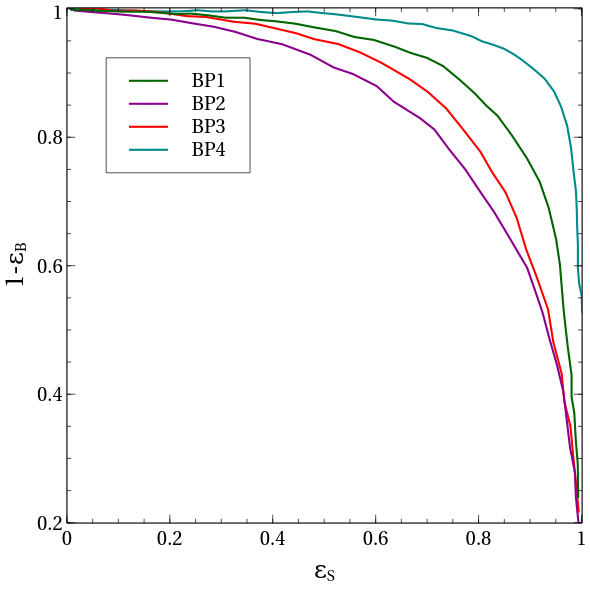}}}
\caption{(a) Variation of significance with BDT-score for $3 l  + 2b +  \slashed{E_T}$ channel, (b) ROC curves for chosen benchmark points for $3 l + 2b + \slashed{E_T}$ channel. }
\label{ROC-BDTScore-3l2bv}
\end{figure}

Here, $p_{T}^{l l}$, $M_{ll}$ and $\Delta \phi_{ll}$ are constructed from the two SFOS leptons which satisfy $Z$-mass criteria discussed at the beginning of this section, while the remaining third lepton refers to the $l_3$ in the kinematic variable $\Delta\phi_{Z l_3}$. $M_{ll}$ and $M_{2bl\nu}$ are the reconstructed $Z$-boson mass and charged Higgs mass respectively. The $p_{T,vis}^{H^\pm}$ and $p_{T,vis}^{H^\pm Z}$ are the visible transverse momentum of the charged Higgs and total system ($H^\pm$ and $Z$-boson) respectively. $\Delta \phi_{H^\pm Z}$ is the azimuthal angle separation between the charged Higgs and $Z$-boson system, whereas $M_{CT}(H^+ Z)$ is the contransverse mass of the $H^\pm Z$-system, constructed as defined in section~\ref{sec:2b2jlnu}. The other kinematic variables have their usual meaning. Among all these kinematic variables, $M_{bb},~\Delta R_{bb},~p_{T}^{l l},~M_{2bl\nu} ,~M_{CT}(H^+ Z), ~p_{T,vis}^{H^\pm},~
p_{T, vis}^{H^\pm Z}$ are the most important to maximise the signal significance. The results are summarised in Table.~\ref{3l2bnu:tab1}. In addition, the signal significances considering $2 \%$ and $5\%$  systematic uncertainty have been relegated in Table \ref{3l2bnu:tab1}. Also, We show the 
normalised distributions of the  kinematic variables having best discriminatory power between the signal and the backgrounds in Fig.~\ref{3l2bnu:fig1}.

\begin{table}[htpb!]
\resizebox{16cm}{!}{
\begin{tabular}{|c|c|c|c|c|c|c|}
\hline
 &  \hspace{5mm} {\texttt{NTrees}} \hspace{5mm} & \hspace{5mm} {\texttt{MinNodeSize}} \hspace{5mm} & \hspace{5mm} {\texttt{MaxDepth}}~~ \hspace{5mm} & \hspace{5mm} {\texttt{nCuts}} ~~\hspace{5mm} & \hspace{5mm} {\texttt{KS-score for}}~~\hspace{5mm}  & \hspace{5mm} {\texttt{BDT Score}} \hspace{5mm} \\
 & & & & & {\texttt{Signal(Background)}} &\\
\hline
\hline
\hspace{5mm} BP1 \hspace{5mm} & 100 & 3 \% & 2 & 20 & 0.169(0.097) & 0.1 \\ \hline
\hspace{5mm} BP2 \hspace{5mm} & 30 & 4 \% & 2 & 20 & 0.144(0.553) & -0.04 \\ \hline
\hspace{5mm} BP3 \hspace{5mm} & 95 & 3 \% & 2 & 20 & 0.136(0.363)  & 0.3 \\ \hline
\hspace{5mm} BP4 \hspace{5mm} & 90 & 3 \% & 2 & 20 & 0.148(0.028) & 0.3 \\ \hline
\end{tabular}}
\caption{Tuned BDT parameters for BP1, BP2, BP3 and BP4 for the $3l + 2b +\met$ channel.}
\label{BDT-param4}
\end{table}

\begin{center}
\begin{table}[htb!]
\centering
\scalebox{0.75}{%
\begin{tabular}{|c|c|c|}\hline
\multicolumn{3}{|c|}{$M_{H^+}=200$ GeV} \\ \hline
 & Process  & Yield at $15~ab^{-1}$ \\ \hline \hline
\multirow{8}{*}{Background}   
 & $t\bar{t}h$                & $5041$ \\  
 & $t\bar{t}Z$                & $15490$ \\ 
 & $t\bar{t}W$                & $855$ \\ 
 & $WZZ$                      & $51$ \\ 
 & $ZZZ$                     & $15$ \\ 
 & $Zh$                           & $10$ \\
 & $Wh$                               & $3.5$\\  
 & $b\bar{b}h$                      & $0.045$\\   
 \cline{2-3} 
 & Total                      & $21466$ \\ \hline
\multicolumn{2}{|c|}{Signal }           & $146$ \\\hline 
\multicolumn{2}{|c|}{Signal significance ($0\%$ $\sigma_{sys\_un}$)}           & $\sim 1$ \\\hline
\multicolumn{2}{|c|}{Signal significance ($2\%~(5\%)$ $\sigma_{sys\_un}$)}           & $0.32~(0.13)$  \\\hline
\end{tabular}}
\quad
\scalebox{0.75}{%
\begin{tabular}{|c|c|c|}\hline
\multicolumn{3}{|c|}{$M_{H^+}=300$ GeV} \\ \hline
 & Process  & Yield at $15~ab^{-1}$ \\ \hline \hline
\multirow{8}{*}{Background}   
 & $t\bar{t}h$                & $13664$ \\  
 & $t\bar{t}Z$                & $44515$ \\ 
 & $t\bar{t}W$                & $2020$ \\ 
 & $WZZ$                      & $84$ \\ 
 & $ZZZ$                     & $22$ \\ 
 & $Zh$                           & $13$ \\
 & $Wh$                               & $5.8$\\  
 & $b\bar{b}h$                      & $0.8$\\   
 \cline{2-3} 
 & Total                      & $60325$ \\ \hline
\multicolumn{2}{|c|}{Signal }           & $335$ \\\hline 
\multicolumn{2}{|c|}{Signal significance ($0\%$ $\sigma_{sys\_un}$)}           & $1.4$ \\\hline
\multicolumn{2}{|c|}{Signal significance ($2\%~(5\%)$ $\sigma_{sys\_un}$)}          & $0.27~(0.11)$  \\\hline
\end{tabular}}

\bigskip

\scalebox{0.75}{%
\begin{tabular}{|c|c|c|}\hline
\multicolumn{3}{|c|}{$M_{H^+}=500$ GeV} \\ \hline
 & Process  & Yield at $15~ab^{-1}$ \\ \hline \hline
\multirow{8}{*}{Background}   
 & $t\bar{t}h$                & $215$ \\  
 & $t\bar{t}Z$                & $4824$ \\ 
 & $t\bar{t}W$                & $59$ \\ 
 & $WZZ$                      & $12$ \\ 
 & $ZZZ$                     & $1.7$ \\ 
 & $Zh$                           & $0.26$ \\
 & $Wh$                               & $0.02$\\  
 & $b\bar{b}h$                      & $0.01$\\   
 \cline{2-3} 
 & Total                      & $5112$ \\ \hline
\multicolumn{2}{|c|}{Signal }           & $198$ \\\hline 
\multicolumn{2}{|c|}{Signal significance ($0\%$ $\sigma_{sys\_un}$)}           & $2.8$ \\\hline
\multicolumn{2}{|c|}{Signal significance ($2\%~(5\%)$ $\sigma_{sys\_un}$)}           & $1.6~(0.74)$  \\\hline
\end{tabular}}
\quad
\scalebox{0.75}{%
\begin{tabular}{|c|c|c|}\hline
\multicolumn{3}{|c|}{$M_{H^+}=1$ TeV} \\ \hline
 & Process  & Yield at $15~ab^{-1}$ \\ \hline \hline
\multirow{8}{*}{Background}   
 & $t\bar{t}h$                & $322$ \\  
 & $t\bar{t}Z$                & $2582$ \\ 
 & $t\bar{t}W$                & $33$ \\ 
 & $WZZ$                      & $7.8$ \\ 
 & $ZZZ$                     & $0.63$ \\ 
 & $Zh$                           & $0.05$ \\
 & $Wh$                               & $0.001$\\  
 & $b\bar{b}h$                      & $0.004$\\   
 \cline{2-3} 
 & Total                      & $2945$ \\ \hline
\multicolumn{2}{|c|}{Signal }           & $249$ \\\hline 
\multicolumn{2}{|c|}{Signal significance ($0\%$ $\sigma_{sys\_un}$)}           & $4.5$ \\\hline
\multicolumn{2}{|c|}{Signal significance ($2\%~(5\%)$ $\sigma_{sys\_un}$)}           & $3.0~(1.5)$  \\\hline
\end{tabular}}
\caption{The signal and background yields at $15~{\rm ab}^{-1}$ for $M_{H^+}$ = 200 GeV, 300 GeV, 500 GeV and 1 TeV along with signal significances for the $3l + 2b +\met$ channel after the BDTD analysis. The signal yields are computed for the ($F$,BR($H^+ \to t \bar{b}$)) = (0.4,0.4) reference point.}
\label{3l2bnu:tab1}
\end{table}
\end{center}

\begin{figure}
\centering
\includegraphics[scale=0.37]{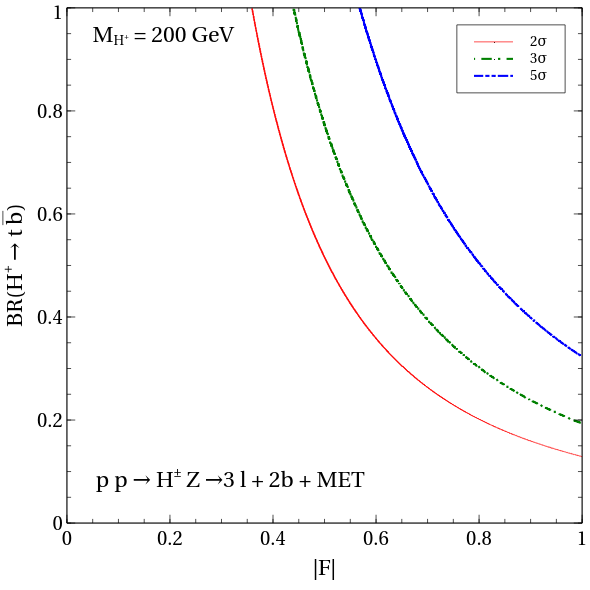}
\includegraphics[scale=0.37]{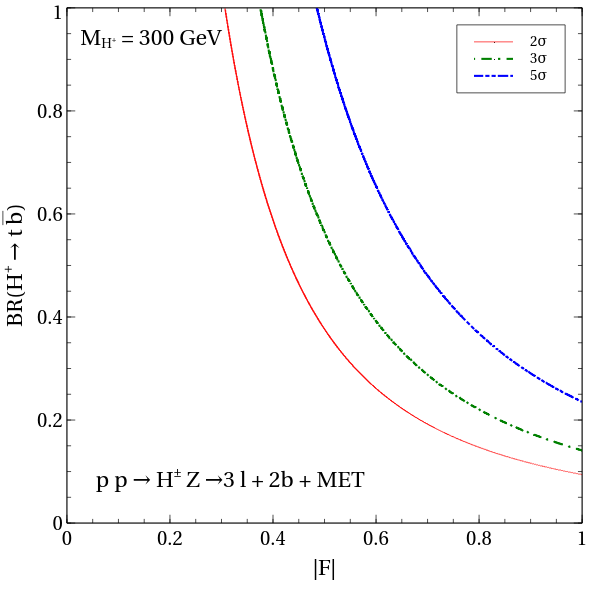}\\
\includegraphics[scale=0.37]{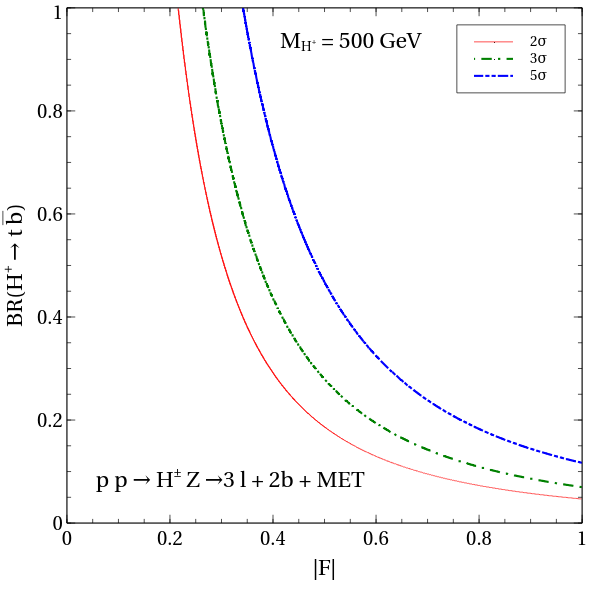}
\includegraphics[scale=0.37]{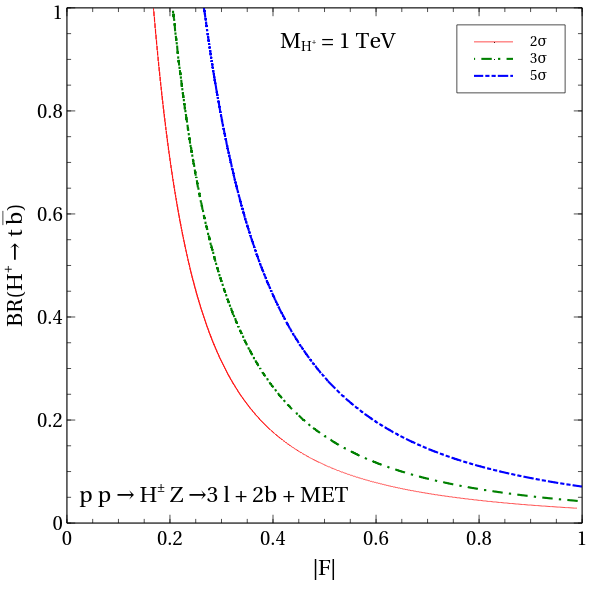} \\
\caption{The 2$\sigma$, 3$\sigma$ and 5$\sigma$ contours
for the $p p \to Z H^\pm,~H^+ \to t \bar{b}$ signal in the $|F|$-BR($H^+ \to t \bar{b}$) plane for different $M_{H^+}$.}
\label{contour:lllnubb}
\end{figure}

After computing the signal significance for our proposed signal ($ p p \to 3 l + 2b + \met$) corresponding to the BPs and a specific value of the coupling $F = 0.4$ and BR$(H^{\pm} \to t b$) = 40\%, we present the contours of constant signal significance in the $|F|$-BR$(H^+ \to t \bar{b}$) plane in Fig.~\ref{contour:lllnubb}. A straightforward comparison with Fig.~\ref{contour:2b2jlnu} establishes that for 
$H^\pm \to t b$ too, the prospects of the $Z H^\pm$ production process is much weaker compared to the VBF process in searching for an $H^+$. We quote a few numbers to put this into perspective. For $F$ = 0.4, the VBF process offers 5$\sigma$ observability for 
BR$(H^+ \to t \bar{b})$ $\gtrsim$ 15$\%$. This correspondingly weakens to 
BR$(H^+ \to t \bar{b})$ $\gtrsim$ 70$\%$ in case of 
$Z H^\pm$.


\section{Comparing the VBF sensitivities} \label{combined}

\begin{figure}[htpb!]
\centering 
\includegraphics[height = 7 cm, width = 7.5 cm]{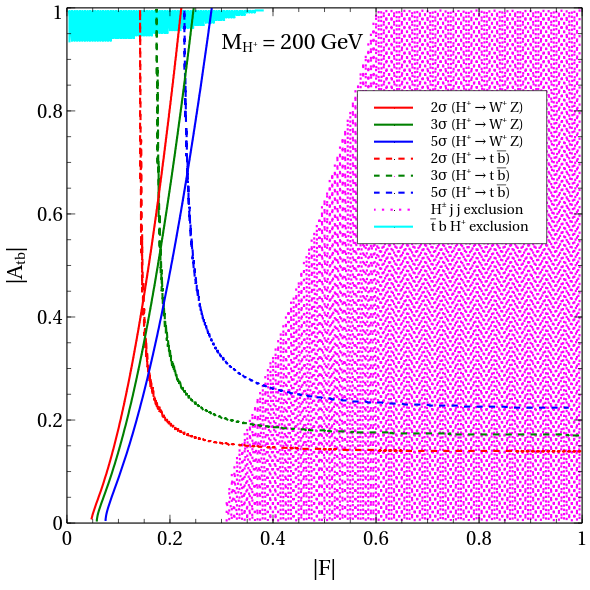}~~~~
\includegraphics[height = 7 cm, width = 7.5 cm]{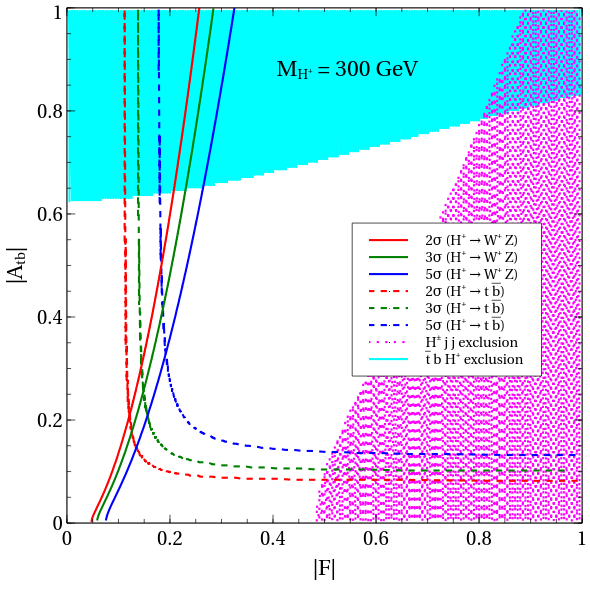} \\
\includegraphics[height = 7 cm, width = 7.5 cm]{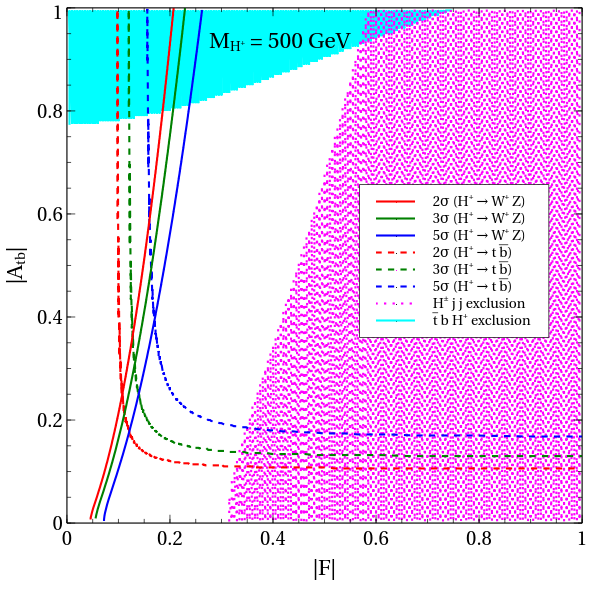}~~~~  
\includegraphics[height = 7 cm, width = 7.5 cm]{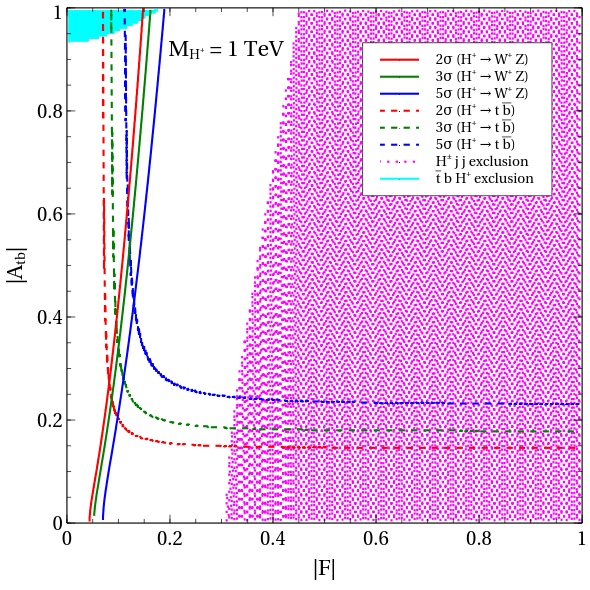} 
\caption{2$\sigma$,3$\sigma$ and 5$\sigma$ contours in the $|F|$-$|A_{tb}|$ plane for different $M_{H^+}$. The magenta (light blue) region is ruled out at 95$\%$ confidence level by the $pp \to H^\pm j j,~H^\pm \to W^\pm Z$ ($pp \to t b H^\pm,~H^\pm \to t b$) search.}
\label{fig:F-Atb}
\end{figure}

In this section, we compare the sensitivities of the various channels in the $|F|$-$|A_{tb}|$ plane. The magenta (sky blue) area in Fig.\ref{fig:F-Atb} is excluded by the $H^+jj ~(tbH^\pm)$ search. Since the  
$p p \to j j H^\pm$ process statistically outperforms the $p p \to Z H^\pm$ process, we choose it to compare the sensitivities of the $H^\pm \to t b, W^\pm Z$ decay channels corresponding to the former. The dashed (solid) red, green and blue lines in Fig.\ref{fig:F-Atb} respectively correspond to the 2$\sigma$-exclusion, 3$\sigma$-discovery and 5$\sigma$-discovery contours for the $2b + 1 l + 2j+\met$ ($3l+2j+\met$)
channel. These contours are drawn by appropriately scaling the results of the multivariate analysis of either channel. 
An inspection of Fig.~\ref{fig:F-Atb} reveals that the 
$2b + 1 l+2j+\met$ final state is useful in probing high values of $|F|$ for low $|A_{tb}|$ for all $M_{H^+}$. In case of $M_{H^+}$ = 200 GeV, $|A_{tb}| \simeq 0.3$ can be probed at 
5$\sigma$ statistical yield for $|F| \gtrsim 0.28$. 
A higher statistical yield corresponding to a higher 
$M_{H^+}$ expectedly enhances the sensitivity in the 
$|F|-|A_{tb}|$ plane. This is concurred by an inspection of Fig.~\ref{fig:F-Atb}. An $H^+$ of mass 500 GeV is seen to offer a 5$\sigma$ observability when 
$|A_{tb}| \simeq 0.3$ and $|F| \gtrsim 0.2$.  

On the other hand, the $3l+2j+\met$ final state comes handy in probing simultaneously low values of 
$|F|$ and $|A_{tb}|$. The reason to this can be traced back to the fact that $\sigma_{3l+2j+\met} \propto \frac{|F|^4}{|F|^2 + |A_{tb}|^2}$ and therefore increasing (decreasing)
$|F|$ while keeping the significance fixed requires increasing (decreasing) $|A_{tb}|$ too. This is how the lower left corner of the $|F|-|A_{tb}|$ plane gets accessible, a region the $2b + 1 l +2j+\met$ signal is blind to.
For example, $|F| \simeq |A_{tb}| \simeq$ 0.1 for $M_{H^+}$ = 200 GeV prospects a 5$\sigma$-discovery potential. 
A complementarity between the two search channels is thus seen. 

\begin{figure}[htpb!]
\centering 
\includegraphics[height = 7 cm, width = 7.5 cm]{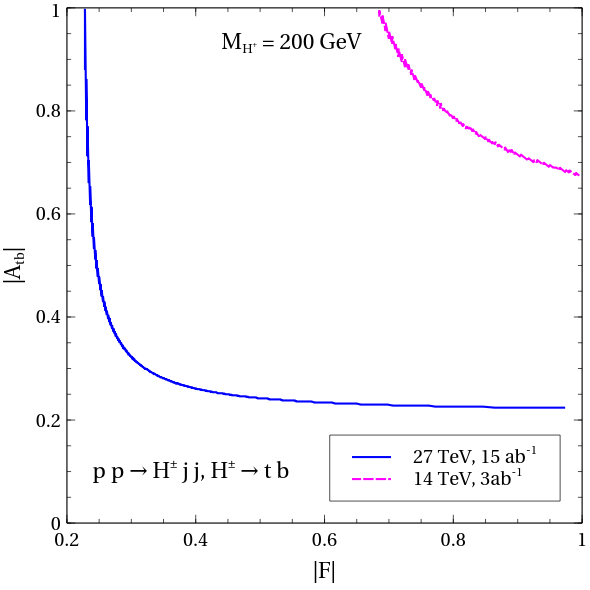}~~~~
\includegraphics[height = 7 cm, width = 7.5 cm]{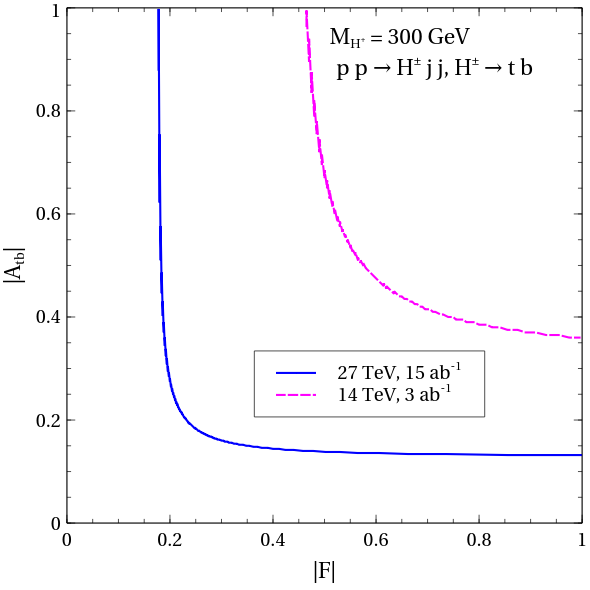} \\
\includegraphics[height = 7 cm, width = 7.5 cm]{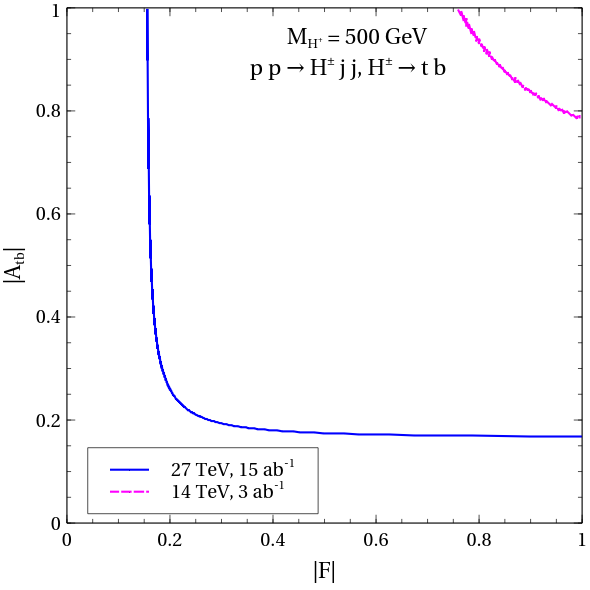}~~~~  
\includegraphics[height = 7 cm, width = 7.5 cm]{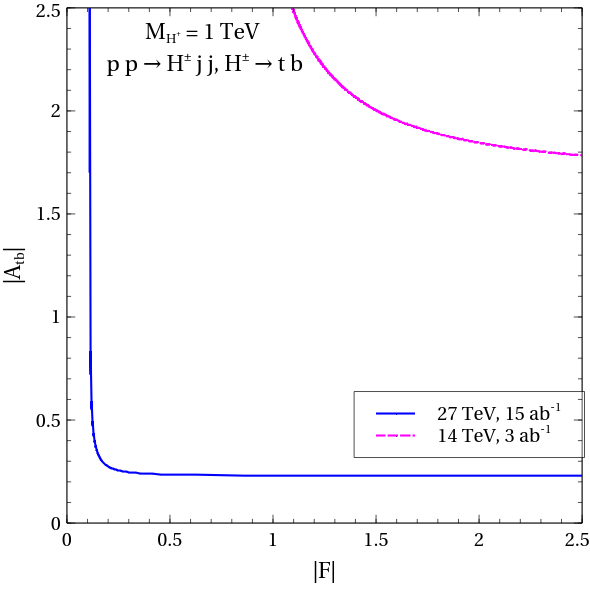} 
\caption{The 5$\sigma$-discovery contours corresponding to the $2b + 1 l +2j+\met$ channel in the $|F|-|A_{tb}|$ plane for the HL-LHC (magenta curve) and 27 TeV collider with 15 ab$^{-1}$ integrated luminosity (blue curve).}
\label{f:compare_tb}
\end{figure}

\begin{figure}[htpb!]
\centering 
\includegraphics[height = 7 cm, width = 7.5 cm]{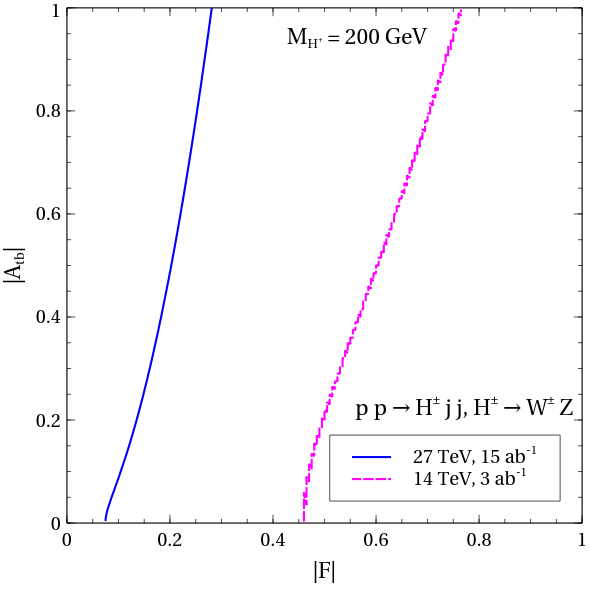}~~~~
\includegraphics[height = 7 cm, width = 7.5 cm]{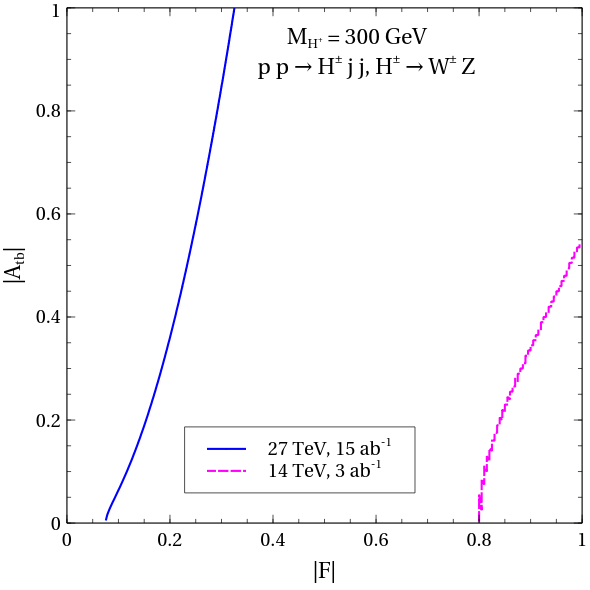} \\
\includegraphics[height = 7 cm, width = 7.5 cm]{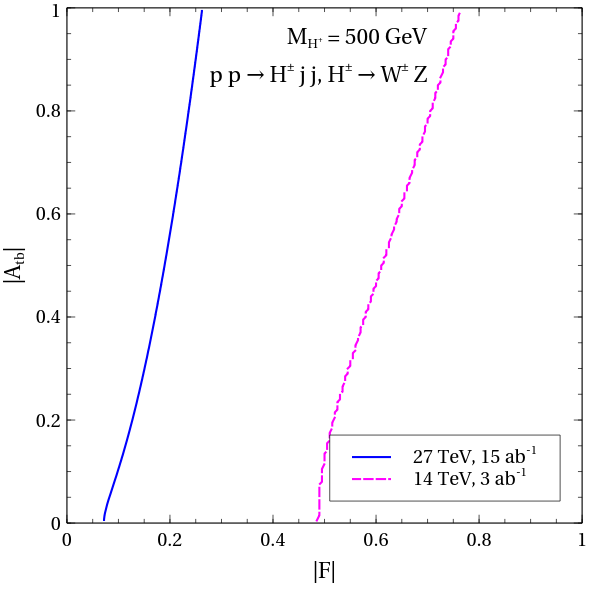}~~~~  
\includegraphics[height = 7 cm, width = 7.5 cm]{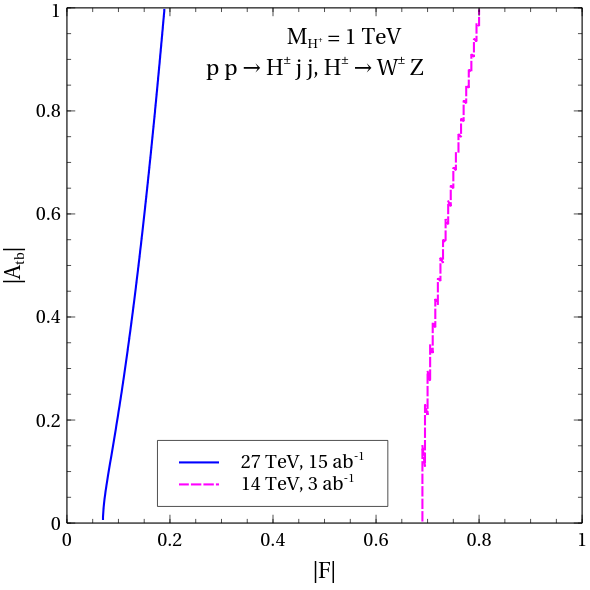} 
\caption{The 2$\sigma$ discovery contours corresponding to the 
$ 3 l +2j+\met$ channel in the $|F|-|A_{tb}|$ plane for the HL-LHC (magenta curve) and 27 TeV collider with 15 ab$^{-1}$ integrated luminosity (blue curve).}
\label{f:compare_wz}
\end{figure}

In order to see how the present analyses for a 27 TeV $pp$ collider fares against the 14 TeV HL-LHC, we extrapolate the 13 TeV, 3 ab$^{-1}$ analyses of the two VBF channels as reported in \cite{Cen:2018okf}, to 14 TeV. The extrapolation procedure makes use of the fact that kinematical distributions, and hence, the cut-efficiencies do not change appreciably upon changing from $\sqrt{s}$ = 13 TeV to 14 TeV. Keeping the integrated luminosity fixed, one can write $N^{14}_{S(B)} \simeq (\sigma_{S(B)}^{14}/\sigma_{S(B)}^{13}) N^{13}_{S(B)}$ in that case, where $N^{13}_{S(B)}$ and $N^{14}_{S(B)}$ denote the number of signal (background) events at 13 TeV and 14 TeV respectively. The subscripts and superscripts of a cross section 
$\sigma$ follow a similar notation. The contours in the 
$|F|-|A_{tb}|$ plane for 13 TeV can thus be extrapolated to 14 TeV (pink curve) as shown in Figs.\ref{f:compare_tb} and \ref{f:compare_wz}. The 5$\sigma$ discovery contours corresponding to the $2b+1l+2j+\met$ channel are compared for 14 TeV and 27 TeV in Fig.\ref{f:compare_tb}. The 27 TeV collider at 15 ab$^{-1}$ is seen to improve the sensitivity of the channel by a great margin. For example, the lowest value of $|A_{tb}|$ the HL-LHC can probe at 5$\sigma$ is 
$\simeq$ 0.67 and one requires a large $|F|$ = 1 for it. The senstivity for higher $M_{H^+}$ is also seen to be much better for 27 TeV.

Fig.\ref{f:compare_wz} compares the 2$\sigma$ exclusion contours corresponding to the $3l + 2j + \met$ channel. For $M_{H^+}$ = 200 and $|A_{tb}|$ = 0.1, the HL-LHC demands $|F| \simeq$ 0.5 for 2$\sigma$ exclusion as opposed to the much lower 
$|F| \simeq$ 0.1 in case of 27 TeV. The exclusion by a 27 TeV collider is seen to be much stronger than 14 TeV for the other masses also. In all, it is clearly established that a 27 TeV $pp$ collider with 15 ab$^{-1}$ integrated luminosity leads to a significant improvement of the VBF senstivities.

The sensitivity we obtain in terms of the generic parameters $|F|$ and $|A_{tb}|$ can be straightforwardly translated to a realistic model containing singly charged boson(s) coupling dominantly to $(t,b)$ and $(W^\pm,Z)$. 
As discussed before, a prime example is the 
Georgi-Machacek (GM) model \cite{Georgi:1985nv} whose scalar sector comprises a real triplet $\xi$ and a complex triplet $\Delta$ over and above the regular doublet. The triplets can be expressed as,
\besub
\bea
\Delta = 
\begin{pmatrix}
\frac{\delta^+}{\sqrt{2}} && \delta^{++} \\
\frac{v_\Delta + h_\delta + i z_\delta}{\sqrt{2}}
 && -\frac{\delta^+}{\sqrt{2}}
\end{pmatrix},~~~~~~~~~~
\xi = 
\begin{pmatrix}
\frac{v_\xi + h_\xi}{\sqrt{2}} && \xi^{+} \\
\xi^{-} && -\frac{v_\xi + h_\xi}{\sqrt{2}}
\end{pmatrix},
\eea
\eesub
Here $v_\Delta(v_\xi)$ refers to the vacuum expectation value (VEV) of the triplet $\Delta(\xi)$. With $v_d$ denoting the VEV of the scalar doublet, one has $\sqrt{v_d^2 + 2 v_\Delta^2 + 4 v_\xi^2}$ = 246 GeV. A custodial-symmetric scalar potential such as in the GM model demands 
$v_\xi = v_\Delta/\sqrt{2}$ in which case there is no constraint on $v_\Delta$ from the $\rho$-parameter. 

Counting the singly charged 
boson coming from the scalar doublet, the total number of singly charged scalars in the GM model becomes three. The corresponding $3\times3$ mass matrix is diagonalised by the action of a mixing angle $\a$ giving rise to a charged goldstone $G^+$, and, the physical scalars $H_3^+$ and $H_5^+$. Most importantly, both couple to $W^\pm,Z$ and the third generation quarks for an arbitrary $\a$. More precisely,  
\besub
\bea
F = \frac{g v_\Delta}{c_W M_W} \text{sin}\a,~~~~~~
A_{tb} = \frac{2 v_\Delta}{v_d}V_{tb} \text{cos}\a
~~~~~~~~~~\text{for $H^+_3$,} \\
F = -\frac{g v_\Delta}{c_W M_W} \text{cos}\a,~~~~~~
A_{tb} = \frac{2 v_\Delta}{v_d}V_{tb} \text{sin}\a
~~~~~~~~~~\text{for $H^+_5$.}
\eea
\eesub 
$c_W$ being the cosine of Weinberg angle.

That the $H^+ W^- Z$ interaction is proportional to the triplet VEV is a generic feature of models featuring scalar triplets. The sensitivity contours in the $|F|$-$|A_{tb}|$ plane can therefore be mapped to the $v_\Delta$-|sin$\a$| plane for the GM model for a given charged boson.

\begin{figure}[htpb!]
\centering 
\includegraphics[height = 7 cm, width = 7.5 cm]
{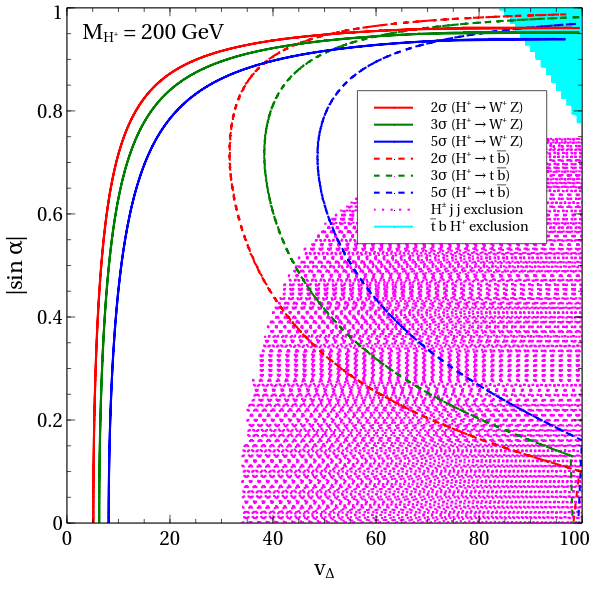}~~~~
\includegraphics[height = 7 cm, width = 7.5 cm]
{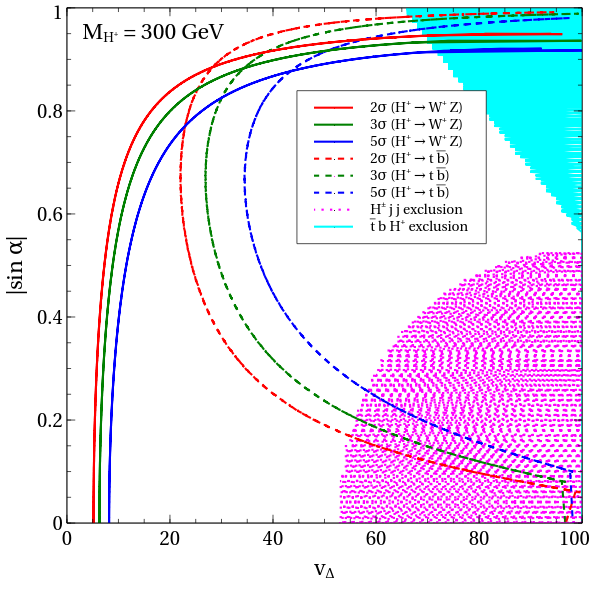} \\
\includegraphics[height = 7 cm, width = 7.5 cm]
{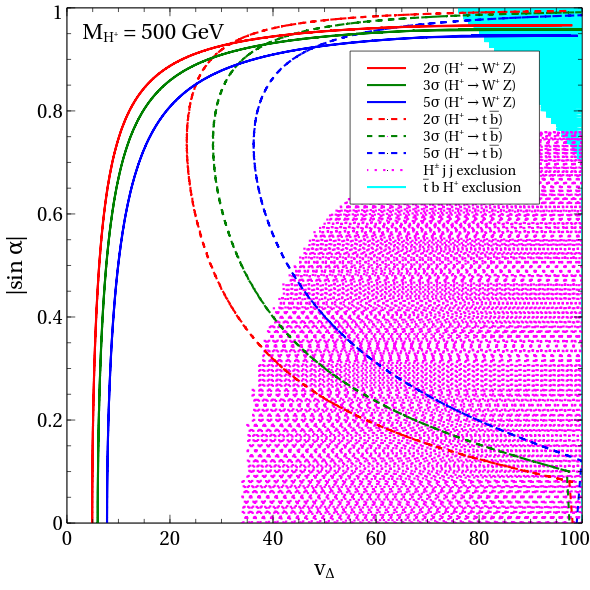}~~~~
\includegraphics[height = 7 cm, width = 7.5 cm]
{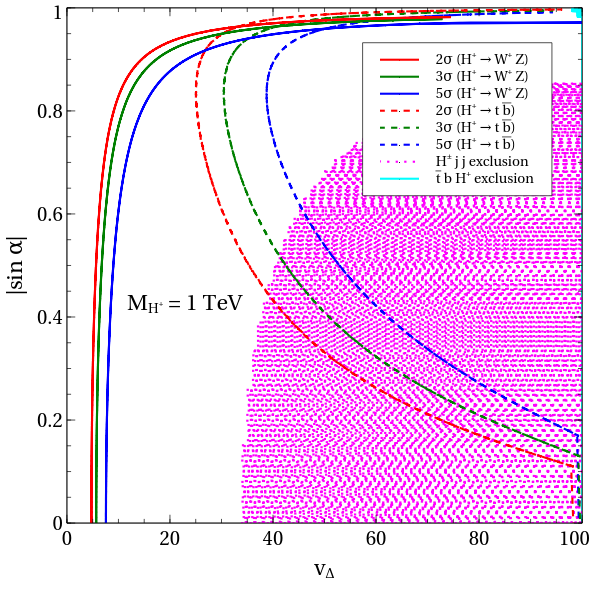}
\caption{2$\sigma$,3$\sigma$ and 5$\sigma$ contours in the $v_\Delta$-|sin$\a$| plane for different $M_{H^+}$. The magenta (light blue) region is ruled out at 95$\%$ confidence level by the $pp \to H^\pm j j,~H^\pm \to W^\pm Z$ ($pp \to t b H^\pm,~H^\pm \to t b$) search.}
 \label{fig:GM}
\end{figure}

Fig.\ref{fig:GM} displays the contours for $H_5^+$\footnote{There exist other GM model parameters~\cite{Keeshan:2018ypw,Cen:2018wye} such as scalar couplings, masses and mixing angles that directly do not enter a parameter region shown in the
$v_\Delta-|\text{sin}\a|$ plane.
It can therefore be inferred that the entire $v_\Delta-|\text{sin}\a|$ plane gets allowed by electroweak precision measurements by a judicious choice of the other model parameters. Hence, the impact of electroweak precision constraints is not separately investigated for this mostly model-independent study.} and the colour-coding remains the same as in Fig.\ref{fig:F-Atb}\footnote{We assume that $H_3^+$ is heavier than $H_5^+$ so that the $H_5^+ \to H_3^+ Z$ decay is disallowed kinematically.}. It is worthwhile here to comment on how the choice of $\a$ impacts the observability. The cross section of the $p p \to H_5^\pm j j,~H_5^\pm \to t b$ process is proportional to 
$\frac{F^2 A_{tb}^2}{F^2 + A_{tb}^2} = \bigg(\frac{4 g^2 v^4_\Delta}{c_W^2 M^2_W v_d^2}\bigg) \bigg\{1/
\Big(\frac{g^2 v^2_\Delta}{c_W^2 M_W^2} c^2_\a 
+ \frac{4 v^2_\Delta}{v_d^2} s^2_\a \Big) \bigg\} s^2_\a c^2_\a$. Therefore, the cross section becomes maximum around maximal mixing, i.e, $\a = \frac{\pi}{4}$ thereby also maximising the observability. We find that a 27 TeV hadronic collider with an integrated luminosity of 15 ab$^{-1}$ excludes an 
$H_5^+$ of mass 200 GeV at 95$\%$ 
confidence level for $v_\Delta \gtrsim 32$ GeV in the vicinity of maximal mixing, a significant improvement over  an earlier 13 TeV study~\cite{Chiang:2018cgb} that reports $v_\Delta \gtrsim 70$ GeV to be the corresponding exclusion limit. The discovery potential also markedly improves upon switching from 13 TeV to 27 TeV. We read from Fig.\ref{fig:GM} that a maximally mixing $H_5^+$ of mass 200 GeV can be discovered at 5$\sigma$ for $v_\Delta \gtrsim$ 48 GeV. For $M_{H^+}$ = 300 GeV, 500 GeV and 1 TeV, the minimum values of 
$v_\Delta$ leading to a 5$\sigma$ observability are $\simeq$ 35 GeV, 37 GeV and 39 GeV respectively. In complementarity with a discussion on maximal mixing, low values of sin$\a$ can be probed taking high $v_\Delta$. For example, taking $v_\Delta$ = 90 GeV implies that the lowest 
|sin$\a$| allowing for a 5$\sigma$-discovery of an $H_5^+$ of mass equaling  200 GeV, 300 GeV, 500 GeV and 1 TeV are $\simeq$ 0.21, 0.12, 0.15 and 0.20 respectively.  

The lower left corner of the $|F|-|A_{tb}|$ plane that the 
$p p \to H_5^\pm j j, ~H_5^\pm \to W^\pm Z$ process can probe
maps to the low $v_\Delta$-low |sin$\a$| region of the GM model. This is concurred by an inspection of Fig~\ref{fig:GM}. 
For |sin$\a$| = 0.2, an $H_5^+$ of mass 200 GeV is excluded 
by the present analysis for $v_\Delta \gtrsim$ 6 GeV. Considering the corresponding bound from the CMS analysis 13 TeV and 36 fb$^{-1}$
stands at $v_\Delta \gtrsim$ 50 GeV, we deem our analysis a considerable improvement over the former and demonstrative of the potential of a 27 TeV $pp$-collider. A 5$\sigma$-discovery is predicted for $v_\Delta \gtrsim$ 9 GeV. The sensitivity of this channel for the higher masses remains approximately the same in this corner of the $v_\Delta$-|sin$\a$| plane as can be read from Fig.~\ref{fig:GM}.

Prior to closing this section, we add that the results we obtained in this study can be interpreted in context of any NP scenario that predicts singly charged Higgs bosons dominantly coupling to the third generation quarks and 
$W^\pm,Z$. The non-minimal variants of the GM model can straightforwardly similarly be examined for observability of the singly charged scalars. However, we would like to comment on the observability in case of multi-doublet models where $F$ is radiatively suppressed. As stated in the introduction, the Type-I 2HDM and its inert or color octet extensions predict 
$F \simeq \mathcal{O}(10^{-2})$ which is below the sensitivity threshold of the present analysis. That said, there is still the option of arbitarily increasing the number of scalar doublets to enhance $F$ and bring it within the sensitivity reach. However, such models are theoretically contorted and tightly constrained by experimental data. Overall, it can therefore be concluded that despite such a huge enhancement in sensitivity compared to the HL-LHC, the 27 TeV $pp$ machine with 15 ab$^{-1}$ integrated luminosity cannot probe the radiative $H^+ W^- Z$ vertex, at least in the minimal multi-doublet models such as 2HDMs and 3HDMs.

\section{Summary and conclusions} \label{summary}
In this work, we have investigated the prospects of observing a gauge-phillic singly charged Higgs boson at the 27 TeV upgrade of the LHC attaining 15 ab$^{-1}$ integrated luminosity. Motivated by certain realistic scenarios, it is assumed that the charged scalar dominantly interacts with ($W^\pm,~Z$) and ($t,~b$) via the generic couplings $F$ and $A_{tb}$ respectively. Two kinematically distinct topologies for $H^\pm$ production, i.e., the VBF process
$pp \to H^\pm j j$ and $p p \to Z H^\pm$ are proposed for detailed study. The $H^\pm$ so produced decays to the $tb$ and $W^\pm Z$ pairs and complete leptonic cascades of the same are chosen for detailed analyses. That is, the following signals were analysed. 
(a) $p p \to H^\pm j j \to W^\pm Z j j \to 3l + 2j + \met$,
(b) $p p \to H^\pm j j \to t b j j \to 2b + 1l + 2j + \met$, 
(c) $p p \to H^\pm Z \to W^\pm Z Z \to 5l + \met$ and
(d) $p p \to H^\pm Z \to t b Z \to 3l + 2b + \met$, where $l = e, \mu$. 
We took 
$M_{H^+}$ = 200 GeV, 300 GeV, 500 GeV and 1 TeV as benchmark mass points for the ensuing analysis. 
In addition to using the conventional cut-based method, 
multivariate techniques stemming from the boosted-decision-tree algorithm were also adopted. We summarise below our key results.

\begin{itemize}

\item The statistical yield of the VBF production process supersedes that of the $p p \to Z H^\pm$ process for both  
$H^\pm \to W^\pm Z$ and $H^\pm \to t b$.

\item The two VBF channels are seen to probe somewhat complementary regions of the $|F|$-$|A_{tb}|$ plane, which clearly is an upshot of the present analysis. The $p p \to H^\pm j j, H^\pm \to W^\pm Z$ signal is found to be able to probe the lower left corner of the 
$|F|$-$A_{tb}$ plane.

\item The generic nature of the study has enabled to interpret the obtained results in context of the Georgi-Machacek model. The relevant parameters for this case are the VEV $v_\Delta$ and the mixing angle $\a$. 
The $p p \to H^\pm j j, H^\pm \to t b$ signal predicts
5$\sigma$ discovery of a 500 GeV charged scalar for a $v_\Delta$ $\simeq$ 38 GeV. With the $p p \to H^\pm j j, H^\pm \to W^\pm Z$, the 5$\sigma$ threshold lowers to $v_\Delta$ $\simeq$ 8 GeV.

\end{itemize}

As a passing remark, our study highlights that the $H^+ W^- Z$ vertex is an important interaction to search for a singly charged Higgs boson. It also establishes that a 
$\sqrt{s}$ = 27 TeV $pp$ collider with an integrated luminosity of 15 ab$^{-1}$ can be a powerful tool to improve the discovery potential of a charged Higgs via such an interaction over the earlier analyses carried out at lower centre-of-mass energies. The present analysis therefore serves as an important case study on the efficacy of a 27 TeV hadronic collider.
\newpage

\acknowledgements
The authors sincerely thank Biplob Bhattacherjee for crucial inputs throughout the course of this work, and, for a careful reading of the manuscript. NC is financially supported by IISc (Indian Institute of Science) through the C.V.Raman postdoctoral fellowship.
NC also acknowledges support from DST, India,
under grant number IFA19- PH237 (INSPIRE Faculty Award). IC acknowledges support from DST, India, under grant number IFA18-PH214 (INSPIRE Faculty Award), and, hospitality extended by IISc while this work was in progress. The work of JL is supported by funding available from the Department of Atomic Energy, Government of India, for the Regional Centre for Accelerator-based Particle Physics (RECAPP), Harish-Chandra Research Institute.

\appendix

\section{Plots}
\label{Plots}
\begin{figure}[tp!]{\centering
\subfigure[]{
\includegraphics[width=2.8in,height=2.55in]{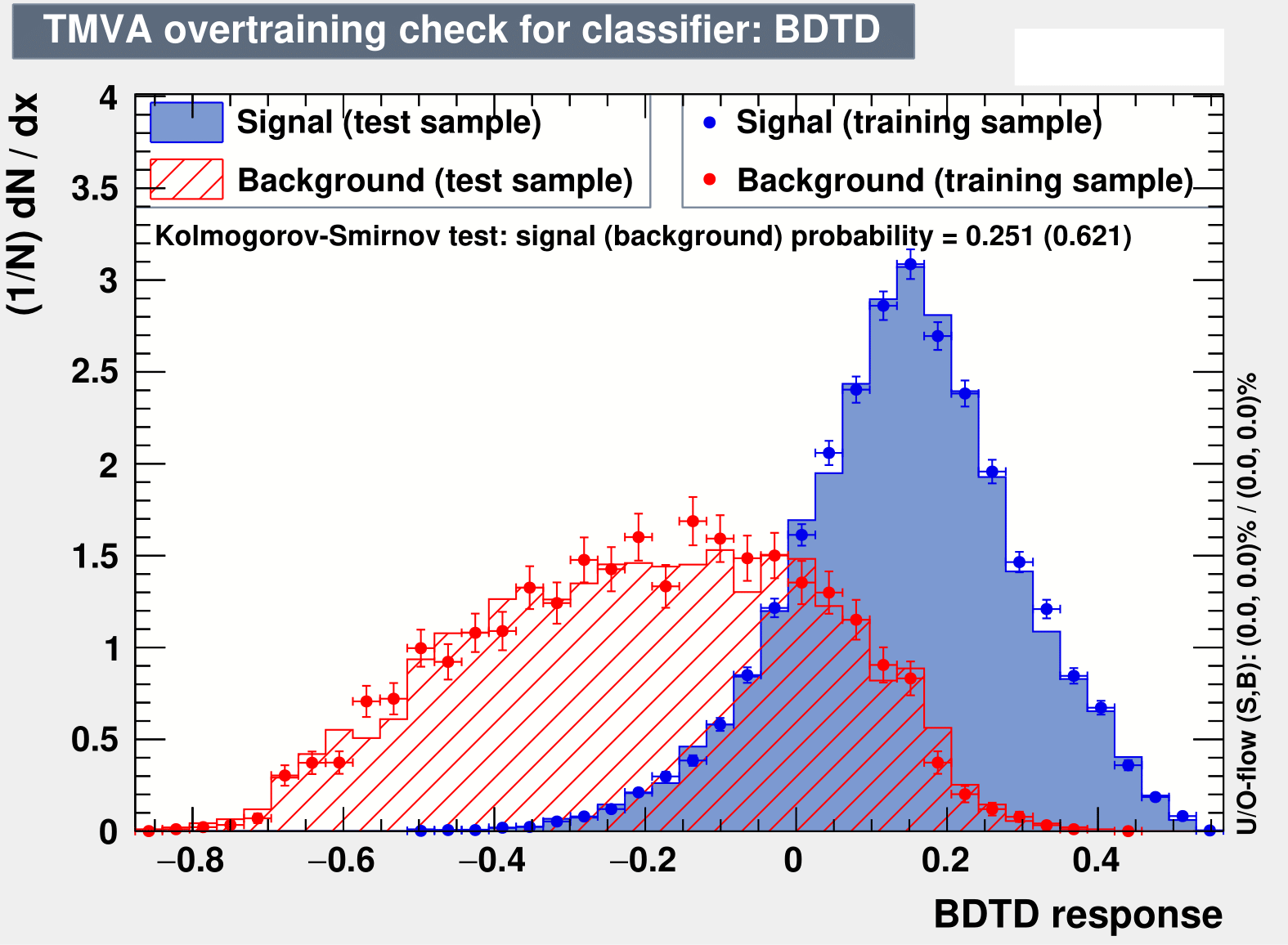}}
\subfigure[]{
\includegraphics[width=3in,height=2.55in]{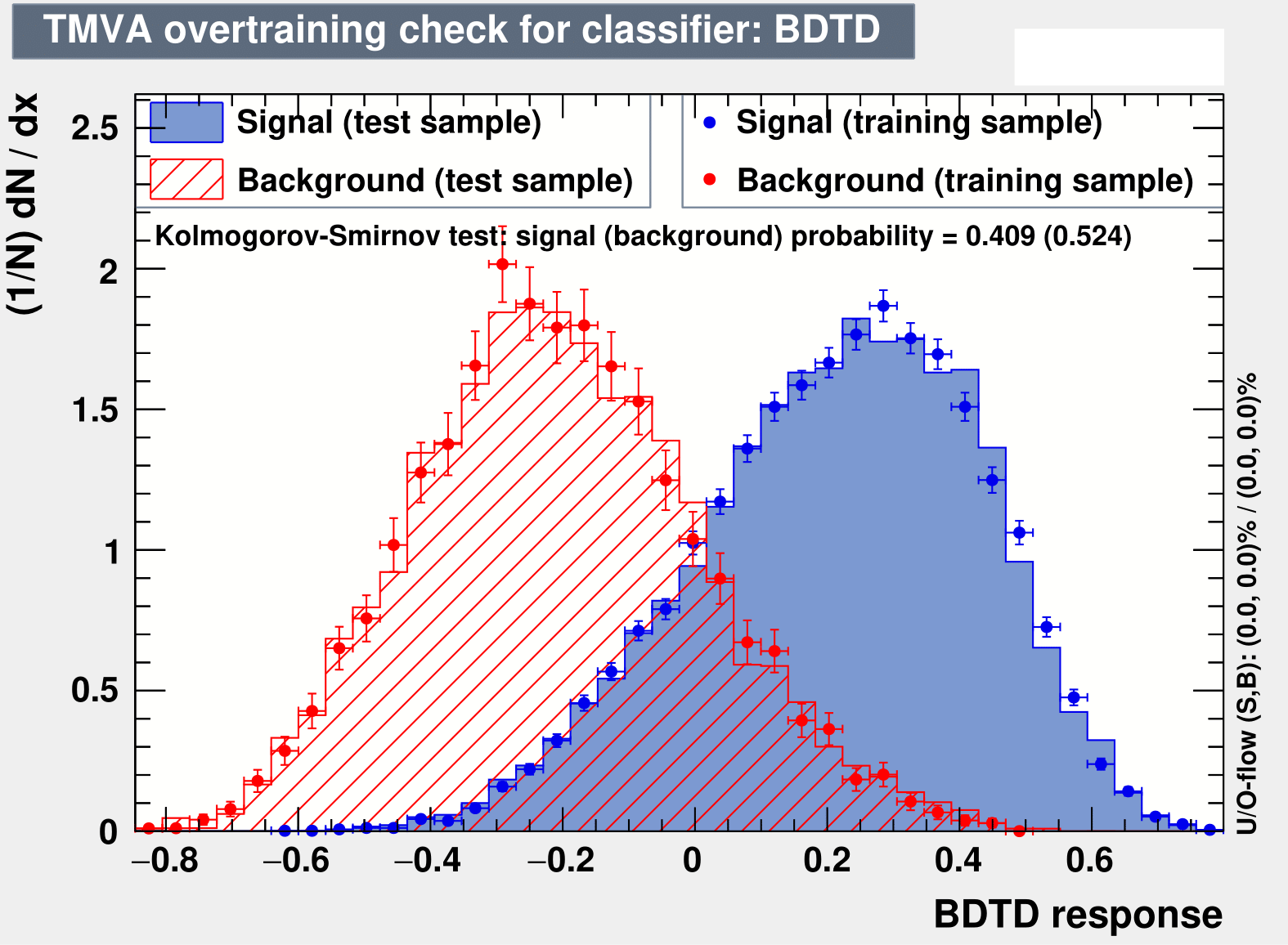}} \\
\subfigure[]{
\includegraphics[width=3in,height=2.55in]{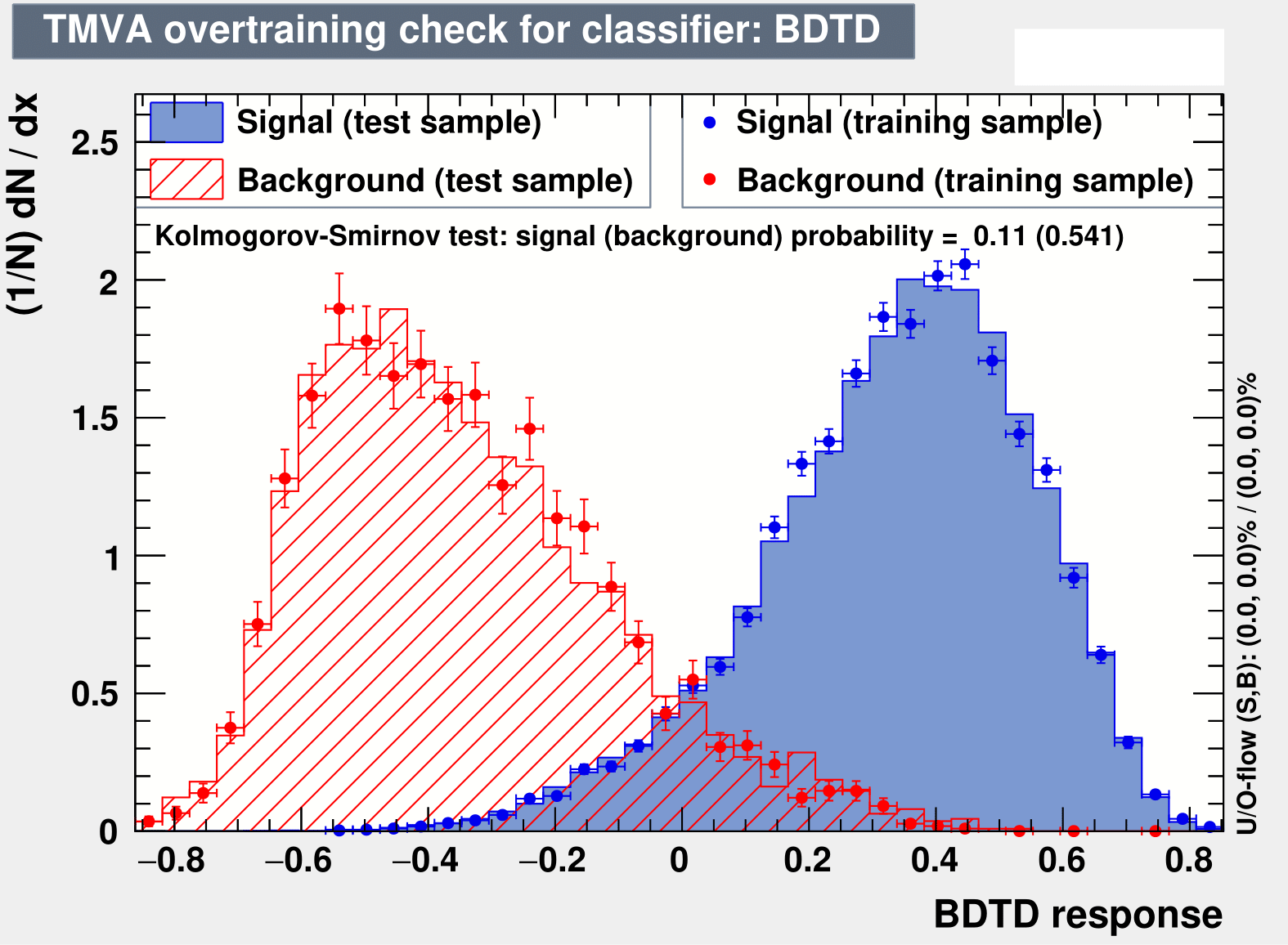}}
\subfigure[]{
\includegraphics[width=3in,height=2.55in]{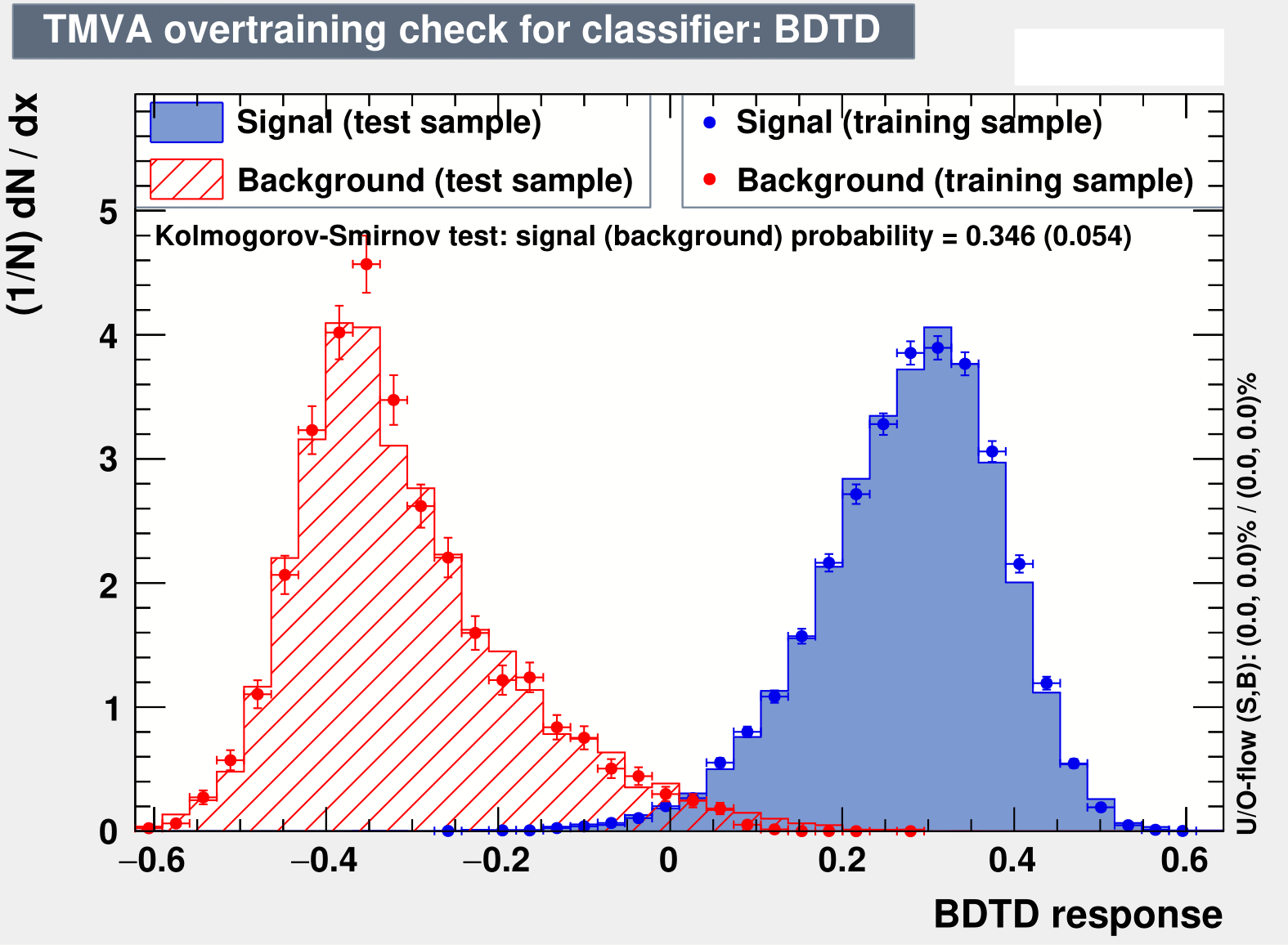}}}
\caption{ KS-scores corresponding to BP1, BP2, BP3 and BP4 for $3 l  + 2 j + \slashed{E_T}$ channel.}
\label{KSscore-3lvjj}
\end{figure}

We depict below the distribution of signal and backgrounds along with the KS-scores for each BP corresponding to the four channels.

\begin{figure}[tp!]{\centering
\subfigure[]{
\includegraphics[width=2.8in,height=2.55in]{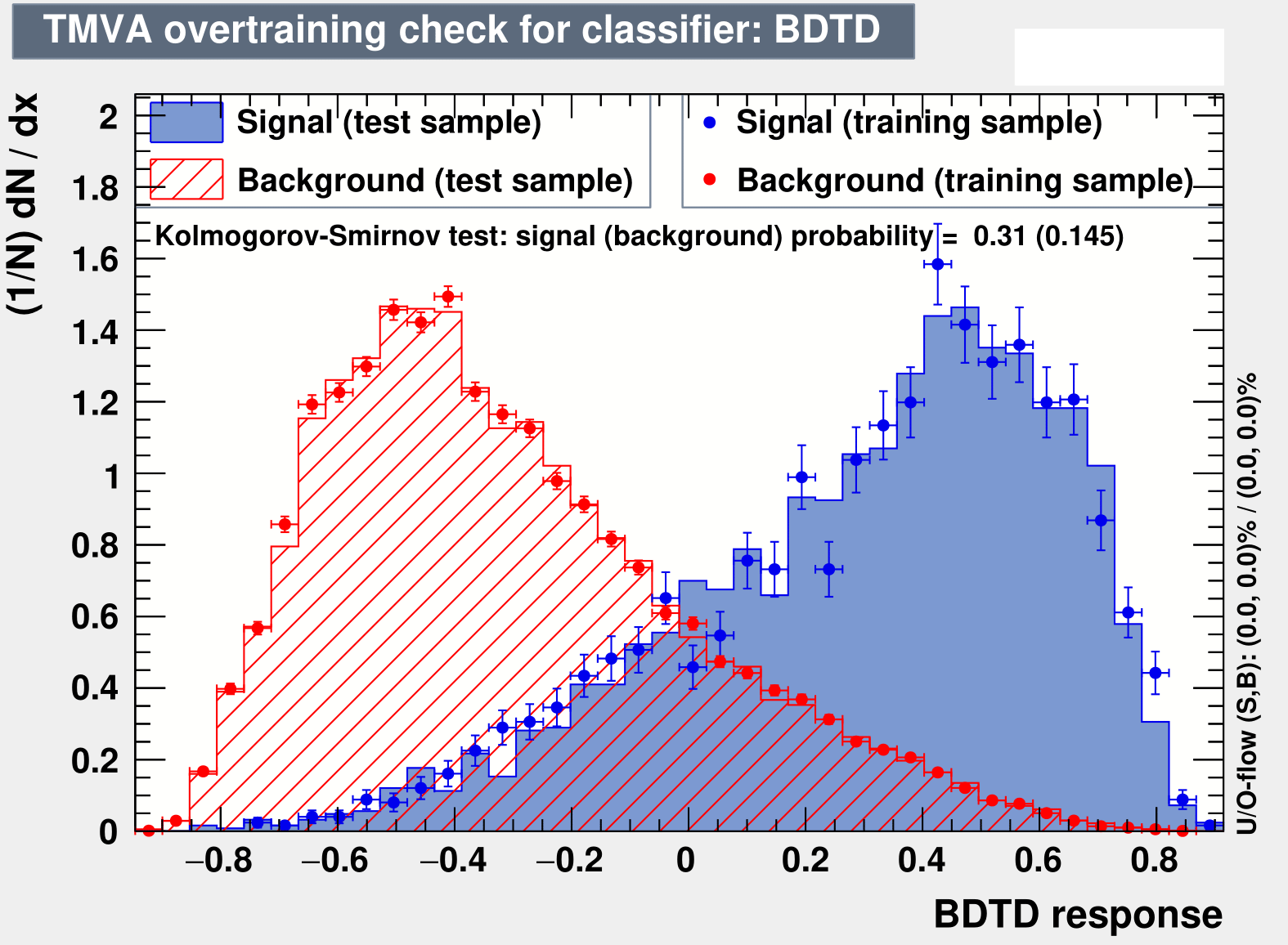}}
\subfigure[]{
\includegraphics[width=3in,height=2.55in]{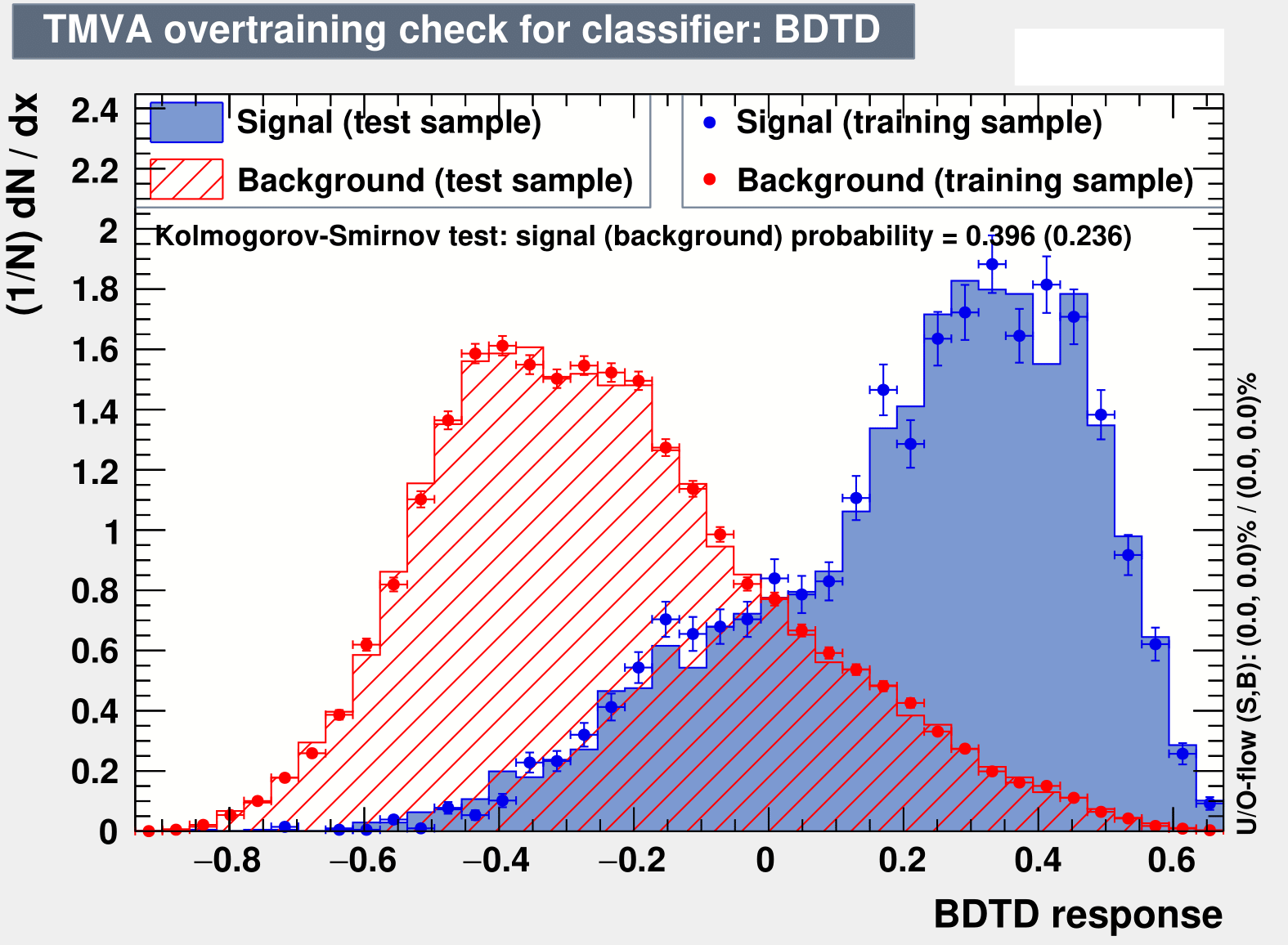}} \\
\subfigure[]{
\includegraphics[width=3in,height=2.55in]{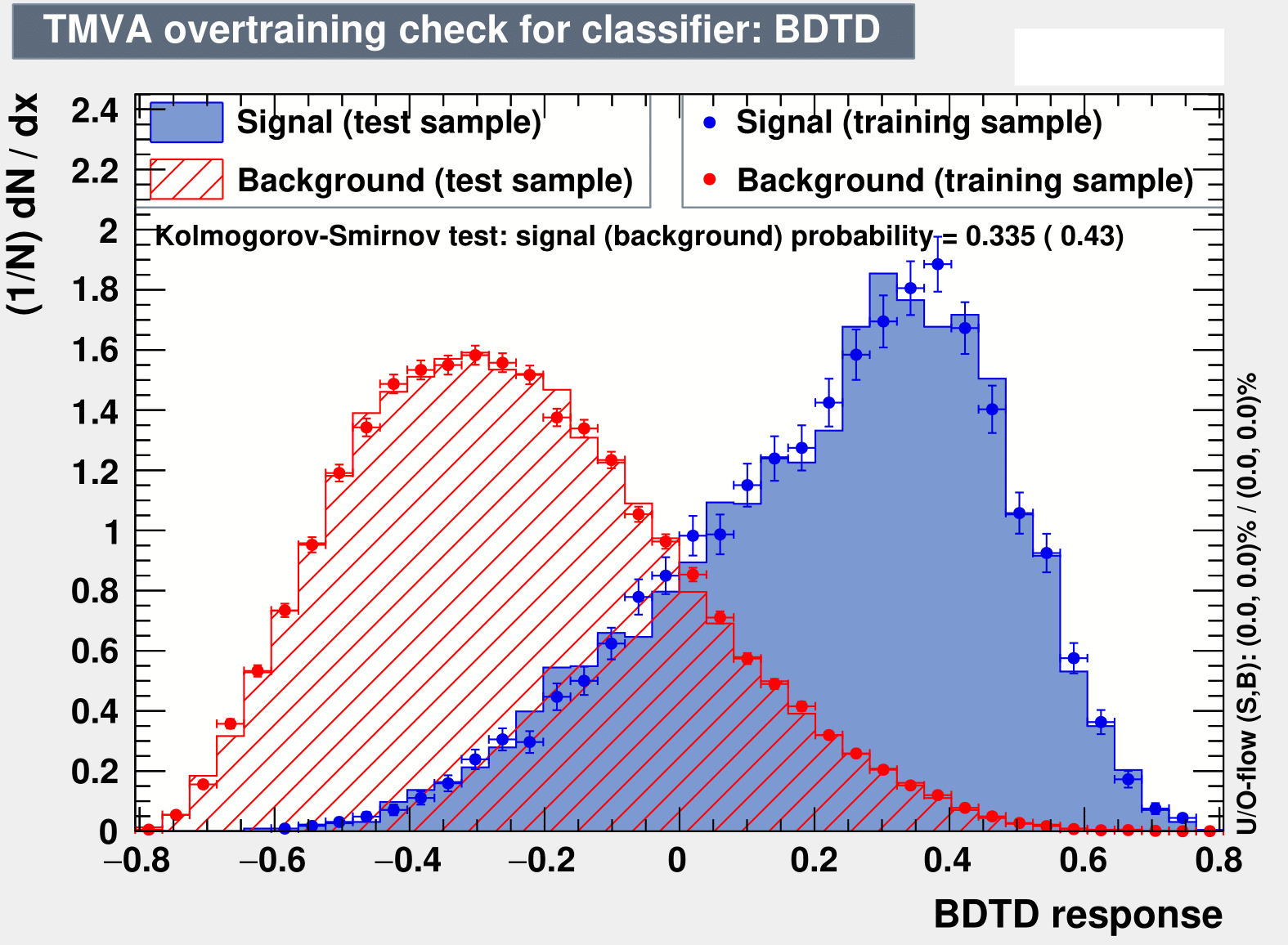}}
\subfigure[]{
\includegraphics[width=3in,height=2.55in]{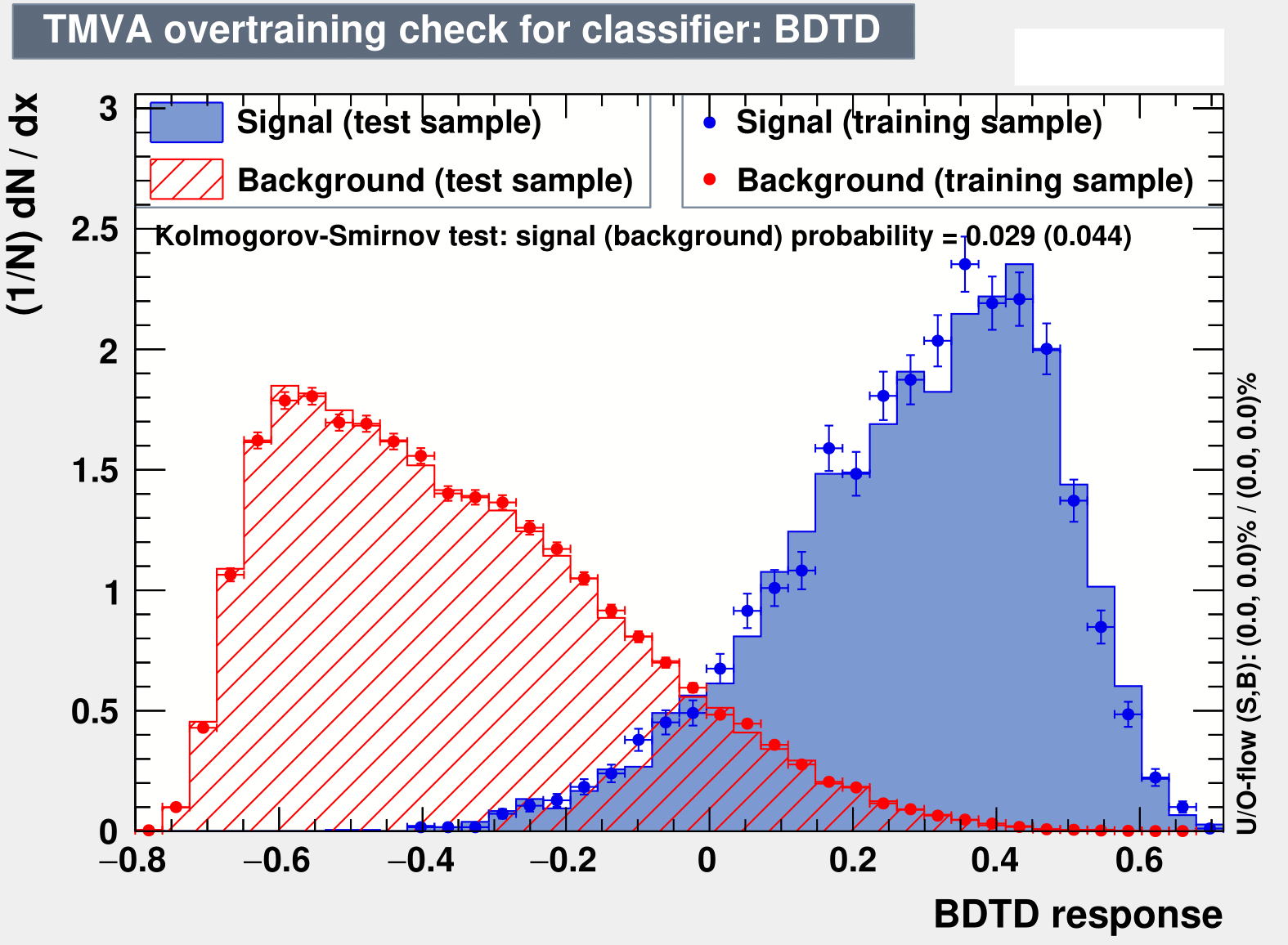}}}
\caption{ KS-scores corresponding to BP1, BP2, BP3 and BP4 for $2 b +   1l + 2 j + \slashed{E_T}$ channel.}
\label{KSscore-2b1l2jv}
\end{figure}
\begin{figure}[tp!]{\centering
\subfigure[]{
\includegraphics[width=2.8in,height=2.55in]{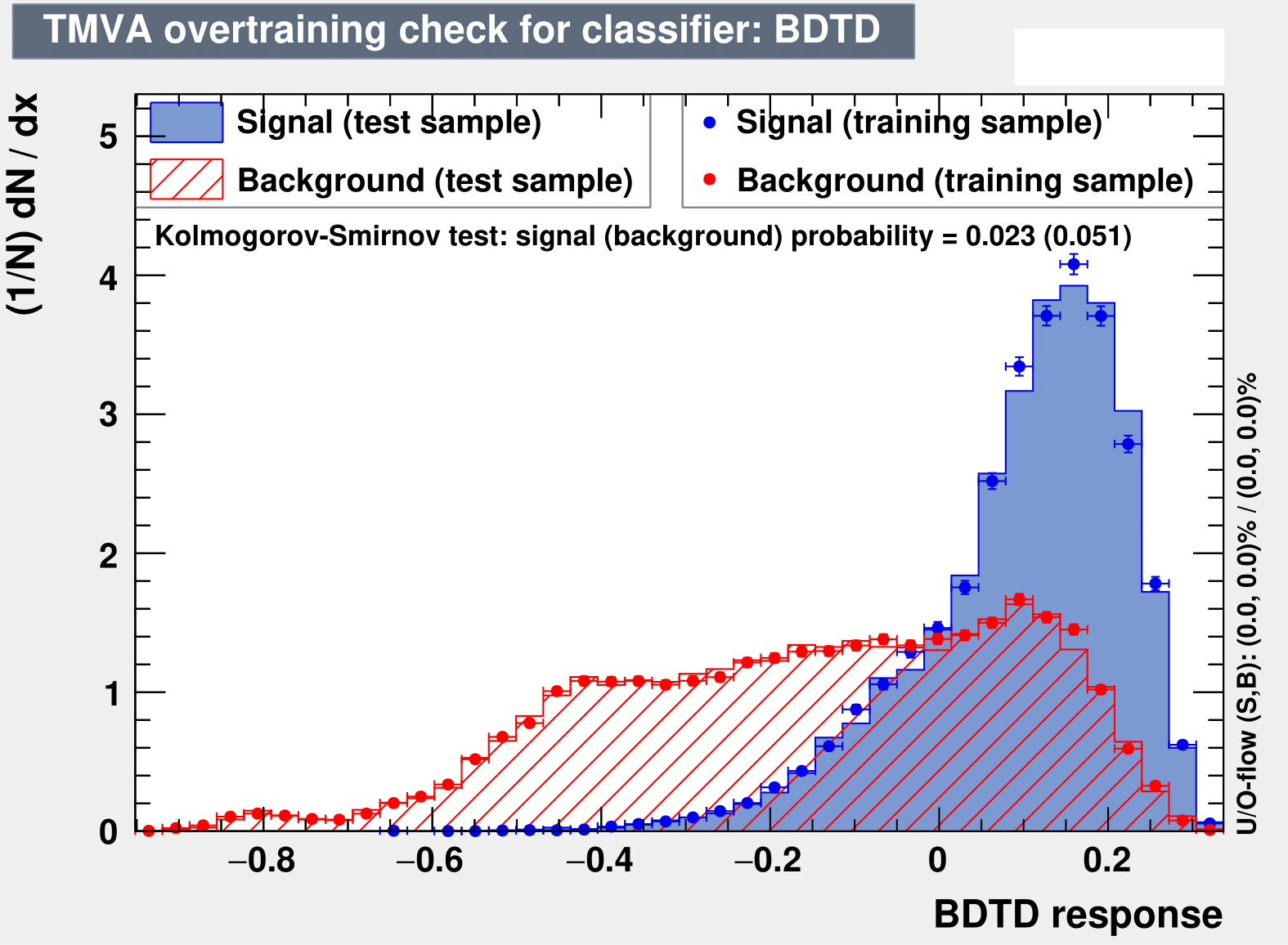}}
\subfigure[]{
\includegraphics[width=3in,height=2.55in]{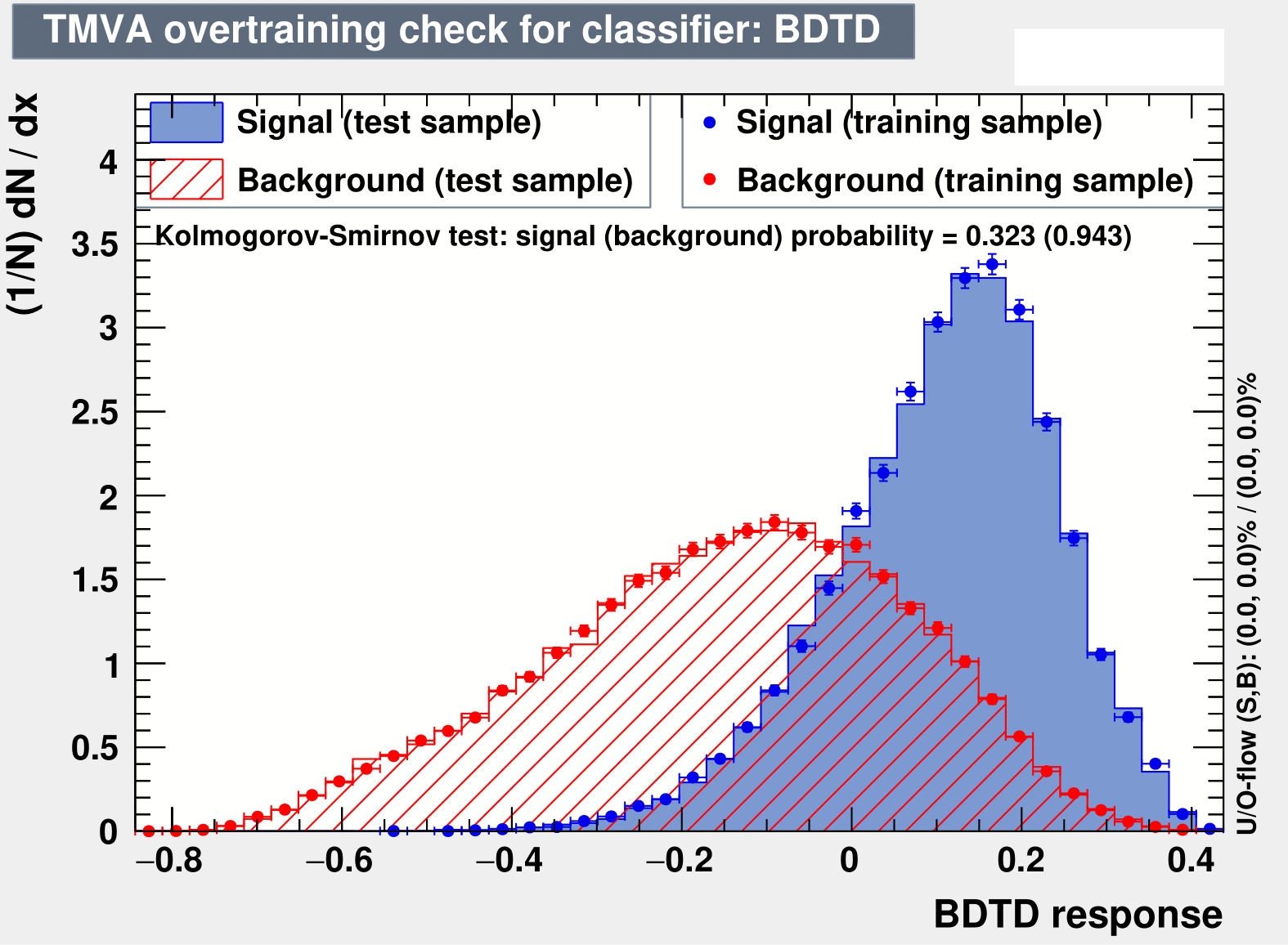}} \\
\subfigure[]{
\includegraphics[width=3in,height=2.55in]{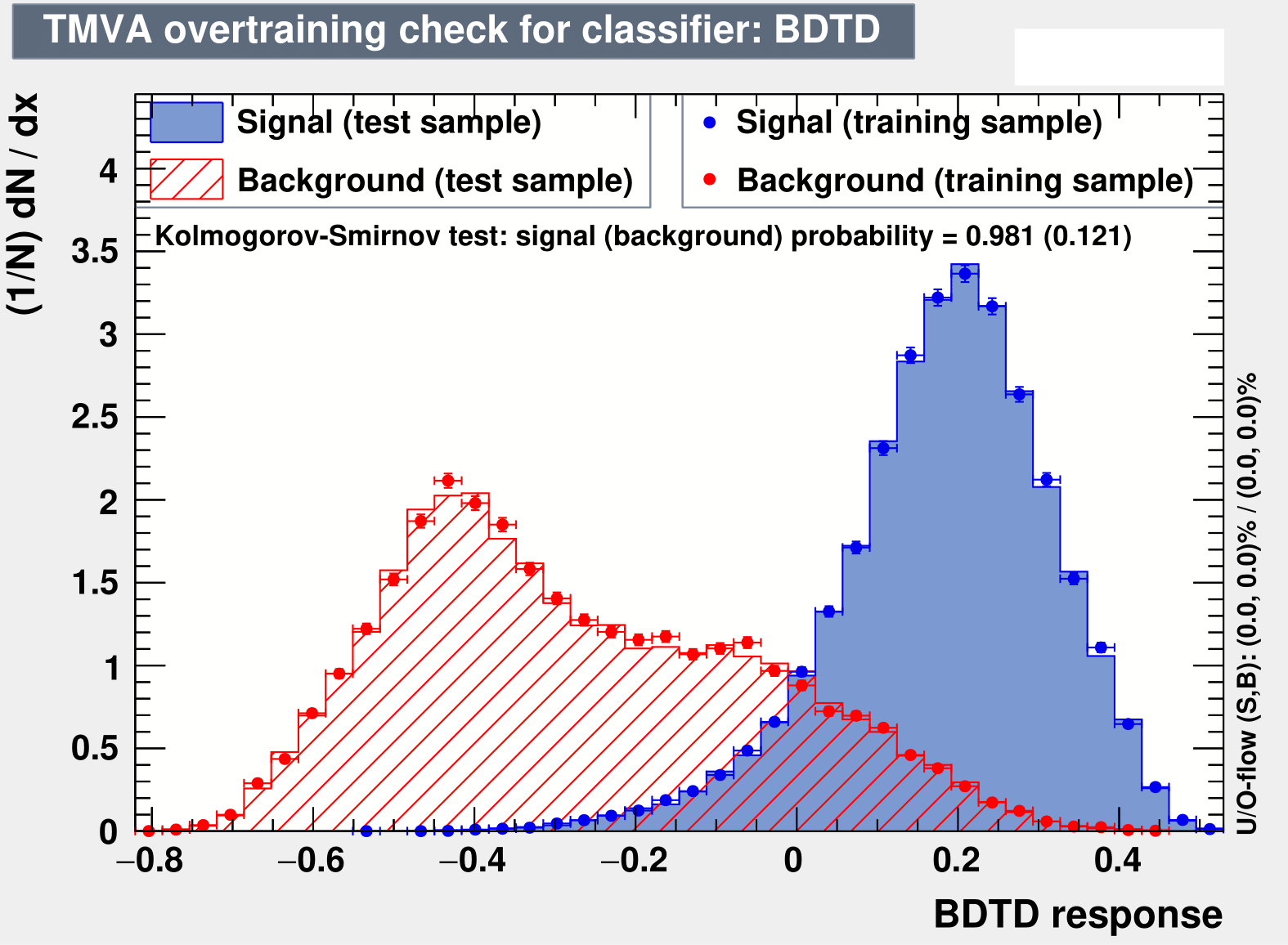}}}
\caption{ KS-scores corresponding to BP1, BP2 and BP3 for $5 l  + \slashed{E_T}$ channel.}
\label{KSscore-5lv}
\end{figure}
\begin{figure}[tp!]{\centering
\subfigure[]{
\includegraphics[width=2.8in,height=2.55in]{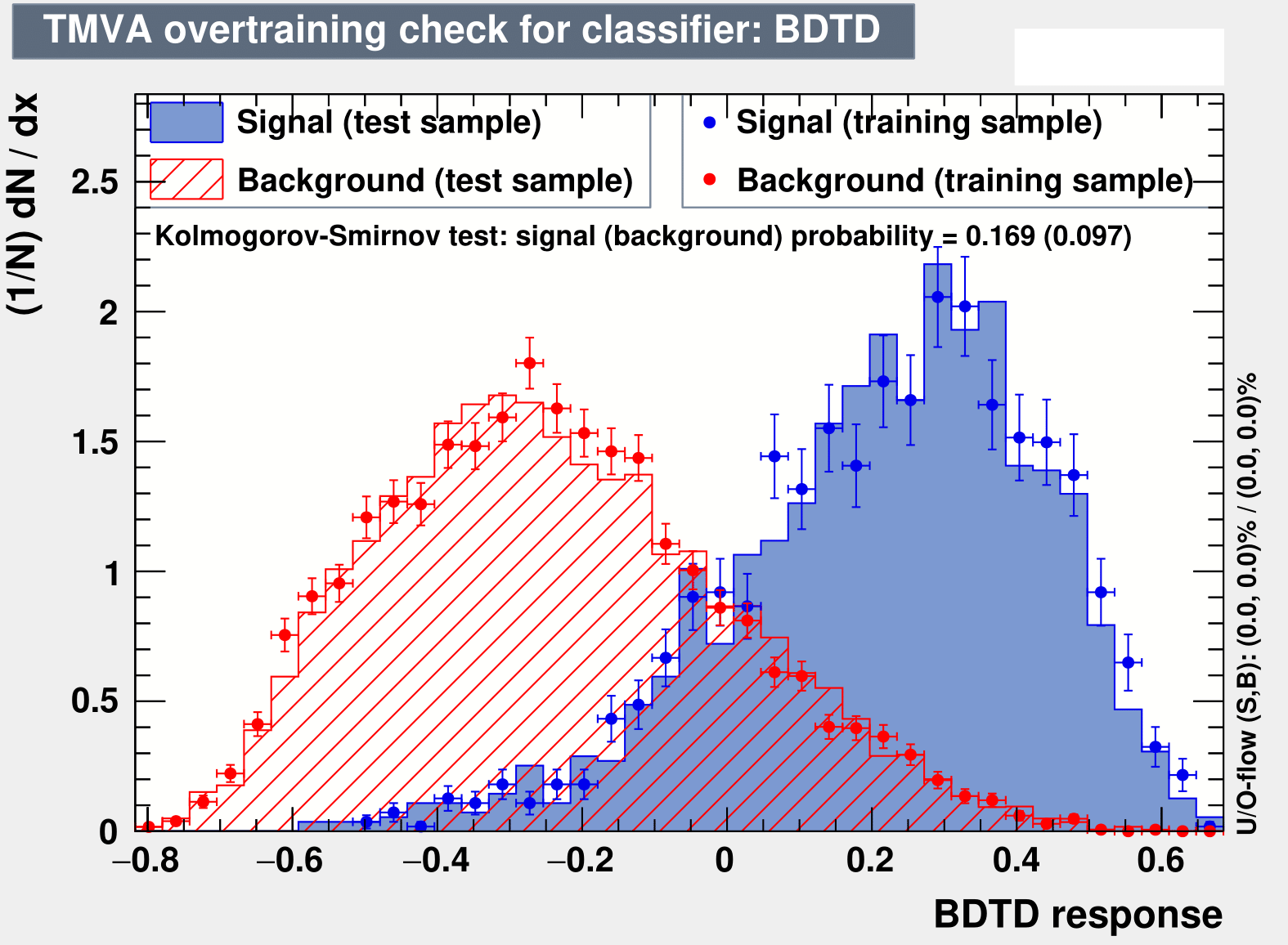}}
\subfigure[]{
\includegraphics[width=3in,height=2.55in]{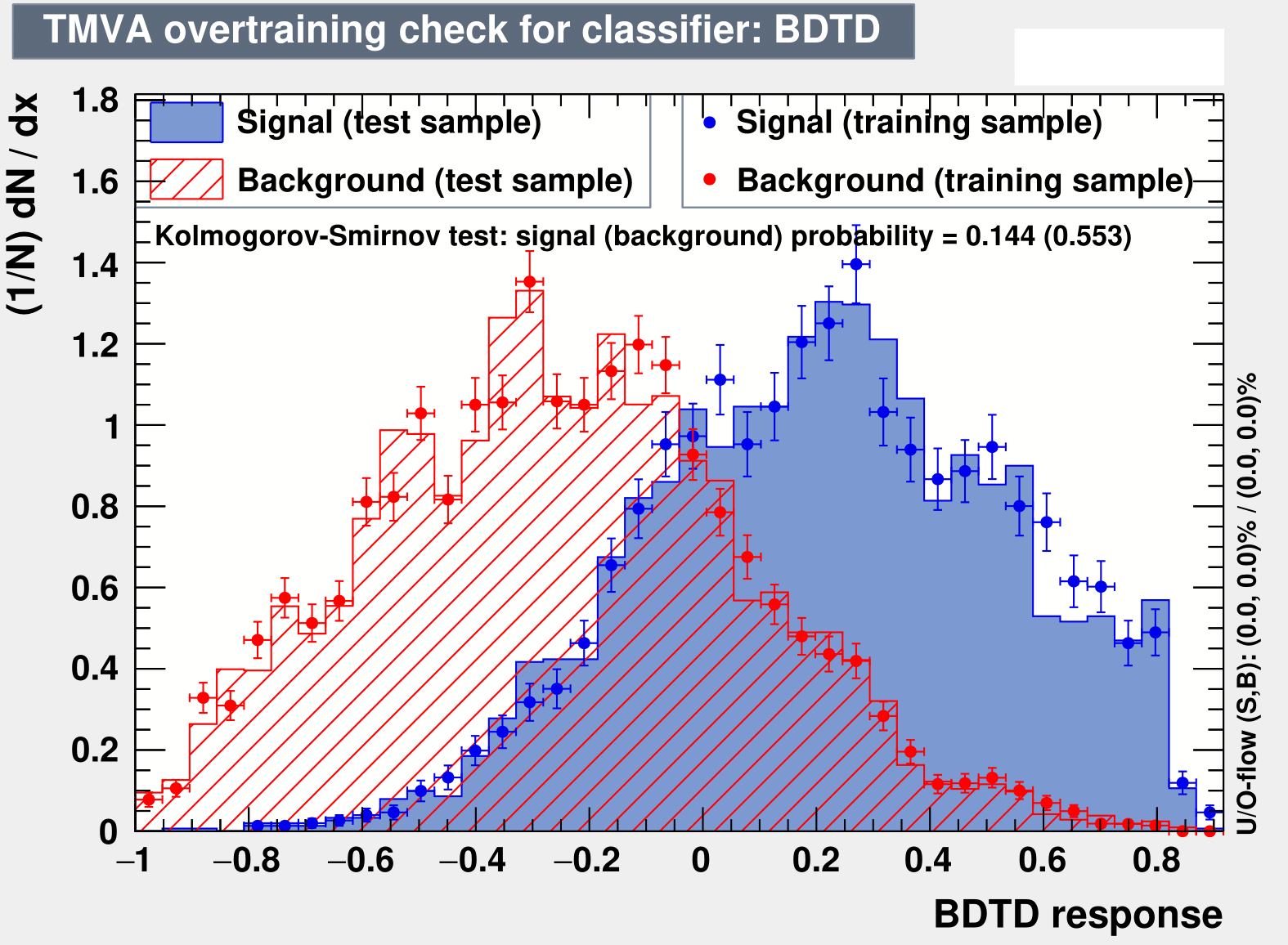}} \\
\subfigure[]{
\includegraphics[width=3in,height=2.55in]{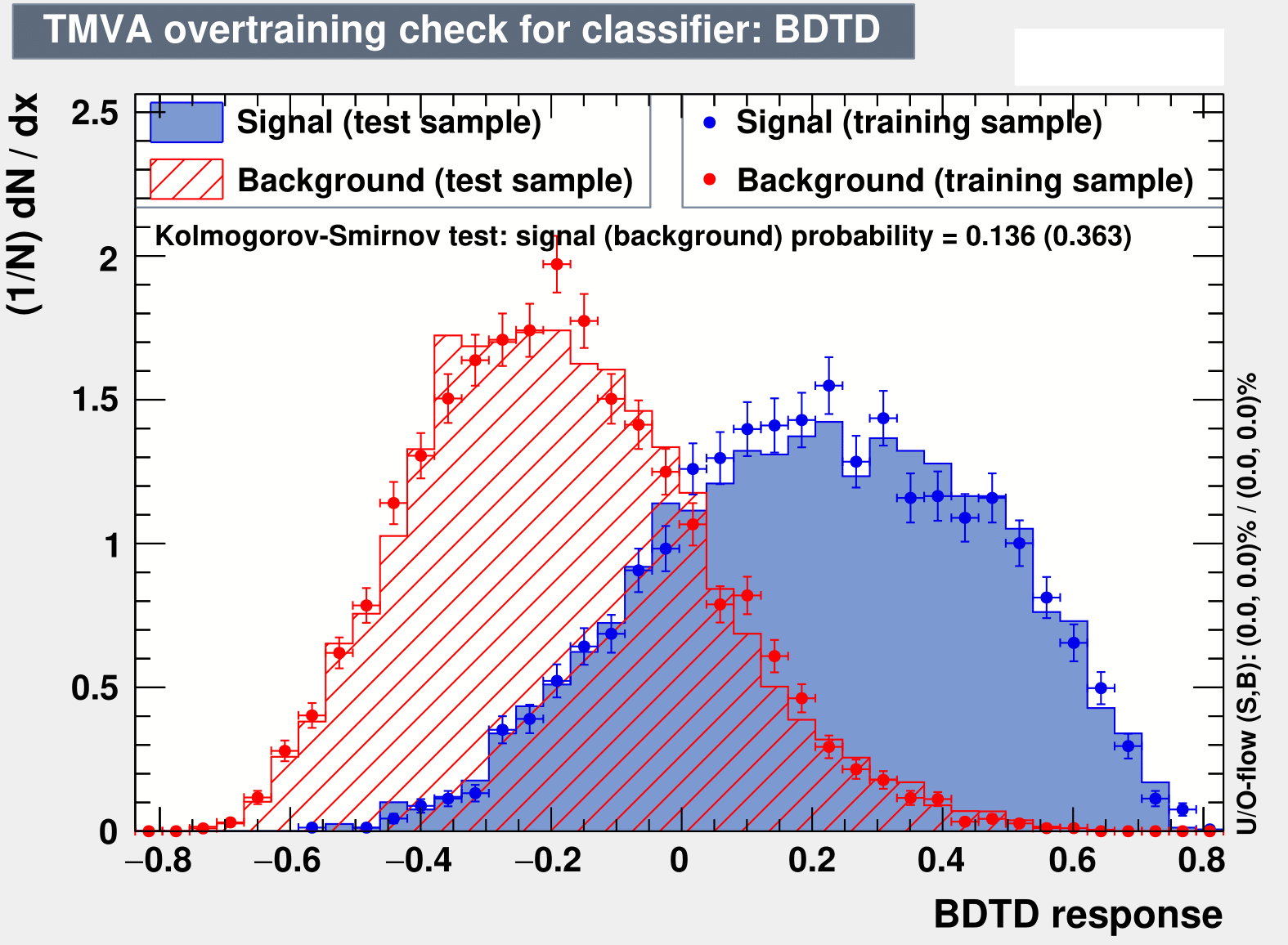}} 
\subfigure[]{
\includegraphics[width=3in,height=2.55in]{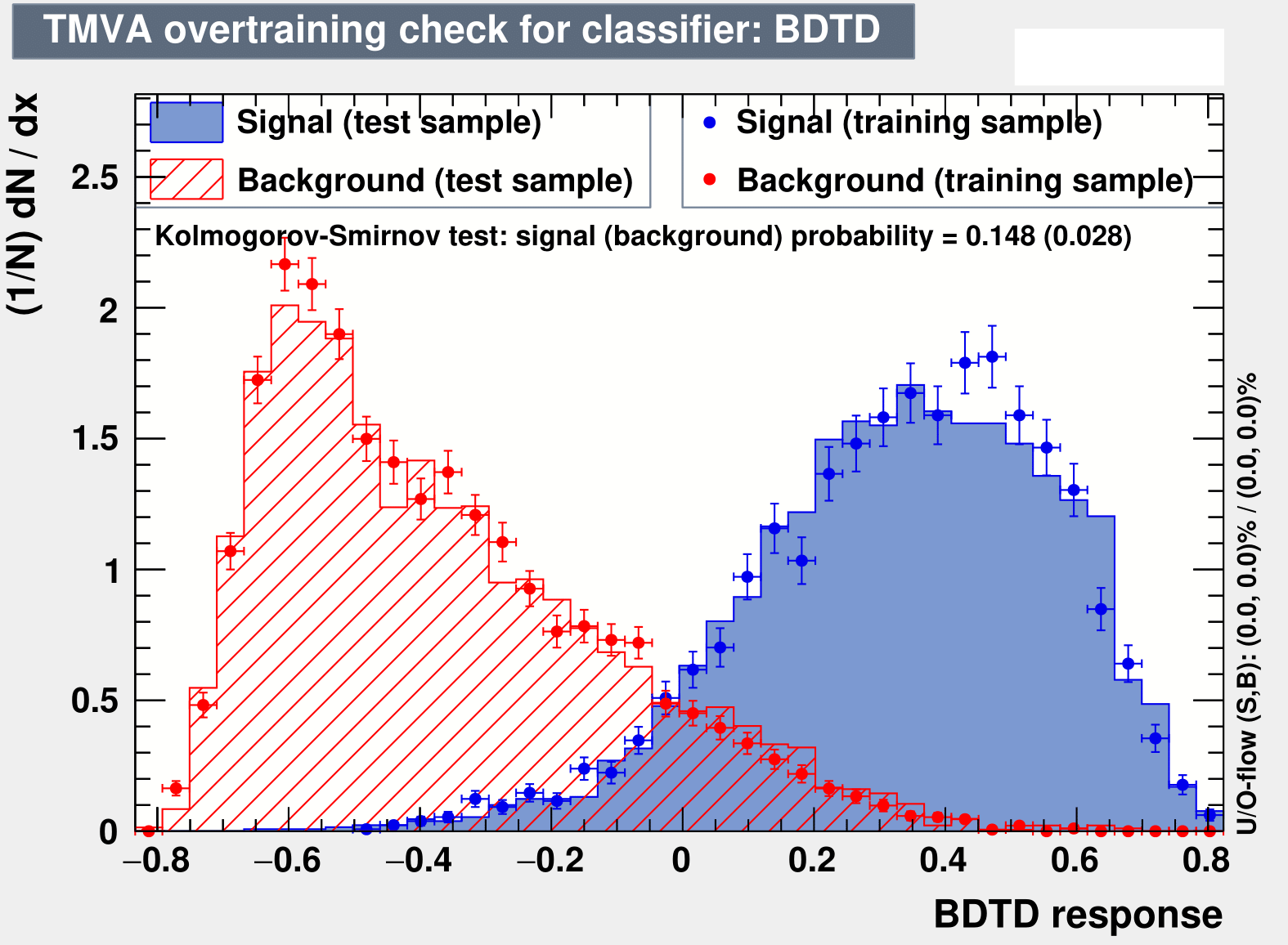}}}
\caption{ KS-scores corresponding to BP1, BP2 and BP3 for $3 l  + 2b + \slashed{E_T}$ channel.}
\label{KSscore-3l2bv}
\end{figure}

\bibliography{ref} 
\end{document}